\pdfoutput=1
\documentclass[12pt,a4paper]{article}

\usepackage{natbib}

\usepackage{ucs}
\usepackage[utf8x]{inputenc}
\usepackage[T1]{fontenc}

\usepackage{geometry}
\usepackage{fullpage}

\usepackage{xspace}
\usepackage{enumerate}

\usepackage{amsmath}
\usepackage{amsthm}
\usepackage{stmaryrd} \usepackage{stix} 

\usepackage{tikz}  \usetikzlibrary{intersections}
\usetikzlibrary{calc}
\usetikzlibrary{patterns}
\usetikzlibrary{decorations.markings}
\usetikzlibrary{decorations.pathmorphing}
\usetikzlibrary{decorations.pathreplacing}

\usepackage{subcaption}

\usepackage{h_general}
\usepackage{h_vocabulary}
\usepackage{h_graphics}

\usepackage{Add}
\usepackage{comment}
\newcommand{\TilePolice}[1]{\textsf{#1}\xspace}

\newcommand{\GluePolice}[1]{\textsf{#1}\xspace}

\newcommand{\Tile}[3][last tile]{\draw (#2) node[inner sep=0cm,minimum size=.8\unitlength,draw] (#1) {\TilePolice{#3}} ;
}

\newcommand{\LinkTile}[5]{\draw (#4) --
  	node[sloped,pos=.15,above,inner sep=0.1em] {\tiny #1}
    node[sloped,pos=.85,inner sep=0.1em,above] {\tiny #2}
    node[sloped,above,inner sep=0.2em] {\scriptsize \GluePolice{#3}}
    (#5) ;}

\newcommand{\LinkBotTop}[3][]{\LinkTile{\South}{\North}{#1}{#2}{#3}}

\newcommand{\LinkLeRi}[3][]{\LinkTile{\West}{\East}{#1}{#2}{#3}}

\newcommand{\ArcTile}[3]{\draw[->] (#1) --
    node[sloped,above,inner sep=0.2em] {\scriptsize #2}
    (#3) ;}

\newcommand{\ArcSouth}[2]{\ArcTile{#1}{\South}{#2}
}
\newcommand{\ArcEast}[2]{\ArcTile{#1}{\East}{#2}
}
\newcommand{\ArcWest}[2]{\ArcTile{#1}{\West}{#2}
}

\newcommand{\GlueWidth}{.25}

\newcommand{\GlueTop}[2][]{\begin{scope}[shift={(#2)}]
    \begin{scope}[shift={(0,.5)}]
      \draw (0,0) node[fill=LightBrown,inner sep=0cm,minimum width=2*\GlueWidth\unitlength,minimum height=.2\unitlength,draw] {\tiny \GluePolice{#1}} ;
    \end{scope}
  \end{scope}
}

\newcommand{\GlueRi}[2][]{\begin{scope}[shift={(#2)}]
    \begin{scope}[shift={(.5,0)}]
      \draw (0,0) node[fill=LightBrown,inner sep=0cm,minimum height=2*\GlueWidth\unitlength,minimum width=.2\unitlength,draw] {\tiny \GluePolice{#1}} ;
    \end{scope}
  \end{scope}
}

\newcommand{\SEED}[1]{\begin{scope}[shift={(#1)}]
  \newcommand{\Rad}{.325}
  \draw[ultra thick,densely dotted] (\Rad,\Rad) -- (\Rad,-\Rad) -- (-\Rad,-\Rad) -- (-\Rad,\Rad) -- cycle;
  \end{scope}
}

\newcommand{\GlueLe}[2][]{\begin{scope}[shift={(#2)}]
    \begin{scope}[shift={(-1,0)}]
      \GlueRi[#1]{0,0}
    \end{scope}
  \end{scope}
}

\newcommand{\GlueBot}[2][]{\begin{scope}[shift={(#2)}]
    \begin{scope}[shift={(0,-1)}]
      \GlueTop[#1]{0,0}
    \end{scope}
  \end{scope}
}

\newcommand{\TasAutom}[2]{
  \begin{tikzpicture}
    \newcommand{\ControlEW}{.5,.75}
    \newcommand{\ControlLoopS}{1.5,-1.5}
    \Tile[A]{0,2}{A}
    \Tile[S]{0,4}{\Seed}
    \SEED{0,4}
    \ArcSouth{S}{A}
    \Tile[Bl]{-2,2}{B}
    \ArcWest{A}{Bl}
    \Tile[Al]{-4,2}{A}
    \ArcWest{Bl}{Al}
    \draw[->] (Al) .. controls +(\ControlEW) and +(-\ControlEW) .. 
    node[above,inner sep=0.2em] {\scriptsize \West} (Bl) ;
    \Tile[Cl]{-4,0}{C}
    \ArcSouth{Al}{Cl}
    #1\draw[->] (Cl) .. controls +(-\ControlLoopS) and +(\ControlLoopS) .. 
    node[above,inner sep=0.2em] {\scriptsize \South} (Cl) ;
    \Tile[Br]{2,2}{B}
    \ArcEast{A}{Br}
    \Tile[Ar]{4,2}{A}
    \ArcEast{Br}{Ar}
    \draw[->] (Ar) .. controls +(-\ControlEW) and +(\ControlEW) ..
    node[above,inner sep=0.2em] {\scriptsize \East} (Br) ;
    \Tile[D]{2,0}{D}
    \ArcSouth{Br}{D}
    #2\draw[->] (D) .. controls +(-\ControlLoopS) and +(\ControlLoopS) .. 
    node[above,inner sep=0.2em] {\scriptsize \South} (D) ;
  \end{tikzpicture}
}

\pgfdeclarepatternformonly{NE lines}{\pgfqpoint{-1pt}{-1pt}}{\pgfqpoint{8pt}{8pt}}{\pgfqpoint{6pt}{6pt}}{
  \pgfsetlinewidth{0.4pt}
  \pgfpathmoveto{\pgfqpoint{0pt}{0pt}}
  \pgfpathlineto{\pgfqpoint{6.2pt}{6.2pt}}
  \pgfusepath{stroke}
}

\pgfdeclarepatternformonly{NW lines}{\pgfqpoint{-1pt}{-1pt}}{\pgfqpoint{8pt}{8pt}}{\pgfqpoint{6pt}{6pt}}{
  \pgfsetlinewidth{0.4pt}
  \pgfpathmoveto{\pgfqpoint{0pt}{6pt}}
  \pgfpathlineto{\pgfqpoint{6.2pt}{-0.2pt}}
  \pgfusepath{stroke}
}

\newcommand{\Ea}[1]{-- ++ (#1,0) 
}
\newcommand{\No}[1]{-- ++ (0,#1) 
}
\newcommand{\We}[1]{-- ++ (-#1,0)
}
\newcommand{\So}[1]{-- ++ (0,-#1)
}

\makeatletter
\def\IfNextChar#1#2#3#4{\ifx#1#4\@empty\@empty
  \expandafter\@firstoftwo
  \else
  \expandafter\@secondoftwo
  \fi
  {#2}{#3}}
\makeatother

\def\IfPathWest#1{\IfNextChar{W}{ - - ++ (-1,0)\IfPathSouth}{}{#1}}\def\IfPathEast#1{\IfNextChar{E}{ - - ++ (1,0)\IfPathSouth}{\IfPathWest#1
  }{#1}}\def\IfPathNorth#1{\IfNextChar{N}{ - - ++ (0,1)\IfPathSouth}{\IfPathEast#1
  }{#1}}\def\IfPathSouth#1{\IfNextChar{S}{ - - ++ (0,-1)\IfPathSouth}{\IfPathNorth #1
  }{#1}}
\newcommand{\PreparePath}[2]{\expandafter\edef\csname Path#1\endcsname{\IfPathSouth #2.}
}

\newcommand{\ProtoTooth}[5][7]{\PreparePath{CombPTm}{#3}
  \PreparePath{CombPTb}{#4}
  \draw[-] (C1) ++(#2) \PathCombPTm \CoorNode{C1-1} \CoorNode{D} ;
  \begin{scope}[struct-z2]
    \foreach \i in {1,2,...,#1} {
      #5
      \draw (C1-1) \PathCombPTb \CoorNode{C1-1} \CoorNode{D} ;
    };    
  \end{scope}
}
\newcommand{\CombBB}[5][7]{\PreparePath{CombM}{#3}
  \PreparePath{CombP}{#4}
  \draw[struct-z0] (0,0) ++(#2) \PathCombM \CoorNode{C1} \CoorNode{D} ;
  \begin{scope}[struct-z1]
    \foreach \i in {1,2,...,#1} {
      #5
      \draw (C1) \PathCombP \CoorNode{C1} \CoorNode{D};
    };
  \end{scope}
}
\newcommand{\Deco}[2]{
  \PreparePath{Deco}{#2}
  \draw[-] (D) ++(#1)  \PathDeco ;
}

\newcommand{\TransientPathStyleAdd}[1]{\tikzstyle{TransientPathStyleAddStyle}=[#1]}
\TransientPathStyleAdd{}
\newcommand{\TransientPath}[4][above]{\draw[transient,TransientPathStyleAddStyle] (#2) -- node[#1] {$#3$} (#4) ;}

\newcommand{\PeriodicPathStyleAdd}[1]{\tikzstyle{PeriodicPathStyleAddStyle}=[#1]}
\PeriodicPathStyleAdd{}
\newcommand{\PeriodicPath}[4][above]{\draw[periodic,PeriodicPathStyleAddStyle] (#2) -- node[#1] {$#3$} (#4) ;}
\newcommand{\PeriodicPathRight}[4][right]{\draw[periodic,PeriodicPathStyleAddStyle] (#2) -- (#4) node[#1] {$#3$} ;}

\newcommand{\TileDot}[3][]{\begin{scope}[shift={(#2)}]\path (0,0) \CoorNode{#1} ; \draw (0,0) node [rectangle,fill,inner sep=.1\unitlength] (#1) {} ;\end{scope}}

\JDLvocabulary{\RealSet}{\JDLvocabularyMathBBXspace{R}}{}{Set of all real numbers}
\JDLvocabulary{\RationalSet}{\JDLvocabularyMathBBXspace{Q}}{}{Set of all rational numbers}
\JDLvocabulary{\IntegerSet}{\JDLvocabularyMathBBXspace{Z}}{}{Set of all integers}
\JDLvocabulary{\NaturalSet}{\JDLvocabularyMathBBXspace{N}}{}{Set of all natural numbers}
\providecommand{\IntegerInterval}[2]{\JDLvocabularyMathXspace{{\llbracket}#1,#2{\rrbracket}}}

\JDLvocabulary{\DomainFunction}{\JDLvocabularyMathXspace{\mathsf{dom}}}{}{Domain of a path}
\newcommand{\Domain}[1]{\JDLvocabularyMathXspace{\DomainFunction(#1)}}

\JDLvocabulary{\BoundaryFunction}{\JDLvocabularyMathXspace{\mathsf{boundary}}}{}{Boundary of a sub set of $\IntegerSet^2$}

\JDLvocabulary{\InsideFunction}{\JDLvocabularyMathXspace{\mathsf{inside}}}{}{Inside of a sub set of $\IntegerSet^2$}

\JDLvocabulary{\SubGraphOf}{\JDLvocabularyMathXspace{\sqsubseteq}}{}{Sub-graph inclusion (vertices and edges)}
\JDLvocabulary{\SubGraphCup}{\JDLvocabularyMathXspace{\sqcup}}{}{Union of sub-graphs (vertices and edges)}

\JDLvocabulary{\CupDis}{\ensuremath{\uplus}}{}{Disjoint union}
\JDLvocabulary{\CupSym}{\ensuremath{\triangle}}{}{Symmetric difference}

\newcommand{\Vect}[1]{\JDLvocabularyMathXspace{\overrightarrow{#1}}}
\newcommand{\Backward}[1]{\JDLvocabularyMathXspace{\overline{#1}}}
\newcommand{\Reverse}[1]{\JDLvocabularyMathXspace{{{#1}^R}}}

\newcommand{\RevBack}[1]{\begin{tikzpicture}\path (0,0) node[inner sep=0em,text depth=0] (RB) {\ensuremath{#1}} ;
     \draw[<<<-] ([yshift=.2em]RB.north west) -- ([yshift=.2em]RB.north east) ;
  \end{tikzpicture}\xspace}

\JDLvocabulary{\ZZOrigin}{\JDLvocabularyTextXspace{{\rm\textbf{o}}}}{}{The origin of $\IntegerSet^2$, $=(0,0)$}

\JDLvocabulary{\East}{\JDLvocabularyTextXspace{{\rm\textbf{e}}}}{}{Vector for East, $=(1,0)$}
\JDLvocabulary{\North}{\JDLvocabularyTextXspace{{\rm\textbf{n}}}}{}{Vector for North, $=(0,1)$}
\JDLvocabulary{\South}{\JDLvocabularyTextXspace{{\rm\textbf{s}}}}{}{Vector for South, $=(0,-1)$}
\JDLvocabulary{\West}{\JDLvocabularyTextXspace{{\rm\textbf{w}}}}{}{Vector for West, $=(-1,0)$}

\JDLvocabulary{\EmptyWord}{\JDLvocabularyMathXspace{\varepsilon}}{}{Empty word}

\JDLvocabulary{\RotateFunction}{\JDLvocabularyMathXspace{\mathsf{rot}}}{}{Rotations of a finite free path}
\newcommand{\Rotate}[1]{\JDLvocabularyMathXspace{\RotateFunction(#1)}}

\JDLvocabulary{\TAS}{\JDLvocabularyMathCalXspace{T}}{}{Tile Assembly System}
\JDLvocabulary{\GlueSet}{\JDLvocabularyMathXspace{\Sigma}}{}{Set of glues}
\JDLvocabulary{\TileSet}{\JDLvocabularyMathXspace{T}}{}{Set of tile types}
\JDLvocabulary{\Assembly}{\JDLvocabularyMathXspace{\alpha}}{}{Some assembly of \TAS}
\JDLvocabulary{\AssemblyOther}{\JDLvocabularyMathXspace{\beta}}{}{Some other assembly of \TAS}
\JDLvocabulary{\Path}{\JDLvocabularyMathXspace{\pi}}{}{Some path}

\JDLvocabulary{\DirectionSet}{\JDLvocabularyMathXspace{D}}{}{The set of directions, $\{\East,\North,\South,\West\}$.}

\JDLvocabulary{\IMinPath}{\JDLvocabularyMathXspace{p}}{}{Any Infinite minimal path in the assembly starting on the seed}
\JDLvocabulary{\IMinPathSuf}{\JDLvocabularyMathXspace{\hat{p}}}{}{1st Suffix of $p$ where all glues and tiles used appear infinitely often}

\JDLvocabulary{\VIMinPath}{\JDLvocabularyMathXspace{V_{\IMinPath}}}{}{Tiles infinitely used by an infinite path p}
\JDLvocabulary{\EIMinPath}{\JDLvocabularyMathXspace{E_{\IMinPath}}}{}{Oriented glues infinitely used by an infinite path p}
\JDLvocabulary{\GIMinPath}{\JDLvocabularyMathXspace{G_{\IMinPath}}}{}{Oriented graph corresponding to the infinitely repeated part of an infinite path p}

\JDLvocabulary{\VIMinPathSuf}{\JDLvocabularyMathXspace{V_{\IMinPathSuf}}}{}{Tiles infinitely used by an infinite path p}
\JDLvocabulary{\EIMinPathSuf}{\JDLvocabularyMathXspace{E_{\IMinPathSuf}}}{}{Oriented glues infinitely used by an infinite path p}
\JDLvocabulary{\GIMinPathSuf}{\JDLvocabularyMathXspace{G_{\IMinPathSuf}}}{}{Oriented graph corresponding to the infinitely repeated part of an infinite path p}

\JDLvocabulary{\IMinPathCycle}{\JDLvocabularyMathXspace{c}}{}{Cycle infinitely repeated in p}

\JDLvocabulary{\NonCausalFunction}{\JDLvocabularyMathXspace{\mathsf{non\_causal}}}{}{Non-causal vertices for a vertex in $\IntegerSet^2$}
\newcommand{\NonCausal}[1]{\JDLvocabularyMathXspace{\NonCausalFunction(#1)}}

\JDLvocabulary{\CoGrowFunction}{\JDLvocabularyMathXspace{\mathsf{coGrow}}}{}{Co-growth of a path inside regions}
\newcommand{\CoGrow}[4]{\JDLvocabularyMathXspace{\CoGrowFunction({#1},{#2},{#3},{#4})}}

\JDLvocabulary{\CoGrowStart}{\JDLvocabularyMathXspace{\boxtimes}}{}{Co-growth starting tile}

\JDLvocabulary{\GridFunction}{\JDLvocabularyMathXspace{\mathsf{grid}}}{}{Subgraph of $\IntegerSet^2$ forming a grid}
\newcommand{\Grid}[3]{\JDLvocabularyMathXspace{\GridFunction(#1,#2,#3)}}

\JDLvocabulary{\CombFunction}{\JDLvocabularyMathXspace{\mathsf{comb}}}{}{Subgraph of $\IntegerSet^2$ forming a comb}
\newcommand{\Comb}[3]{\JDLvocabularyMathXspace{\CombFunction(#1,#2,#3)}}

\JDLvocabulary{\AlphaMax}{\JDLvocabularyMathXspace{\alpha_{max}}}{}{The unique maximal assembly of \TAS}
\JDLvocabulary{\Seed}{\JDLvocabularyMathXspace{\mathsf{\sigma}}}{}{The seed tile type}

\JDLvocabulary{\AnyTile}{\JDLvocabularyMathXspace{\mathsf{t}}}{}{Any tile type in $T$}

\JDLvocabulary{\LineSEast}{\JDLvocabularyMathXspace{\ZZOrigin\East}}{}{The (one-side) infinite line $(0,0).\East^{\omega}$}
\JDLvocabulary{\LineSWest}{\JDLvocabularyMathXspace{\ZZOrigin\West}}{}{The (one-side) infinite line $(0,0).\West^{\omega}$}

\JDLvocabulary{\PastFunction}{\JDLvocabularyMathXspace{\mathsf{past}}}{}{Past on a path from a vertex}
\newcommand{\PastFrom}[1]{\JDLvocabularyMathXspace{\PastFunction(#1)}}
\JDLvocabulary{\NearFutureFunction}{\JDLvocabularyMathXspace{\mathsf{future}}}{}{Future on a path from a vertex}
\newcommand{\NearFutureFrom}[1]{\JDLvocabularyMathXspace{\NearFutureFunction(#1)}}

\JDLvocabulary{\Quipu}{\JDLvocabularyMathXspace{Q}}{}{Some quipu}
\JDLvocabulary{\QuipuZero}{\JDLvocabularyMathXspace{\Quipu_{0}}}{}{Initial quipu for a filtration}
\newcommand{\QuipuI}{\JDLvocabularyMathXspace{\Quipu_{i}}}
\newcommand{\QuipuIpO}{\JDLvocabularyMathXspace{\Quipu_{i+1}}}
\JDLvocabulary{\QuipuOther}{\JDLvocabularyMathXspace{\Quipu'}}{}{Some other quipu}
\JDLvocabulary{\QuipuLim}{\JDLvocabularyMathXspace{\Quipu_{\infty}}}{}{Infinite tree-quipu in the limit}
\JDLvocabulary{\Root}{\JDLvocabularyMathXspace{r}}{}{Root vertex of a quipu}
\JDLvocabulary{\VertexLabel}{\JDLvocabularyMathXspace{\eta}}{}{Vertex labelling with tile types of a quipu}
\JDLvocabulary{\ArcLabel}{\JDLvocabularyMathXspace{\lambda}}{}{Arc labelling with directions for a quipu}
\JDLvocabulary{\AlphaQuipu}{\JDLvocabularyMathXspace{\alpha_{\Quipu}}}{}{Assembly generated by a quipu \Quipu}
\JDLvocabulary{\AlphaQuipuOther}{\JDLvocabularyMathXspace{\alpha_{\QuipuOther}}}{}{Assembly generated by a quipu \QuipuOther}
\JDLvocabulary{\AlphaQuipuZero}{\JDLvocabularyMathXspace{\alpha_{\QuipuZero}}}{}{Assembly generated by \QuipuZero}
\JDLvocabulary{\AlphaQuipuI}{\JDLvocabularyMathXspace{\alpha_{\QuipuI}}}{}{Assembly generated by $\Quipu_{i}$}
\JDLvocabulary{\AlphaQuipuLim}{\JDLvocabularyMathXspace{\alpha_{\QuipuLim}}}{}{Assembly generated by \QuipuLim}

\JDLvocabulary{\PseudoComb}{\JDLvocabularyMathXspace{\mu}}{}{Pseudo-comb}
\JDLvocabulary{\PseudoCombOther}{\JDLvocabularyMathXspace{\mu'}}{}{Pseudo-comb}

\JDLvocabulary{\CoverFunction}{\JDLvocabularyMathXspace{\mathsf{cover}}}{}{Part of $\IntegerSet^2$ covered by a quipu vertex}
\newcommand{\CoverBy}[2][]{\JDLvocabularyMathXspace{\CoverFunction_{#1}(#2)}}

\JDLvocabulary{\ZoneZero}{\JDLvocabularyMathXspace{Z_0}}{}{Finite zone around the seed}
\JDLvocabulary{\ZoneOne}{\JDLvocabularyMathXspace{Z_1}}{}{Backbone zone}
\JDLvocabulary{\ZoneTwo}{\JDLvocabularyMathXspace{Z_2}}{}{Proto-teeth zone}

\JDLvocabulary{\SetCombRepresentative}{\JDLvocabularyMathCalXspace{S}}{}{set of comb representatives}

\begin{document}

\title{Deterministic 2-Dimensional Temperature-1 Tile Assembly Systems Cannot Compute}

\author{J{\'e}r{\^o}me \textsc{Durand-Lose}\footnote{Universit\'e d'Orl\'eans, ENSI de Bourges, LIFO \'EA 4022, F-45067 Orl\'eans, France
    and LIX, UMR 7161, CNRS, Ecole Polytechnique, Palaiseau, France}
  \and Hendrik Jan \textsc{Hoogeboom}
  \footnote{Leiden University, The Netherlands}
  \and Nata\v sa \textsc{Jonoska} \footnote{University of South Florida, Department of Mathematics and Statistics, Tampa FL, 33620, USA}
}

\date{\today}

\maketitle

\begin{abstract}
  We consider non cooperative binding in so called `temperature 1', in  deterministic (here called {\it confluent}) tile self-assembly systems (1-TAS) and prove the  standing conjecture that such systems 
  do not have universal computational power. We call a TAS whose maximal assemblies contain at least one 
  ultimately periodic assembly path {\it para-periodic}.  We observe that a confluent 1-TAS has at most one maximal producible assembly,  \AlphaMax, that can be 
  considered a union of path assemblies, and we show that such a system is always para-periodic. This result is obtained through a superposition and a combination of two paths that produce a new path with desired properties, a technique that we call {\it co-grow} of two paths.  Moreover we provide a characterization of an \AlphaMax of a confluent 1-TAS as one of two possible cases, so called, a grid or a disjoint union of combs. 
  To a given \AlphaMax we can associate a finite labeled graph, called {\it quipu},
  such that the union of all labels of paths in the quipu equals \AlphaMax, therefore giving a finite description for \AlphaMax. This finite description implies that \AlphaMax is a union of semi-affine subsets of $\IntegerSet^2$ and since such a finite description can be algorithmicly generated from any 1-TAS, 1-TAS cannot have universal computational power.
\end{abstract}

\renewcommand{\abstractname}{Keywords}

\begin{abstract}
  Tile assembly system; Directed (confluent) system; Non-cooperation; Ultimately periodic; Para-periodic; Quipu; Universal computation.
\end{abstract}

\section{Introduction}
\label{sec:intro}

The abstract Tile self-assembly model (aTAM) was introduced by Winfree in 1998~\cite{Winf98} as a theoretical model that describes DX DNA self-assembly processes.
The DX molecule can be designed with four sticky ends such that their assembly forms a surface area~\cite{Ferong} as if tiled with square tiles.
Hence motivated by Wang tiles, the abstract tile assembly model is based on square tiles with colored edges (`glues', simulating the DNA sticky ends).
Starting from a seed assembly (or a seed tile) the assembly grows through matching glues tile attachments.
Unlike Wang tiles, the glues have strength and when the matching glues are strong enough, the tiles can attach to the growing structure although there may be mismatched glues on other sides of the tile.
It was observed that two or three weaker matching glues can achieve bonding of the tile similar to the bonding with a higher strength single glue.
The notion of glue strength is captured in the model through a parameter called `temperature'. If all tiles have uniform strength sticky ends allowing attachment, then it is said that the model describes `temperature 1' bonding. When there are two or more strengths on the sticky ends, the temperature can be higher than one, say 2, and in this case at least two weaker glues of a tile have to match with the exposed glues of the tiles in the growing assembly to achieve tile attachment. This is called `cooperative' bonding.
Such a bond cooperation is not needed (although can appear) when tiles have uniform strength on their sticky ends, or when the model runs at `temperature 1'. 

There are several experimental assemblies of DNA-based tile arrays that show that aTAM can carry out computation. These include a binary counter using four DX-based tiles \cite{Evans2015}, binary addition by TX molecules \cite{LaBean2000}, Sierpinski triangle as a pattern on a substrate~\cite{Sierpinski,Fujibayashi2008}, and transducer simulations by TX molecules \cite{Banani}.

In his thesis~\cite{Winf98}, Winfree showed that the abstract tile assembly model at temperature 2 can assemble (simulate) a trace of computation
of any Turing machine, thereby proving that the model has universal computational power.
At the same time it was conjectured that temperature 1 systems have strictly lower computational power. 
In~\cite{rothemund+winfree00stoc}, it was also observed that temperature 2 aTAM can assemble certain structures, such as squares, with much smaller number of tiles compared to temperature 1 systems, indicating difference in the two models.

The theoretical proofs for universal computational power of aTAM in temperature 2, as well as many other observations for structure assemblies rely on so called local determinism in the assembly, in other words, for every two producible assemblies there is a larger assembly that contains both as sub-assemblies. This property is sometimes called `directed assembly' or `determinism of the system' \cite{Patitz-survey}. 
In order to avoid the ambiguity and guided by similar notions in other systems, here this property is called {\it confluence}.

The standing question about the computational power of systems with non-cooperative binding (temperature 1 systems) has initiated several studies of these systems. It has been observed that even small modifications of the model can provide universal computational power. 
For example, by considering three-dimensional tile assemblies that can add another layer of tiles (essentially having one array above the other) it was shown that confluent non-cooperative assembly system has universal computational power~\cite{cook+fu+schweller11soda}. 
It was also observed that by allowing one tile with a repelling glue 
(glue with strength $-1$), the confluent non-cooperative system becomes computationally universal~\cite{negative-glue}. On the other side, if one allows certain tiles to be be added to the system in stages, then again, the system gains universal computing power~\cite{stage-temp1-assembly}.
It was also observed that if 1-TAS tiles are equipped with signals that activate glues stepwise, the system can simulate any temperature 2 TAS~\cite{Daria2015}, that is, including the one that provides intrinsic universality of temperature 2 TAS~\cite{IntrUniver}.
In this simulation, $2\times 2$ tiles in the signal-equipped system correspond to a single tile in the temperature 2 system. 

Infinite ribbon construction (or snake tilings) in non-cooperative binding systems were also given attention~\cite{adleman+kari+kari+reishus02focs,adleman+kari+kari+reishus+sosik09siam,brijder+hoogeboom09tcs,kari02dlt}. It was observed that non-determinism, or non-confluence, also adds power to the system. First it was shown that it is undecidable whether one can obtain an infinite ribbon (snake tiling) with a given non-deterministic set~\cite{adleman+kari+kari+reishus+sosik09siam}.
This was achieved with simulating special type of Wang tiles and a space filling curve. In this case, the notion of 
a `directed' system implies that the design of the tiles is accompanied with arrows that guide the direction of the assembly rather than the system being deterministic (or confluent). 
On the other side, one can use the snake tiles, and the space filling curve, to obtain a non-confluent system that can generate recognizable picture languages~\cite{brijder+hoogeboom09tcs}.
And because recognizable picture languages contain the rectangular shapes that can be obtained from Wang tiles, which are known to have universal computational power~\cite{Wang1975}, we have that non-determinism, i.e., non-confluence of the system and a defined condition on acceptable assembly provide universal computing power.

The limitation of a confluent non-cooperative binding was first observed through so-called `pumpable' paths~\cite{doty+patitz+summers11tcs}. An infinite path is pumpable if there is a segment of the path that can be made into ultimately periodic. 
By assuming that a system can have every infinite path `pumpable', it was observed that only a limited number of structures can be constructed with this system. 
Recently, it was proved that confluent non-cooperative binding cannot simulate a trace of bounded Turing machine computation whose halting appears on the boundary of the computation~\cite{meunier+woods17stoc}.
The paper also shows that such a system cannot be intrinsically universal, that is, there is no
temperature-1 confluent aTAM that can simulate any other such a system.

With the current paper we prove the conjecture that temperature 1 confluent systems have computational power strictly lower than the universal Turing machine.
We attain this result by showing that the terminal maximal assembly of a non-cooperative confluent system can only be one of two possibilities: a grid, or a comb-like (a finite union of finite paths, ultimately periodic paths and periodically-repeated ultimately periodic paths).
Both of these cases have the property that they can be completely determined within a finite region of $\IntegerSet^2$ containing the origin.
Therefore the structure of the maximal assembly can be generated in a finite time,
and for each tile type, the set of vertices in $\IntegerSet^2$ that contain that tile forms
a semi-affine set (it is a union of sets of the form $\Vect{a}+\Vect{b}\NaturalSet+\Vect{c}\NaturalSet$).
Hence, if the system were able to simulate Turing computation, something would not be decidable.

The result is achieved by first proving that every infinite maximal assembly must contain an ultimately periodic path and we call this property {\it para-periodicity}.
We observe that a system is para-periodic if and only if there is an infinite path in the maximal assembly that does not intersect an infinite periodic path in the grid of $\IntegerSet^2$. 
We consider finite portions of paths, which we call \emph{off-the-wall paths}, that are bounded by a line in $\IntegerSet^2$.
These paths provide either an ultimately periodic path or an infinite path that does not intersect an infinite periodic path in the grid of $\IntegerSet^2$. Two main notions are used in these proofs: (a) left and right regions in the plane separated by a bi-infinite path and (b) superposition of two paths, that we call {\it co-grow} to obtain a new path that takes the `right-most' way of the two. 
The co-grow of two paths is possible if they are in a well defined region, and because the system is confluent, the intersection of two paths is always at a vertex that can be associated with only one tile type. 

Ultimately periodic paths in the assembly allow assembly of two structures: grids and combs.
In a \emph{grid}, two non co-linear periodic paths are repeated in the assembly and partition $\IntegerSet^2$ into finite identical `parallelogram' faces.
In a comb, an ultimately periodic path, called a backbone, is a starting point for a periodic sequence of ultimately periodic paths.
Both structures can be decorated with finite paths that, if part of the ultimately periodic parts, are also repeated infinitely often.

We introduce a special automaton called \emph{quipu} that can generate unions of finite paths, ultimately periodic paths and combs. The quipu can be constructed for any given maximal terminal assembly of a confluent non-cooperative system.
Being the system para-periodic, if the maximal terminal assembly has an infinite path finitely intersecting the paths generated by the quipu, 
then there is also an ultimately periodic one that is not generated by the quipu.
At every step of the quipu construction an ultimately periodic path with a minimal length of its period and transient part is added to the quipu in a form of a cycle.
If this addition shows that the assembly is a grid, then the construction stops.
Otherwise, we show that the quipu construction stops in finite time.

The induction process that generates the quipu can be turned into an algorithm because the condition for period length minimality corresponds to a breadth first search and the termination of the process is detected by incorporating all decorations as they are found.

The outline of the paper is as follows.
In \RefSection{sec:definitions}, the model and the basic mathematical tools and notations are introduced. Here we introduce two main concepts, the left and the right region of the plane separated by a bi-infinite path, and the co-grow of two paths.
In \RefSection{sec:nice}, it is proven that any confluent non-cooperative assembly system is para-periodic.
The two main structures that a maximal terminal assembly can attain, grids and combs, are defined in \RefSection{sec:structures}.
This section also contains several lemmas that help in construction of the quipu.
Quipus are defined in \RefSec{sec:quipu} with some basic operations.
\RefSection{sec:quipu-extension} provides various extension steps to incorporate into a quipu finite and ultimately period paths in \AlphaMax.
In \RefSection{sec:quipu-finite}, we prove that there is a quipu that covers \AlphaMax and that this quipu can be generated in finite time.
We end with some concluding remarks in \RefSection{sec:conclusion}.

\section{Definitions}
\label{sec:definitions}

The set of integers from $a$ to $b$ is denoted \IntegerInterval{a}{b} ($a$ and $b$ can be infinite).
The two dimensional integer lattice $\IntegerSet^2$ is considered as a two dimensional grid, a periodic graph whose vertices are the 
elements of $\IntegerSet^2$ and two vertices $x$ and $y$ are connected with an edge if $||x-y||=1$.
A \emph{path} in $\IntegerSet^2$ is a simple path without repetition of the vertices and it can be finite or (bi-)infinite.
A \emph{cycle} is a simple path whose first and last vertex are the same. 
The set of vertices visited by a path \Path is the domain of \Path and is denoted $\Domain{\Path}$.
We denote with $\Path_i$ the $i$th vertex visited by $\Path$ and for $a\le b$ we denote  with $\Path_{\IntegerInterval{a}{b}}$ the segment $\Path_a\Path_{a+1}\cdots\Path_{b}$ of a path \Path.
We allow for one, or both $a, b$, to be infinite.
The origin of ${\IntegerSet^2}$ is \ZZOrigin ($=(0,0)$).
The intersection of two paths is the set of vertices visited by both paths.

The set of unit vectors $\DirectionSet=\left\{
  \East{=}(1,0),
  \North{=}(0,1),
  \South{=}(0,-1),
  \West{=}(-1,0)
\right\}$ is called the set of \emph{directions}.
The vectors \East, \North, \South, and \West correspond to the west, east, south and north directions, respectively.

A graph $H$ that is a sub-graph of a graph $G$ is denoted by $H\SubGraphOf G$. 
We consider paths as graphs and the same notation is used for paths and subpaths.
Let $G$, $H$ and $H'$ be graphs such that $H\SubGraphOf G$ and $H'\SubGraphOf G$, $H\SubGraphCup H'$ denotes the graph union of $H$ and $H'$ in $G$.

The set of finite (resp. forward infinite, backward infinite, bi-infinite) sequences of elements, words, over alphabet $\DirectionSet=\{\West,\East,\South,\North\}$ is denoted $\DirectionSet^{*}$ (resp. $\DirectionSet^{\omega}$, $^{\omega}\DirectionSet$, $^{\omega}\DirectionSet^{\omega}$).
The union of $\DirectionSet^{*}$, $\DirectionSet^{\omega}$, $^{\omega}\DirectionSet$ and $^{\omega}\DirectionSet^{\omega}$ is denoted $\DirectionSet^{\IntegerSet}$.
The empty sequence is \EmptyWord.
The reverse of a word is defined by $\Reverse{(d_1\cdot d_2\cdots d_k)}=d_k\cdots d_2\cdot d_1$; in particular, the reverse of a forward infinite sequence is backward infinite.

\paragraph{Free Paths and Paths in $\IntegerSet^2$.}
We say that two paths $\Path$ and $\Path'$ are \emph{equivalent} if $\Path'$ is a translation of $\Path$. The equivalence class of $\Path$, denoted $[\Path]$, is called a \emph{free path} associated with $\Path$. The equivalence class is uniquely determined with a sequence of unit vectors $\kappa$, 
a word over $\DirectionSet$, 
(i.e., an element of $\DirectionSet^{\IntegerSet}$) such that $\kappa_i=d$ if and only if $\Path_i+d=\Path_{i+1}$. We intermittently use the notion `free path' and a word notation $\kappa$ to represent both $[\Path]$ and a word in $\DirectionSet^{\IntegerSet}$.
The null free path is $\EmptyWord$.
If $m_1$ is not forward infinite and $m_2$ is not backward infinite, then $m_1m_2$ designate their concatenation and represents a free path only if it corresponds to a path (and not a walk).
If $m$ is a finite free path, then $m^{\omega}=m m m\cdots$ is its infinite (forward) repetition, $^{\omega}m=\cdots m m m$ is its infinite backward repetition and $^{\omega}m^{\omega}=\cdots m m m\cdots$ is its bi-infinite repetition.
If $m=d_1d_2\cdots d_k$ is finite, its length is $k$ and is denoted $|m|$.
The set of cyclic rotations of a finite free path is $\Rotate{d_1d_2\cdots d_k}=\{d_{i}d_{i+2}\cdots d_k d_1 d_2\cdots d_{i-1}|1\leq i \leq k\}$.
The language operators $\cdot$, $\vee$, $^*$ and $^{\omega}$ are used in regular expressions to denote sets of words in $\DirectionSet^{\IntegerSet}$.
These sets are sets of free paths whenever all words within the sets represent free paths.

If the free path $m=m_1\cdots m_k\in \DirectionSet^*$ is finite, the associated \emph{displacement vector}, \Vect{m} (of $\IntegerSet^2$) is defined as the sum of its elements
$\Vect{m}=m_1+\cdots +m_k$.
Two finite free paths are \emph{collinear} if their associated displacement vectors are collinear.
There are infinitely many free paths associated to a given displacement vector of $\IntegerSet^2$.

A sequence of vertices (elements of ${\IntegerSet^2}$) that represents a path $\Path$ can also be expressed as an \emph{attachment/grounding vertex} $A\in \IntegerSet^2$ with a non forward infinite free path $b$ and a non backward infinite free path $f$ as $b.A.f$.
That is, $b.A.f$ is an instance of a free path $bf$.
If any of these free paths, $b$ or $f$, is null, the notation simplifies to $b.A$, $A.f$ or just $A$.
Extending this notation, a path can also be denoted as a sequence of paths and free-paths; i.e.,  
$q=f_1.\Path_1. f_2.\Path_2.\cdots$ where $f_i$'s are free paths and $\Path_i$'s are paths. In this case $q$ is the unique path instance in the equivalence class $f_1[\Path_1]f_2[\Path_2]\cdots$ `grounded' by the vertices in the domains of $\Path_1,\Path_2,\cdots$.

If $A$ is a vertex of ${\IntegerSet^2}$ and $\Vect{v}$ a vector in $\IntegerSet^2$, then $A+\Vect{v}$ is the vertex $A$ translated by \Vect{v}.
For any path, $bm.A=b.(A-\Vect{m}).{m}$ where $b\in D^*\cup {^{\omega}{D}}$ and $m\in D^*$. 
The \emph{opposite} of a free path $m=d_1d_2\cdots d_k$ is $\Backward{m}=\Backward{d_1}\,\Backward{d_2}\cdots \Backward{d_k}$ where
$\Backward{\East}=\West$,
$\Backward{\North}=\South$, 
$\Backward{\South}=\North$, and
$\Backward{\West}=\East$.
Please note that for any free path $m$, $\Reverse{\Backward{m}}=\Backward{\Reverse{m}}$.
We use the notation \RevBack{m} to denote \Reverse{\Backward{m}}.

We point out the different notations used in the text. For a free path $\alpha=\North\North\East$ we have $\Reverse{\alpha}=\East\North\North$, and $\Backward{\alpha}=\Backward{\North}\Backward{\North}\Backward{\East}=\South\South\West$ while $\RevBack{\alpha}=\West\South\South$.
In particular \RevBack{\alpha} traverses the {free} path $\alpha$ in reverse.

A free path $\alpha=mp^{\omega}$ for $m,p\in \DirectionSet^*$ and $p\neq\EmptyWord$ is called an \emph{ultimately periodic} free path, the prefix $m$ is called the \emph{transient part} of $\alpha$ and $p^{\omega}$ is the \emph{periodic part} of $\alpha$.
Similarly, the path $\Path=A.mp^{\omega}$ is an \emph{ultimately periodic} path, $A.m$ is the \emph{transient part} of $\Path$ and $(A+\Vect{m}).p^{\omega}$ is the \emph{periodic part} of $\Path$.
A \emph{periodic} (free) path is a (free) path whose transient path is $\epsilon$.

Let be $\Vect{\nu}$ any non-null vector of $\RationalSet^2$ and $a$ and $b$ be any two real numbers such that $a+||\Vect{\nu}||\leq b$.
The $(a,b,\Vect{\nu})$ ribbon is the subset of $\IntegerSet^2$ of the points ${x}$ such that $a\leq \Vect{\nu}\cdot\Vect{x}\leq b$ (here $\Vect{\nu}\cdot\Vect{x}$ indicates the dot product).
The condition on $\Vect{\nu}$, $a$ and $b$ ensures that ribbon forms a connected periodic sub-graph of $\IntegerSet^2$.
Please note that $\Vect{\nu}$ is perpendicular to the ribbon.
The vertices in the ribbon such that $a\leq \Vect{\nu}\cdot\Vect{x}\leq a+{||\Vect{\nu}||}$ form one boundary of the ribbon.
The boundary on the other side is formed by the vertices such that  $b-{||\Vect{\nu}||}\leq \Vect{\nu}\cdot\Vect{x}\leq b$.
These boundaries are bi-infinite periodic paths.

\begin{lemma}[double means pumpable]
  \label{lem:double-pumpable}
  Let $m$ be a non null finite free path.
  If $m^{2}=mm$ is a free path then so is $^{\omega}m^{\omega}$.
\end{lemma}

\begin{proof}
  Let $m$ be a free path such that $m^{2}$ is a free path. This implies that $\Vect{m}\not=\Vect{0}$.
  Suppose that $k$ is such that $\ZZOrigin.m^{k}$ is a path, but $\ZZOrigin.m^{k+1}$ is not a path but a walk.

  Let $a$ and $b$ be such that all vertices of the walk $\ZZOrigin.m^{k+1}$ are included in the $(a,b,\Vect{m})$ ribbon and $b-a$ is minimal.
  Observe that all vertices of the walk $^{\omega}m.\ZZOrigin.m^\omega$ are included in the ribbon. 

  By the minimality of $b-a$, there are integers $i$ and $j$ with $\Vect{m'}=\Vect{m_1}+\cdots+\Vect{m_{i-1}}$ and a sub-path $\Vect{m'}.m_{\IntegerInterval{i}{j}}$ of $\ZZOrigin.m$ that splits the ribbon in two, that is, $m_i$ is at one boundary  of the ribbon, $m_j$ is on the other boundary of the ribbon, and none of the vertices $m_r$ are on the boundary for $i<r<j$.
  Notice that $i=j$ can only happen if $\ZZOrigin.m^\omega$ is a horizontal line lying on the $x$-axis, or a vertical line lying on the $y$-axis, in which case $m^\omega$ is a free path.
  Therefore, any path within the ribbon passing from one part of the ribbon to the other must intersect with a vertex from $\Vect{m'}.m_{\IntegerInterval{i}{j}}$. 
  Moreover, for every integer $l$, the vertices of $l\Vect{m}.m$ are included in the ribbon and $(l\Vect{m}+\Vect{m'}).m_{\IntegerInterval{i}{j}}$ splits the ribbon in two parts.

  Being $\ZZOrigin.m^{k+1}$ a walk, and not a path, $k\Vect{m}.m$ it must intersect $\ZZOrigin.m^k$. This means that it either intersects the tail of $(k{-}1)\Vect{m}.m$ or it must cross the sub-path $((k{-}1)\Vect{m}+\Vect{m'}).m_{\IntegerInterval{i}{j}}$.
  Hence it has to intersect $(k{-}1)\Vect{m}.m$, implying that $(k{-}1)\Vect{m}.m^2$ is not a path, but a walk. This is in contradiction with our assumption that $m^2$ is a free path.
\end{proof}

\paragraph{Regions.}
A \emph{region}, R, is a connected subgraph of $\IntegerSet^2$.
For a vertex $v\in \IntegerSet^2$, the \emph{neighborhood of $v$}, $\mathcal{N}(v)$, is the subgraph of $\IntegerSet^2$ induced by the nine vertices at $x$-distance and $y$-distance at most $1$ from $v$.
A \emph{boundary} vertex for a region $R$ is a vertex $v\in R$ such that $\mathcal{N}(v) \not\SubGraphOf R$.
The \emph{boundary of $R$}, denoted $\partial R$, is the subgraph of $R$ induced by the sets of its boundary vertices.
The interior of $R$, denoted $\mathring{R}$ is the complement of $\partial R$ in $R$, that is $\mathring{R}=R\setminus\partial{R}$.

Let $\Path$ be a cycle or a bi-infinite path.
Then $\Path$ defines two regions in the plane whose intersection is $\Path$ itself. We distinguish these two regions as `left' and `right' as described below.

Since all paths are in dimension $2$, a path $\Path$ can be considered oriented by orienting the edges from $\Path_i$ to $\Path_{i+1}$ so that its left side can be defined.
We consider the cyclic counter clockwise orientation of the directions $(\West,\North,\East,\South)$ and we say that $d$ is \emph{to the left} of $d'$ if $d$ precedes $d'$ in the orientation, and $d$ is \emph{to the right} of $d'$ if $d$ succeeds $d'$.
A vertex $v$ in $\IntegerSet^2$ is \emph{directly to the left of $\Path$} if there is $d$ to the left of $d'$ such that $v+d=\Path_i$ for some $i$ and $\Path_i+d'=\Path_{i+1}$.
In this case the edge $(v,\Path_i)$ is \emph{directly to the left} of $\Path$.
Similarly we define a vertex, and an edge \emph{directly to the right of $\Path$}. 
The \emph{left region} of a path $\Path$, consists the subgraph of $\IntegerSet^2$ with vertices $v$ that are either in 
\Domain{\Path}, or there is a path $v.m$ for some free path $m$ that ends at $\Path$ with an edge directly to the left of $\Path$ and does not intersect with $\Path$ in any other vertex.
The \emph{right region} of $\Path$ is defined similarly. Because $\Path$ is bi-infinite or a cycle, the left and right regions are well defined, and their intersection is $\Path$.

A path $\Path$ is \emph{inside a region} $R$ if $\Path\SubGraphOf R$.
It is \emph{strictly inside} $R$ if $\Path\SubGraphOf \mathring{R}$.

\paragraph{Co-growth.}
\label{sec:co-growth}

In this section we describe a method of super imposing two paths to form a new path that is in some sense the `right-most' portion of both. We call this combined new path as `co-growth' of both. 
This co-growth algorithm takes as input:
\begin{itemize}
\item $f_1$ and $f_2$ two forward only infinite free paths such that 
  $f_1(0)=f_2(0)$, meaning, both paths $f_1$ and $f_2$ start in the same direction
  and
\item $b_1$ and $b_2$ two backward only infinite free paths
\end{itemize}
such that $b_1f_1$ and  $b_2f_2$ are also free paths. The free paths $b_1$ and $b_2$ are used to define the two regions producing the boundary that limits the extension of the co-growth.
Let $R_1$ ($R_2$, resp.) be the right region of the bi-infinite path $b_1.\ZZOrigin.f_1$ ($b_2.\ZZOrigin.f_2$, resp.).
These paths and regions are fixed for this rest of the current section.

\begin{definition}[co-Growth]
  A forward possibly infinite free path $f$ is \emph{the right co-growth of the free paths $b_1$, $f_1$, $b_2$ and $f_2$}, denoted $f=\CoGrow{b_1}{f_1}{b_2}{f_2}$, if it satisfies: $\ZZOrigin.f \SubGraphOf R_1\cap R_2$ and for all prefixes $pd$  of $f$ with $d\in\DirectionSet$: $(\ZZOrigin+\Vect{p}).d\SubGraphOf \ZZOrigin.f_1$ or $(\ZZOrigin+\Vect{p}).d\SubGraphOf \ZZOrigin.f_2$.  
\end{definition}

If paths, instead of free paths, are used as arguments for co-growth, the then the co-growth is taken with the associated free paths.
The left co-growth of $b_1$, $f_1$, $b_2$ and $f_2$ is defined in the similar way by considering the left regions of these paths. In the rest of our exposition we use `co-growth' and assume the right co-growth, unless otherwise stated.

\begin{figure}[hbt]
  \small\def\XMax{5}
  \def\XMin{-8}
  \def\YMax{10}
  \def\YMin{-4}
  \def\BB{plot [smooth] coordinates { (\XMin,0) (-3,\YMin/2) (-1,\YMin/2) (0,0) } }
  \def\B{plot [smooth] coordinates { (1,\YMin) (-5,-\YMin/3) (-3,-{\YMin}/3) (0,0) } }
  \def\FF{--plot [smooth] coordinates { (0,0)  (2,\YMax)  } }
  \def\Fpre{--plot [smooth] coordinates { (0,0) (3,3) (-6,\YMax)  } }
  \def\F{ .. controls +(3.3,2) .. (vv)}
  \newcommand{\FILL}[1]{\fill[red] (\XMin,\YMin) --#1 -- (2,\YMax) -- (\XMin,\YMax) ;}
  \centering\SetUnitlength{.9em}
  \begin{tikzpicture}
    \fill[notR2] \BB \FF -- (\XMin,\YMax) -- (\XMin,\YMin) -- cycle ;
    \draw[path,b2f2] \BB ;
    \draw[path,name path=f,b2f2] (0,0) \FF ;
    \draw[path,b1f1] \B ;
    \path[name path=ff] (0,0) \Fpre ;
    \SetIntersect{f}{ff}{vv}
    \path (tmp-2) \CoorNode{vv} ;
    \fill[notR1] \B \F -- (-6,\YMax) -- (\XMin,\YMax) -- (\XMin,\YMin) -- cycle ;
    \draw[path,name path=ff,b1f1] (0,0 ) \F -- (-6,\YMax) ;
    \begin{scope}[v]
      \draw (.4,-.1)  .. controls +(3.3,2) .. ([shift={(.4,0)}]vv) -- (2+.4,\YMax) node [above right] {$f$};
    \end{scope}
    \path (0,0) node[below left] {\ZZOrigin} node[below right] {\CoGrowStart} ;
    \path (\XMin,-1) node[left] {$b_2$} ;
    \path (1,\YMin) node[below] {$b_1$} ;
    \path (2,\YMax) node[above left] {$f_2$} ;
    \path (-6,\YMax) node[above] {$f_1$} ;
    \path (\XMin,\YMin) +(1.3,1.3) node[fill=white] {$\overline{R_1}$} ;
    \path (0,\YMax) +(-1.3,-1.3) node[fill=white] {$\overline{R_2}$} ;
    \path (\XMax,\YMin) +(-1,2) node {$R_1\cap R_2$} ;
  \end{tikzpicture}
  \caption{Co-growth. The dashed (green) line indicates $\CoGrow{b_1}{f_1}{b_2}{f_2}$.}
  \label{fig:cog:regions}\end{figure}
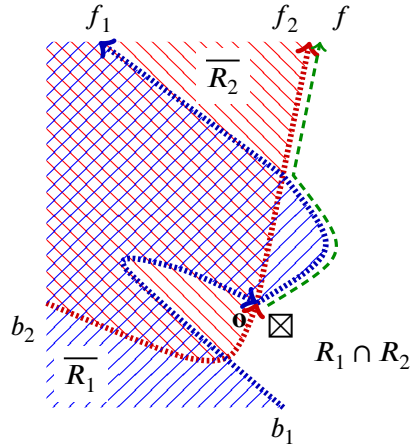

The `growing direction' of the co-growth at any vertex coincides with the direction of at least one of the paths at the same location, and it always takes the `right-most' path of the two paths at each vertex.
The co-growth  produces a `maximal' free path $f$ in the boundary of the region $R_1\cap R_2$.
Paths, regions and expected free path $f$ is illustrated in \RefFig{fig:cog:regions}.
The symbol \CoGrowStart is used to indicate where co-growth is started.

A way to obtain the path $f=\CoGrow{b_1}{f_1}{b_2}{f_2}$ is to start with $\epsilon$ from the origin and follow the directions of both paths $f_1$ and $f_2$ until one of the paths takes a direction different from the other.
At that point one follows the path that takes the right direction of the other, until the point when both paths intersect.
In a sense, in between any two intersections of paths $\ZZOrigin.f_1$ and $\ZZOrigin.f_2$ one follows the path that is strictly to the right of the other. 

\begin{lemma}
  \label{inf-co-grow}If there is a forward infinite path $\Path$ in $R_1\cap R_2$ starting at \ZZOrigin, then $f$ is infinite and $\Path$ is to the right of both paths  $b_1.\ZZOrigin.f$ and by $b_2.\ZZOrigin.f$\,.
\end{lemma}

\begin{proof}
  Observe that by definition $\ZZOrigin.f$ visits the intersections of $\ZZOrigin.f_1$ and $\ZZOrigin.f_2$\,.
  In between intersections, $\ZZOrigin.f$ is on the boundary of either $R_1$ or on the boundary of $R_2$. Hence, $\ZZOrigin.f$ is on the boundary of $R_1\cap R_2$, and $R_1\cap R_2$ is to the right of $f$. Hence $\Path$ must be to the right of $b_1.\ZZOrigin.f$ and $b_2.\ZZOrigin.f$\,. 

  Because $\ZZOrigin.\Path$ is forward infinite, the region $R_1\cap R_2$ is infinite, and therefore its boundary $\ZZOrigin.f$ is infinite.
\end{proof}

\paragraph{Tile Assembly System.}
Let \GlueSet be an a finite set called an \emph{alphabet} whose elements are symbols that we will also call \emph{glues}.
A \emph{tile type} is a map $t: \DirectionSet \to \GlueSet$.
We use notation $t_d$ for the value (glue) of $t$ in direction $d$. 

A \emph{(temperature-1) tile assembly system} (TAS) is a pair $\TAS =(\TileSet,\Seed)$ where $\TileSet$ is a finite set of tiles types and \Seed is a tile type (not necessarily in \TileSet) called the \emph{seed}.

For a TAS with set of tile types \TileSet and seed \Seed, an \emph{assembly} over \TAS is a partial map $\Assembly:\IntegerSet^2\to \TileSet\cup\{\Seed\}$ where $\alpha^{-1}(\Seed)$ is empty or a singleton if \Seed  is not in \TileSet.
The domain of \Assembly is the set of points of $\IntegerSet^2$ for which $\Assembly$ is defined, and is denoted $\Domain{\Assembly}$.
The \emph{binding graph} of $\Assembly$ is a subgraph of the lattice $\IntegerSet^2$ with vertices $\Domain{\Assembly}$ such that for $v,v'\in \Domain{\Assembly}$ there is an edge with endpoints $v$ and $v'$ if and only if $v+d=v'$ for some direction unit vector $d$ and 
$\Assembly(v)_d=\Assembly(v')_{-d}$.
The assembly $\Assembly$ is \emph{stable} if its binding graph is connected. 
An assembly $\Assembly$ is \emph{producible} in \TAS if it is stable and $\Assembly(0,0)$ is the seed.
If \Seed does not belong to \TileSet, it appears in $\alpha$ only at the origin.

Note that although neighboring vertices $v,v'\in\IntegerSet^2$ with $v+d=v'$ may be in the domain of $\alpha$ it may happen that $\alpha(v)_d\not=\alpha(v')_{-d}$. In this case the tiles $\alpha(v)$ and $\alpha(v')$ mismatch in direction $d$, and the binding graph of $\alpha$ has no edge between $v$ and $v'$. 

A stable assembly $\Assembly$ over \TAS is said to be \emph{an assembly path} if its binding graph is a path $\Path$ with $\Path_0=(0,0)$.

We extend the notions for ultimately periodic paths to ultimately periodic assembly paths. We say that 
and assembly path $\Assembly_{\Path}$ is ultimately periodic if $\Path=\ZZOrigin.mp^\omega$ is ultimately periodic and for all $i$ 
$\Assembly(\Vect{m}+i\Vect{p})=\Assembly(\Vect{m})$.

We introduce the following property for tile assembly systems that we show holds for all confluent systems and helps to characterize the assemblies obtained in these systems.

\begin{definition}
  Two assemblies $\Assembly$ and $\Assembly'$ are \emph{compatible} if for all $v\in \Domain{\Assembly}\cap \Domain{\Assembly'}$ we have that $\Assembly(v)=\Assembly'(v)$. 
\end{definition}

\begin{definition}[confluent]
  A tile assembly system \TAS is \emph{confluent} if every two producible assemblies $\Assembly$ and $\Assembly'$ are compatible.
\end{definition}

An assembly $\beta$ is \emph{maximal} for a system \TAS if for any other assembly $\alpha$ satisfying $\Domain{\beta}\subseteq \Domain{\alpha}$ we have that $\Domain{\alpha}=\Domain{\beta}$.
In a confluent system, any two assembly paths can `coexist' within a larger assembly because all intersections of their domains are mapped to the same tiles by both paths. 

\begin{lemma}
  If \TAS is confluent then there is a unique maximal producible stable assembly $\AlphaMax$ such that for every other stable assembly $\alpha$, $\Domain{\alpha}\subseteq\Domain{\AlphaMax}$.
\end{lemma}

\noindent
\emph{Notation.} In the rest of the paper we assume that \TAS is a fixed confluent tile assembly system that produces an infinite stable maximal assembly denoted \AlphaMax.

\begin{definition}[para-periodic]
  A confluent tile assembly system \TAS is called \emph{para-periodic} if its maximal producible assembly contains an ultimately periodic assembly path. 
\end{definition}

\begin{lemma}
  \label{path-assembly}In a confluent system if the binding graph \Path of an assembly path $\Assembly_{\Path}$ is ultimately periodic, then $\Assembly_{\Path}$ is ultimately periodic.
\end{lemma}

\begin{proof}
  Suppose the binding graph of $\Assembly_{\Path}$, $\Path=\ZZOrigin.mp^\omega$, is ultimately periodic. Observe that because \TAS is finite, there are $i<j$ such that $\Assembly_{\Path}(\Vec{m}+i\Vec{p})=\Assembly_{\Path}(\Vec{m}+j\Vec{p})$.
  Let $q$ be any prefix of $p^{j-i}$. Because of confluence of \TAS, it must be that $\Assembly_{\Path}(\Vec{m}+i\Vec{p}+\Vect{q})=\Assembly_{\Path}(\Vec{m}+j\Vec{p}+\Vect{q})$.
  Hence, $\Assembly_{\Path}$ is ultimately periodic.
\end{proof}

\begin{example}
  \label{example:confl}The tile set in figs. \ref{fig:seed} and \ref{fig:tas} generates only one maximal assembly, the one depicted in \RefFig{fig:assembly}.
  If the seed would have been \TilePolice{D} which can be considered as part of \TAS, instead of \TilePolice{\Seed}, then the system would not have been confluent and the maximal assembly would not be unique.
\end{example}

\begin{figure}[hbt]
  \centering\small\SetUnitlength{1.9em}\begin{tabular}[b]{c}
    \SetUnitlength{2em}\subcaptionbox{seed\label{fig:seed}}{\qquad\begin{tikzpicture}\Tile[S]{0,4}{\Seed}\GlueBot[s]{S}\SEED{0,4}\end{tikzpicture}\qquad\mbox{}}\\[4em]
    \SetUnitlength{2em}\subcaptionbox{tile type set \TileSet\label{fig:tas}}{\begin{tikzpicture}
      \Tile[A]{0,2}{A}
      \GlueTop[s]{A}
      \GlueLe[a]{A}
      \GlueRi[b]{A}
      \GlueBot[c]{A}
      \Tile[B]{2,2}{B}
      \GlueLe[b]{B}
      \GlueRi[a]{B}
      \GlueBot[d]{B}
      \Tile[D]{2,0}{D}
      \GlueTop[d]{D}
      \GlueBot[d]{D}
      \Tile[C]{0,0}{C}
      \GlueTop[c]{C}
      \GlueBot[c]{C}
    \end{tikzpicture}
    }\end{tabular}
  \qquad\subcaptionbox{maximal assembly graph \AlphaMax\label{fig:assembly}}{\begin{tikzpicture}\Tile[S]{8,4}{\Seed}
      \SEED{8,4}
      \begin{scope}{opacity=0}
        \Tile[A]{8,2}{A}
      \end{scope}
      \LinkBotTop[s]{S}{A}
      \foreach \X in {1,2,3} {
        \begin{scope}[shift={(4*\X,0)}]
          \Tile[A]{0,2}{A}
          \ifnodedefinedcurrpic{B}{
            \LinkLeRi[a]{A}{B}}{}
          \Tile[B]{2,2}{B}
          \LinkLeRi[b]{B}{A}
          \Tile[C]{0,0}{C}
          \LinkBotTop[c]{A}{C}
          \Tile[D]{2,0}{D}
          \Tile[D1]{2,-2}{D}
          \LinkBotTop[d]{B}{D}
          \LinkBotTop[d]{D}{D1}
          \Tile[C1]{0,-2}{C}
          \LinkBotTop[c]{C}{C1}
          \draw (0,-2.75) node {\vdots} ;
          \draw (2,-2.75) node {\vdots} ;
        \end{scope}
      };
      \draw (3,2) node {\dots} ;
      \draw (15,2) node {\dots} ;
    \end{tikzpicture}
  }
  \caption{Example of \TAS.}
  \label{fig:tas-example}
\end{figure}

For assembly paths $\{\Assembly_i\}_{i\in I}$ we say that $\AssemblyOther=\cup_{i\in I} \Assembly_i$ if $\AssemblyOther$ has a binding graph that is union of the paths that are binding graphs for $\Assembly_i$ for all $i\in I$, and $\AssemblyOther|_{\Domain{\Assembly_i}}=\Assembly_i$. 
Note this is well defined because all paths are pairwise compatible.

\begin{lemma}
  A stable assembly in a confluent TAS \TAS is a union of assembly paths. 
\end{lemma}

\begin{proof}
  It follows directly from the fact that the binding graph of a stable assembly is connected and \TAS is confluent.
\end{proof}

The assembly of \TAS is the maximum assembly and is defined by the union of all legal paths. 

\begin{corollary}
  The maximal assembly of \TAS, is $\AlphaMax=\cup_{\Assembly \in \Path} \Assembly$ where $\Pi$ denotes the set of all assembly paths.
\end{corollary}

\begin{figure}[hbt]
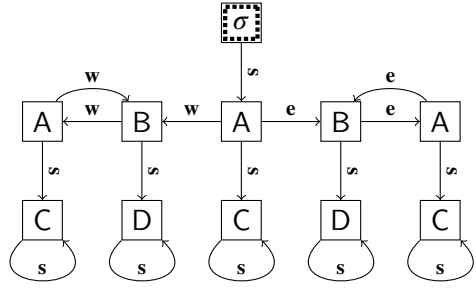

  \centering\small\SetUnitlength{1.7em}\TasAutom{\Tile[Dl]{-2,0}{D}
    \draw[->] (Dl) .. controls +(-\ControlLoopS) and +(\ControlLoopS) .. 
    node[above,inner sep=0.2em] {\scriptsize \South} (Dl) ;
    \ArcSouth{Bl}{Dl}}{
    \Tile[C]{[yshift=-2\unitlength]A}{C}
    \draw[->] (C) .. controls +(-\ControlLoopS) and +(\ControlLoopS) .. 
    node[above,inner sep=0.2em] {\scriptsize \South} (C) ;
    \ArcSouth{A}{C}\Tile[Cr]{[yshift=-2\unitlength]Ar}{C}
    \draw[->] (Cr) .. controls +(-\ControlLoopS) and +(\ControlLoopS) .. 
    node[above,inner sep=0.2em] {\scriptsize \South} (Cr) ;
    \ArcSouth{Ar}{Cr}}{}
  \caption{Automaton generating assembly paths whose union of domains covers \Domain{\AlphaMax}.}
  \label{fig:path-automaton}
\end{figure}

\RefFigure{fig:path-automaton} depicts an `automaton' that generates all the tiles in $\AlphaMax$ from Example~\ref{example:confl}.
It is a compact and finite representation of \AlphaMax in \RefFig{fig:assembly}.
In the rest of the paper, we show that such automata, called \emph{quipu}, can be generated for confluent systems.

In the rest of the paper, to simplify notations, \AlphaMax is often used to refer to its biding graph.

A vertex $y$ in $\IntegerSet^2$ is \emph{non-causal} for vertex $x$ in $\IntegerSet^2$ if either $x$ does not belongs to \AlphaMax or there is a path in \AlphaMax from \ZZOrigin to $x$ that does not contain $y$.
The set of non-causal vertices for $x$ in \AlphaMax is
\begin{math}
  \NonCausal{x}=\left\{
    y\in 
    \IntegerSet^2
    \mid
    \exists m\in\DirectionSet^{*},
    \ZZOrigin.m\SubGraphOf\AlphaMax
    \wedge
    y\not\in \Domain{\ZZOrigin.m}
    \wedge
    \ZZOrigin+\Vect{m}=x
  \right\}
  \cup\{x\}
\end{math}\, ($x$ is included to simplify later expressions).

An assembly $\Assembly$ that has a path as a binding graph may not necessarily be part of an assembly path, i.e., may not necessarily be part of a producible assembly.
If $\Assembly$ starts with a tile $t_0$, it may be the case that every assembly path (that starts from the seed, i.e. the origin) that reaches $t_0$ also forms an obstacle for $\Assembly$.
The non-causal set of vertices for a given point in $x\in \IntegerSet^2$ identifies the region that is free from those obstacles. In particular, we have the lemma. 

\begin{lemma}
  \label{non-causal-extension}Let $x\in\Domain{\AlphaMax}$.
  Suppose $p$ is a free path such that there is an assembly $\alpha_p$ with a binding graph $p$ and start tile $\alpha_p(0)=\AlphaMax(x)$.
  If $\Domain{x.p}\subseteq \NonCausal{x}$ then $x.p\SubGraphOf\AlphaMax$.
\end{lemma}

\begin{proof} 
  Inductively, for $p=\epsilon$ we have $x.p=x$. Suppose there is an assembly $\Assembly_{p'}$ with binding graph $p'=pd$ such that $x.p\SubGraphOf\AlphaMax$ and 
  $y=(x+\Vect{p}).d\in \NonCausal{x}\subseteq \Domain{\AlphaMax}$.
  Let $\ZZOrigin.q$ be an assembly path in $\AlphaMax$ such that $y\not \in \Domain{\ZZOrigin.q}$ and $\ZZOrigin +\Vec{q}=x$.
  Suppose $p_1p_2=p$ is such that $p_1$ is the longest prefix of $p$ where $x+\Vec{p_1}\in \Domain{\ZZOrigin.q}$.
  Let $q'$ be the prefix of $q$ such that $x+\Vec{p_1}=\ZZOrigin+\Vec{q'}$.
  Then $\ZZOrigin.q'p_2d$ is a binding graph of a producible assembly because \TAS is confluent, $\ZZOrigin.q'p_2$ is a path in \AlphaMax, and $p_2d$ is a suffix of an assembly path that exists.
  Therefore, the edge between $x+\Vect{p}$ and $y$ exists, i.e., $x.pd\SubGraphOf\AlphaMax$.
\end{proof}

The following lemma shows that $f=\CoGrow{b_1}{f_1}{b_2}{f_2}$ allows to extend paths within $\AlphaMax$ in a compatible way since \TAS is confluent.

Let $b_1q_1f_1$ and $b_2q_2f_2$ be two bi-infinite free paths with  $b_i$ being backward infinite, and $q_if_i$ forward infinite starting with the same direction ($i=1,2$).
Let $qf=\CoGrow{b_1}{q_1f_1}{b_2}{q_2f_2}$ where $q$ is the maximal portion of the co-growth that consists of segments of $q_1$ or $q_2$ only.
Let $A_1,A_2\in \IntegerSet^2$ and consider the right regions $R_1,R_2$ of $b_1.A_1.q_1f_1$, and $b_2.A_2.q_2f_2$, respectively. We further suppose that $A_i.q\SubGraphOf \AlphaMax$ ($i=1,2$) and $\AlphaMax(A_1)=\AlphaMax(A_2)$. 

\begin{lemma}[co-Growth compatibility]
  \label{lem:co-grow-tile}
  If $R_1\subseteq\NonCausal{A_1}$ and $R_2\subseteq\NonCausal{A_2}$
  then both paths
  $A_1.q$ and
  $A_2.q$ are subgraphs of \AlphaMax.
\end{lemma}

\begin{proof}
  The hypotheses imply that $A_1.q$ is in \NonCausal{A_1} and $A_2.q$ in \NonCausal{A_2}, hence by \RefLem{non-causal-extension} these paths exist in $\AlphaMax$.
  Furthermore, whenever the two paths $q_1$ and $q_2$ intersect, because of the confluence of the \TAS, the tiles in \AlphaMax corresponding to these intersections must coincide.
\end{proof}

\section{Every Confluent TAS is Para-periodic}
\label{sec:nice}

In this section we show that a $\AlphaMax$ in a confluent tiling assembly systems is para-periodic.
In \RefSec{sec:quipu}, para-periodicity allows bootstrapping construction of a finite `automaton' that characterizes $\AlphaMax$.
We start with the following lemma saying that para-periodicity is equivalent to existence of an infinite path in $\AlphaMax$ that does not intersect an ultimately periodic path in the grid $\IntegerSet^2$.
An extended version of this lemma is used extensively later on.
As mentioned, we assume \TAS is confluent.

\begin{lemma}[para-periodic]
  \label{lem:nice}
  The following two properties are equivalent:
  \begin{enumerate}
  \item \AlphaMax contains an ultimately periodic assembly path (i.e. is para-periodic), and
    \label{lem:nice:path}
  \item there is a point $A$ in $\IntegerSet^2$, an infinite periodic free path $p^{\omega}$ and an infinite path $\ZZOrigin.\Path$ in \AlphaMax rooted in the seed in \AlphaMax with $\Domain{A.p^\omega}\cap \Domain{\ZZOrigin.\Path} =\emptyset$.
    \label{lem:nice:avoid}
  \end{enumerate}
\end{lemma}

\begin{proof}
  The property \ref{lem:nice:path} implies the property \ref{lem:nice:avoid} 
  by taking the same period $p$ and choosing a point $A$ sufficiently away from the seed.
  
  The property \ref{lem:nice:avoid} implies the property \ref{lem:nice:path} as follows.
  Without loss of generality (and to ease exposition following a figure) we can consider that $p$ extends eastwards and that the infinite path \Path eventually infinitely extends eastwards north of $A.p^{\omega}$.
  Otherwise, the situation can be rotated and symmetrized appropriately.
  Consider the north vector $\North\in \DirectionSet$.
  Let $\Path_{i}$ be the intersection of \Path and $A.\North^{\omega}$ closest to $A$ (it cannot be in $A.p^{\omega}$) and let $a$ be such that $A+a\Vect{\North}=\Path_{i}$.
  By definition of non-causality $\Domain{\Path_{\IntegerInterval{i+1}{+\infty}}}\subseteq\NonCausal{\Path_i}$.
  Moreover, the right region of $^{\omega}\RevBack{{p}}.A.\North^{a}.\Path_{\IntegerInterval{i}{+\infty}}$ (going backwards on $p^\omega$, then north with $\North^a$ and continue with $\Path_{\IntegerInterval{i}{+\infty}}$) is also included in $\NonCausal{\Path_i}$, because none of these vertices are visited by the path $\Path_{\IntegerInterval{0}{i}}$.
  
  The situation is depicted in \RefFig{fig:nice-eq}(a).

  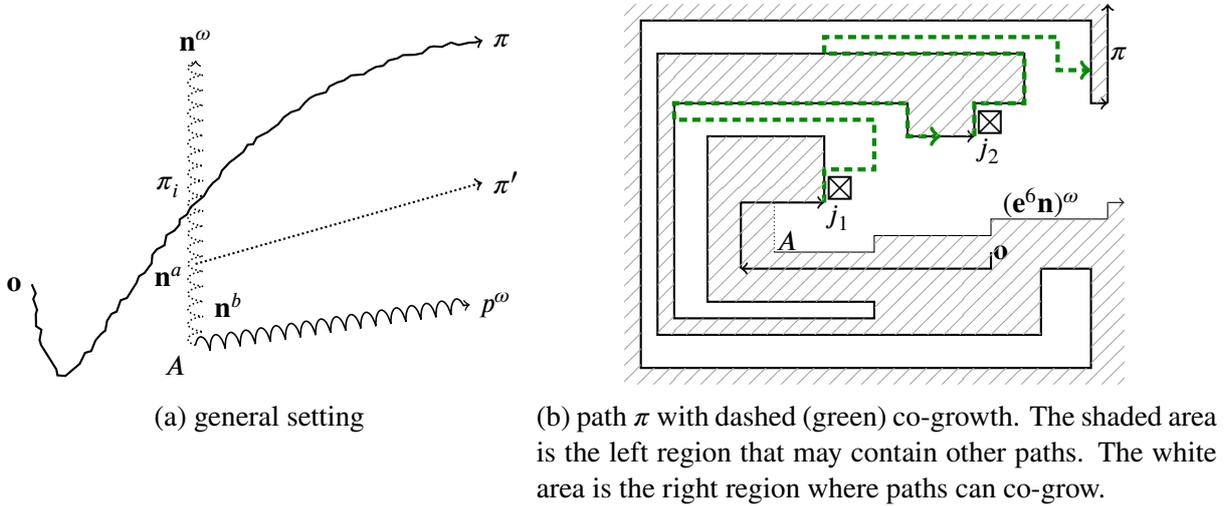
\begin{figure}[hbt]
    \small\SetUnitlength{1.4em}\centerline{\subcaptionbox{general setting \label{fig:nice-eq:a}}{\begin{tikzpicture}
          \path (1,0) \CoorNode{o} node [left] {\ZZOrigin} ;
          \begin{pgfinterruptboundingbox}
            \draw[transient,name path=pi] (o) .. controls (2,-7) and (3,5) .. (12,6) node [right] {\Path} ;
          \end{pgfinterruptboundingbox}
          \path (5,-1.5) \CoorNode{a} node [below left] {$A$} ;
          \draw[path,densely dotted] (a) +(0,2) \CoorNode{b} -- (12,2.5) node [right] {$\Path'$} ;
          \draw[periodic] (a) -- +(6.7,1) node [right] {$p^{\omega}$}; 
          \draw[periodic,densely dotted,name path=n-o] (a) -- +(0,7) node [above] {$\North^{\omega}$} ;
          \SetIntersect{pi}{n-o}{p=i}
          \path (p=i) node [above left] {$\Path_{i}$} ;
          \path (a) -- node [left] {$\North^{a}$} (p=i) ;
          \path (a) -- node [right=.3em] {$\North^{b}$} (b) ;
        \end{tikzpicture}}
      \subcaptionbox{path \Path with dashed (green) co-growth.
        The shaded area is the left region that may contain other paths.
        The white area is the right region where paths can co-grow. \label{fig:nice-eq:b}}{\SetUnitlength{.57em}\newcommand{\VPathOne}{\No{2} \Ea{3} \No{3} \We{12} }\newcommand{\VPathTwo}{\No{1} \Ea{14} \So{2} \Ea{2} }\newcommand{\PathPomega}{-- ++(-1,0) -- ++(0,-1) -- ++(-6,0)
          -- ++(-1,0) -- ++(0,-1) -- ++(-6,0)
          -- ++(-1,0) -- ++(0,-1) -- ++(-6,0) }\newcommand{\PathJoJt}{-- ++(0,4) -- ++(-7,0) -- ++(0,-10) -- ++(10,0) -- ++(0,-1) -- ++(-12,0) -- ++(0,12) \VPathTwo -- ++(2,0) }\newcommand{\PathJtJtt}{\VPathOne -- ++(-10,0) -- ++(0,-17) -- ++(23,0) -- ++(0,4) -- ++(3,0) -- ++(0,-6) -- ++(-27,0) -- ++(0,21) -- ++(27,0) -- ++(0,-5) -- ++(1,0)}\qquad\begin{tikzpicture}[inner sep=0.1ex]
          \path (0,0) coordinate (A) node [above right] {$A$} ;
          \path (16,2) node[above] {$(\East^6\North)^{\omega}$} ;
          \draw[<-] (A) ++(21,3) \PathPomega ;
          \draw[densely dotted] (A) -- +(0,3) \CoorNode{A'} ;
          \draw[path,-] (A') -- ++(-2,0) -- ++(0,-4) \CoorNode{O'} ;
          \draw[path,<-] (O') -- ++( 15,0) -- ++(0,1) node [right] {\ZZOrigin} ;
          \draw[path] (A') -- ++(3,0) \CoorNode{j1} node[below right] {$j_1$} node[above right] {\CoGrowStart} ;
          \draw[path] (j1) \PathJoJt \CoorNode{j2} node[below right] {$j_2$} node[above right] {\CoGrowStart} ;
          \draw[path] (j2) \PathJtJtt \CoorNode{j33} ;
          \draw[path] (j33) -- node[right] {$\Path$} ++(0,6) ;
          \draw[v,ultra thick] (j1) \VPathOne \VPathTwo ; 
          \draw[v,ultra thick] (j2) \VPathOne \VPathTwo ;
          \fill[notR1,pattern color=Grey] (A) -- (A') -- (j1) \PathJoJt \PathJtJtt -- ++(0,6) \CoorNode{Q}
          -- ++(-29,0) \CoorNode{R} -- ++(0,-23) \CoorNode{S} -- ++(30,0) \CoorNode{W} -- (21,3) \CoorNode{F} \PathPomega
          -- cycle ;
        \end{tikzpicture}\qquad\qquad}}
    \caption{Situation for \RefLem{lem:nice}.}
    \label{fig:nice-eq}
  \end{figure}

  Let $\Pi$ be the set of forward (only) infinite paths in \AlphaMax that are in the right region of $^{\omega}\RevBack{{p}}.A.\North^{a}.\Path_{\IntegerInterval{i}{+\infty}}$, extend infinitely to the east, start in $\Domain{A.\North^{a}}$ and do not intersect $A.p^\omega$.
  Further, we require that for every path $(A+\North^b).f'$ ($b\le a$) in $\Pi$, the right region of $^{\omega}\RevBack{{p}}.A.\North^{b}f'$ is included in $\NonCausal{A+\North^{b}}$.
  The set $\Pi$ is not empty because it contains $\Path_{\IntegerInterval{i}{+\infty}}$.
  
  Let $\Path'$ be the path in $\Pi$ such that, except for $\Path'$, there is no other path in $\Pi$ that is inside the right region of $^{\omega}\RevBack{{p}}.A.\North^{b}.\Path'$.
  This path does exist because $\IntegerSet^2$ is discrete and it can be inductively defined from $A.\North^{b}$, taking the rightmost continuation with an infinite path in $\Pi$.

  Observe that the path $\Path_{\IntegerInterval{i}{+\infty}}$ can be quite complicated, as depicted in \RefFig{fig:nice-eq:b}.
  In the figure $p=\East^6\North$ and the unshaded (white) region is the right region of $^{\omega}\RevBack{{p}}.A.\North^{b}.\Path$.
  This region belongs to $\NonCausal{A+\North^b}$. 
  An example of co-grow (as explained below) is displayed with dashed lines showing that this path cannot be the path $\Path'$ defined above.

  We consider the infinitely increasing sequence of indexes $j$ in $\Path'$ such that $\Path_j'$ belongs to $A.\North^kp^{\omega}$ for some positive $k$ and $\Path'_{\IntegerInterval{j}{\infty}}$ does not intersect the path $A.\North^{k-1}p^{\omega}$.
  There are infinitely many such indexes because $\Path'$ is bounded by $A.p^\omega$.
  Since there are infinite such $j$'s, there must be two distinct indices $j_1$, $j_2$ ($j_1<j_2$) such that $\AlphaMax(\Path'_{j_1}) =\AlphaMax(\Path'_{j_2})$, $\Path'_{j_1}$ and $\Path'_{j_2}$ are followed with the same direction in $\Path'$ and have the same horizontal phase with respect to the period $p$.
  The forward path $f$ generated with $\CoGrow{^{\omega}{\RevBack{{p}}}\North^{b}\Path'_{\IntegerInterval{0}{j_1}}}{\Path'_{\IntegerInterval{j_1}{\infty}}}{^{\omega}{\RevBack{{p}}}\North^{b}\Path'_{\IntegerInterval{0}{j_2}}}{\Path'_{\IntegerInterval{j_2}{\infty}}}$ is an infinite free path because both right regions of $(^{\omega}{\RevBack{{p}}}\North^{b}\Path'_{\IntegerInterval{0}{j_1}}).\ZZOrigin.\Path'_{\IntegerInterval{j_1}{+\infty}}$ and 
  $({\RevBack{{p}}}\North^{b}\Path'_{\IntegerInterval{0}{j_2}}).\ZZOrigin.\Path'_{\IntegerInterval{j_2}{\infty}}$ contain the infinite path $\ZZOrigin.p^{\omega}$ (\RefLem{inf-co-grow}).
  The forward infinite path $f$ is the `right most' part of the segments $\Path'_{\IntegerInterval{j_1}{\infty}}$ and $\Path'_{\IntegerInterval{j_2}{\infty}}$, both of which are part of $\AlphaMax$.

  Both paths $\Path'_{\IntegerInterval{0}{j_1}}.f$ and $\Path'_{\IntegerInterval{0}{j_2}}.f$ belong to \AlphaMax (by \RefLem{lem:co-grow-tile}) and belong to $\Pi$, so they have to be left of $\Path'$ by definition of $\Path'$.
  By co-growth, they also have to be right of $\Path'$.
  So that $\Path'=\Path'_{\IntegerInterval{0}{j_1}}.f=\Path'_{\IntegerInterval{0}{j_2}}.f$.
  Since ${j_1}< j_2$, $\Path'$ is ultimately periodic.

  Since $\Path'$ is in \AlphaMax and in \NonCausal{\Path'_0}, there is a path $\Path''$ from the seed to $\Path'_0$. Then $\Path''\Path'$ is an ultimately periodic assembly path.
\end{proof}

\begin{corollary}\label{cor:new-periodic-path}
  Let $\Path$ be an infinite path in \AlphaMax, $mp^\omega$ be a free path and $i$ be a positive number such that $\Domain{\ZZOrigin.\Path}\cap \Domain{\Path_i.mp^\omega}=\{\Path_i\}$.
  Then one of the three following possibilities appears:
  \begin{enumerate}
  \item $\Path$ is ultimately periodic, 
  \item there is an ultimately periodic path in \AlphaMax intersecting $\Domain{\Path}$ on an infinite set, or
  \item there is an ultimately periodic path in \AlphaMax strictly inside the right region of $\RevBack{{ mp^\omega}}.\Path_{\IntegerInterval{i}{\infty}}$
  \end{enumerate}
  and \AlphaMax is para-periodic.
\end{corollary}

\begin{proof}
  The proof follows directly from the proof of \RefLem{lem:nice}. If the first two conditions are not satisfied, then the path $\Path''\Path'$ constructed in the proof of \RefLem{lem:nice} satisfies the third condition.
\end{proof}

\begin{remark}
  We note that all observations in \RefLem{lem:nice} and \RefCor{cor:new-periodic-path} are left-right symmetric, hence,
  hold when left co-growth is applied and the left regions are considered.
\end{remark}

\begin{definition}[off-the-Wall Path]
  Let $y_{0}$ be an integer and $\delta$ a positive integer. 
  A finite path $\Path=\Path_0\cdots\Path_l$ is \emph{$(y_{0},\delta) $-off-the-wall} if there exists a positive integer $i$ less than $l$ such that: 
  \begin{enumerate}
  \item \Path is an assembly path in \AlphaMax rooted in the seed,
  \item there is $x_w\in \IntegerSet$ satisfying 
    $\Path_{l}=(x_{w},y_{0})$
    and
    $\Path_{i}=(x_{w}+\delta,y_{0})$,
  \item 
    $ \Domain{\Path}\ \cap\ \Domain{{^{\omega}\West}.\Path_{i}}
    = \{\Path_{i}\} 
    = \Domain{\Path}\ \cap\ \Domain{\Path_{l}.\West^{\omega}}$,
    and
  \item and \ZZOrigin is left of the bi-infinite path
    \begin{equation}
      \label{eq:nice:bi-infinite}
      ^{\omega}\West
      .\Path_{\IntegerInterval{i}{l}}
      .\West^{\omega}
      \enspace.
    \end{equation}
  \end{enumerate}
\end{definition}

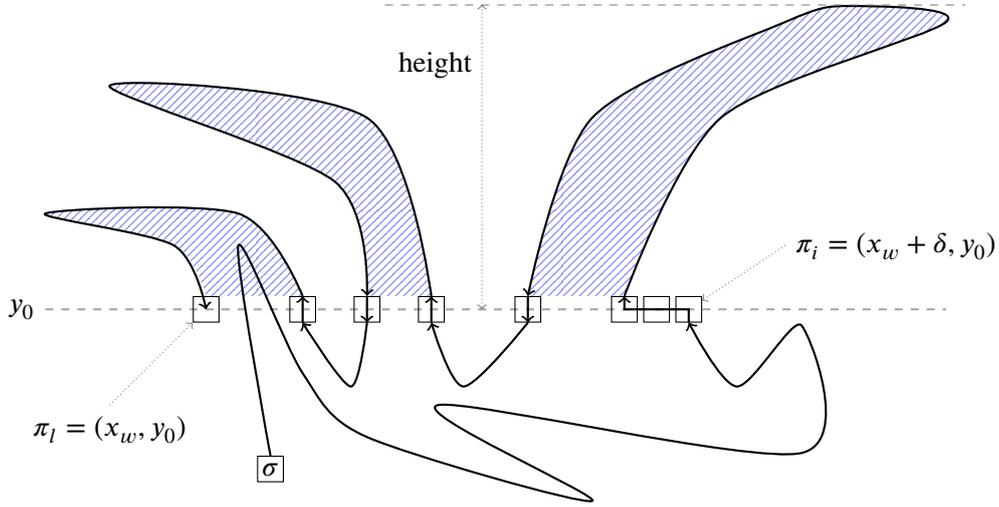
\begin{figure}[hbt]
  \centering\small\SetUnitlength{1.1em}\begin{tikzpicture}
    \draw[wall] (-5,0) node[left] {$y_{0}$} -- (23,0) ;
    \Tile[tL]{0,0}{}
    \Tile[t1]{3,0}{}
    \Tile[t2]{5,0}{}
    \Tile[t3]{7,0}{}
    \Tile[t4]{10,0}{}
    \Tile[t5]{13,0}{}
    \Tile[t6]{14,0}{}
    \Tile[tR]{15,0}{}
    \Tile[S]{2,-5}{\Seed}
    \newcommand{\ONEtoL}{ (t1.north) ($(t1)+(-2,3)$) ($(t1)+(-8,3)$) ($(tL)+(-1,2)$) }
    \newcommand{\THREEtoTWO}{ (t3.north) ($(t3)+(-2,6)$) ($(t3)+(-10,7)$) ($(t2)+(-1,4)$) (t2.north) }
    \newcommand{\FIVEtoFOUR}{ (t5.north) ($(t5)+(3,6)$) ($(t5)+(10,9)$) ($(t5)+(7,9.5)$) ($(t5)+(5,9)$) ($(t5)+(-1,6)$) (t4.north) }
    \begin{scope}[path]
      \begin{scope}[pattern=north east lines, pattern color=LightBlue]
        \fill (t1.north) -- plot [smooth] coordinates { \ONEtoL (tL.north) };
        \fill plot [smooth] coordinates { \THREEtoTWO };
        \fill plot [smooth] coordinates { \FIVEtoFOUR };
      \end{scope}
      \draw plot [smooth] coordinates { (S.north)
        ($(tL)+(1,2)$) ($(t1)+(0,-2)$) ($(t1)+(2,-4)$)
        ($(S)+(10,-1)$) ($.6*(S)+.4*(tR)$) ($.5*(S)+.5*(tR)+(10,-2)$) ($.5*(S)+.5*(tR)+(10,+2)$) ($(tR.south)+(1.5,-2)$)
        (tR.south) };
      \draw (tR.south) -- (tR.center) -- (t6.center) -- (t5.center) -- (t5.north) ;
      \draw plot [smooth] coordinates { \FIVEtoFOUR };
      \draw (t4.north) -- (t4.south) ;
      \draw plot [smooth] coordinates { (t4.south)
        ($.5*(t4.south)+.5*(t3.south)+(-.5,-2)$)
        (t3.south) };
      \draw (t3.south) -- (t3.north) ;
      \draw plot [smooth] coordinates { \THREEtoTWO };
      \draw (t2.north) -- (t2.south) ;
      \draw plot [smooth] coordinates { (t2.south)
        ($.5*(t1.south)+.5*(t2.south)+(.5,-2)$)
        (t1.south) };
      \draw (t1.south) -- (t1.north) ;
      \draw plot [smooth] coordinates { 
        \ONEtoL
        (tL) };
    \end{scope}
    \begin{scope}[densely dotted,->,draw=DarkGrey]
      \draw[-,dashed] (current bounding box.north) +(-3,0) -- (current bounding box.north east) ;
      \draw[<->] (current bounding box.north)-- node[pos=.2,left] {height} (current bounding box.north|-tL) ;
      \draw (tL) +(-3,-3) node[below] {$\Path_{l}=(x_{w},y_{0})$} -- (tL) ; 
      \draw (tR) +(3,2) node[right] {$\Path_{i}=(x_{w}+\delta,y_{0})$} -- (tR) ;
    \end{scope}
  \end{tikzpicture}
  \caption{Off the wall path and parameters.}
  \label{fig:off-the-wall}
\end{figure}

The \emph{$y_0$-wall} is the subset of $\IntegerSet^2$: $ ^{\omega}\West.(0,y_0).\West^{\omega}$.
The segment $\Path_{\IntegerInterval{i}{l}}$ of \Path is called \emph{above-the-wall} part of \Path (even though some of its portions might be 
under `the wall' as illustrated in \RefFig{fig:off-the-wall}).
We say that \Path is \emph{$y_0$-off-the-wall} if there is some positive $\delta$ such that \Path is $(y_0,\delta)$-off-the-wall.

\begin{definition}[height, surface and area above the wall]
  Let \Path be any $(y_{0},\delta)$-off-the-wall path.
  Its height is the maximal $y$-coordinate it reaches above the wall minus $y_0$.
  The \emph{surface above (the wall)} is defined by what is the intersection of the right region of $y_{0}$-wall with the left region of $^{\omega}\West
  .\Path_{\IntegerInterval{i}{l}}
  .\West^{\omega}$.
  The \emph{area above} is the area of the surface above.
\end{definition}

The notions of off-the-wall path, the height above, and surface above are illustrated in \RefFig{fig:off-the-wall} where the surface above is shaded. The area above is the area of the shaded portion.

\begin{definition}[wall valuation]
  Let \Path be some $(y_{0},\delta)$-off-the-wall path.
  The \emph{wall valuation} of the off-the-wall \Path is a sequence of tile types (or absence of) of \Path in $[x_{w},x_{w}+\delta]\times\{y_{0}\}$ together with the directions of the edges of \Path incident to the tiles on the segment 
  $[x_{w},x_{w}+\delta]\times\{y_{0}\}$.
\end{definition}

The following lemma allows to combine and extend off-the-wall paths.

\begin{lemma}[Off-the-wall combination]
  \label{lem:off-the-wall-combination}
  Let
  $\Path=\Path_0\cdots\Path_{i}\cdots\Path_{l}$ and
  $\Path'=\Path_0'\cdots\Path'_{i'}\dots\Path'_{l'}$ be any two off-the-wall assembly paths that have identical wall valuations.
  There exists a free path $f$ such that: the path $\Path_{\IntegerInterval{0}{i}}.f$ (resp. $\Path'_{\IntegerInterval{0}{i'}}.f$) is an off-the-wall path which has the same wall valuation as $\Path$ (resp. $\Path'$) and the area of its surface above contains the area of the surface above $\Path$ (resp.$\Path'$).
\end{lemma}

\begin{proof}
  The portions above-the-wall of \Path and $\Path'$
  are $\Path_{\IntegerInterval{i}{l}}$ and  $\Path'_{\IntegerInterval{i'}{l'}}$ respectively.
  By definition of the off-the-wall path, everything in the right region of \RefEq{eq:nice:bi-infinite} (included) belongs \NonCausal{\Path_{i}} and $\NonCausal{\Path_{i'}'}$.
  We co-grow the bi-infinite paths 
  $^{\omega}\West.\Path_{\IntegerInterval{i}{l}}.\West^{\omega}$ and 
  $^{\omega}\West.\Path'_{\IntegerInterval{i'}{l'}}.\West^{\omega}$ with forward infinite segments $\Path_{\IntegerInterval{i}{l}}.\West^{\omega}$ and $\Path'_{\IntegerInterval{i'}{l'}}.\West^{\omega}$. 
  Let $f=\CoGrow{^{\omega}\West}{\Path_{\IntegerInterval{i}{l}}.\West^{\omega}}{^{\omega}\West}{\Path'_{\IntegerInterval{i'}{l'}}.\West^{\omega}}$.
  The infinite path $\ZZOrigin.\East^{\omega}$ is in both regions (as $\Path_{i}.\East^{\omega}$ and  $\Path'_{i'}.\East^{\omega}$)  so that $f$ is infinite.
  By \RefLem{lem:co-grow-tile} both $\Path_{\IntegerInterval{0}{i-1}}.\Path_i.f$ and $\Path'_{\IntegerInterval{0}{i'-1}}.\Path_{i'}'.f$ are in \AlphaMax. 
  And because the co-grow $f$ takes the right most segment of the two paths, both original areas of the surfaces above are included in the area of the surface above $\Path_{\IntegerInterval{0}{i-1}}.\Path_i.f$ (resp. $\Path'_{\IntegerInterval{0}{i'-1}}.\Path_{i'}'.f$).
  By definition of co-grow $f$ visits all intersections of $\Path$ and $\Path'$ on the wall, in the same directions, the wall valuation of $f$ remains unchanged.
\end{proof}

Above lemma allows to generate paths in \AlphaMax that are compatible with respect to their wall valuations and extend the areas of the surfaces above of the original paths.

\begin{lemma}
  \label{lem:OtW-unbounded-height}
  If there exists an integer $k$ such that the set of areas above $(y_{0},\delta)$-off-the-wall paths with $\delta<k$ is not bounded, then \TAS is para-periodic.
\end{lemma}

\begin{proof}
  There are finitely many possible wall valuations such that $\delta<k$.
  If the set of areas above $(y_{0},\delta)$-off-the-wall paths is not bounded, there is at least one wall valuation such that the set of areas above for off-the-wall paths with that valuation is not bounded. 
  Let $\hat{\Path}$ be a path with such a wall valuation. Let $\hat{\Path}_{\IntegerInterval{i}{l}}$ be the segment of $\hat{\Path}$ that is above the wall. 

  The combination lemma (\RefLem{lem:off-the-wall-combination}) allows to co-grow $\hat{\Path}$ with any other off-the-wall path with the same valuation. Moreover, each co-grow is also in \AlphaMax. They all pass through $\hat{\Path}_i$, and by \RefLem{lem:off-the-wall-combination}, they are all off-the-wall with the same wall valuation.
  Consider the the subgraph $\hat G$ of \AlphaMax that consists of the union of the co-grows of $\hat{\Path}$ above the wall. 
  This subgraph $\hat G$ must be a subgraph of \AlphaMax (because all co-grow paths are subgraphs of \AlphaMax) and it is infinite because the areas of the surfaces above the co-grows are not bounded. 

  By K\"onig's lemma there is an infinite path $\Path'$ in $\hat G$ starting at point (vertex) $A$ above the $y_0$-wall. Because $A$ is in one of the co-grows of $\hat{\Path}$, there is a path $\Path''$ in \AlphaMax from the origin (seed) to $A$ passing through $\hat{\Path}_i$.
  Therefore $\Path=\Path''.\Path'$ is an infinite assembly path in \AlphaMax.
  Then $\hat{\Path}_i.\East^{\omega}$ is a periodic path in $\IntegerSet^2$ that intersects the infinite assembly path $\Path$ rooted in the seed only at $\hat{\Path}_i$.
  By \RefCor{cor:new-periodic-path}, \TAS is para-periodic.
\end{proof}

In the following we observe that condition in \RefLem{lem:nice} is always satisfied in a confluent system \TAS as soon as there exists an infinite path.
To conclude this, we observe that we can always find off-the-wall paths with the same wall valuation and unbounded set of areas above, therefore satisfying \RefLem{lem:OtW-unbounded-height}. 
We concentrate on paths in \AlphaMax intersecting the $x$-axis an infinite number of times.
If there exists an infinite path in \AlphaMax that does not intersect $\ZZOrigin.\East^{\omega}$ (or $\ZZOrigin.\West^{\omega}$) an infinite number of times, then by \RefLem{lem:nice}, \TAS is para-periodic.
We start by observing that the heights of the $0$-off-the-wall paths that intersect both sides of the $x$-axis is unbounded. 

\begin{lemma}\label{lem:infinite-OtW}
  If there is an infinite assembly path \Path rooted in the seed intersecting both $\ZZOrigin.\East^{\omega}$ and $\ZZOrigin.\West^{\omega}$ an infinite number of times then, up to a vertical symmetry, there exist infinitely many $0$-off-the-wall paths and the set of heights of these paths is unbounded. 
\end{lemma}

\begin{proof}
  Let $E$ be the sets of indices $i$ of \Path such that $\Path_i\in\ZZOrigin.\East^{\omega}$ and all the indices of intersection of \Path and $\ZZOrigin.\East^{\omega}$ east of $\Path_i$ are greater than $i$.
  Let $W$ be defined similarly on the west direction.
  Both $E$ and $W$ are infinite because \Path intersects the $x$-axis an infinite number of times on both sides.
  There are infinitely many pairs $(e,w)$ such that: $e\in E$, $w\in W$, $e<w$, $\IntegerInterval{e+1}{w-1}\cap ( E\cup W ) = \emptyset$. An example is depicted in \RefFig{fig:E+W} where the indices $50$ and $130$ form such a pair. 
  
  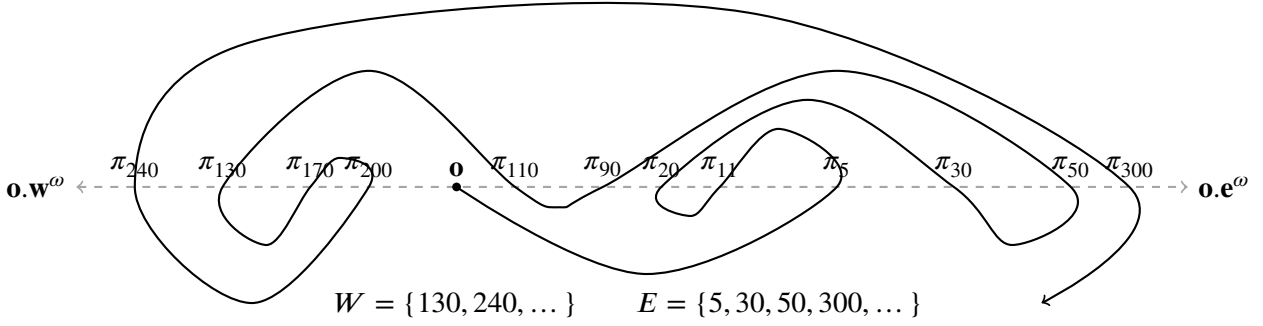
\begin{figure}[hbt]
    \small\SetUnitlength{1em}\newcommand{\PosPart}[2]{\path (#1,0) \CoorNode{p#2} node[above] {$\Path_{#2}$} ; }\newcommand{\MidMv}[4][p]{\path ($.5*(#1#2)+.5*(p#3)$) +(0,#4) \CoorNode{#1#2=#3} ; }\centerline{\begin{tikzpicture}
        \draw[wall,<->] (-17,0) node[left] {$\ZZOrigin.\West^{\omega}$} -- (21,0) node[right] {$\ZZOrigin.\East^{\omega}$} ;
        \path[fill] (-4,0) \CoorNode{O} node[above] {\ZZOrigin} circle (.15) ;
        \PosPart{9}{5}
        \PosPart{5}{11}
        \PosPart{3}{20}
        \PosPart{13}{30}
        \PosPart{17}{50}
        \PosPart{1}{90}
        \PosPart{-2}{110}
        \PosPart{-12}{130}
        \PosPart{-9}{170}
        \PosPart{-7}{200}
        \PosPart{-15}{240}
        \PosPart{19}{300}
        \MidMv[O]{}{5}{-3}
        \MidMv{5}{11}{2}{}{}{}
        \MidMv{11}{20}{-1}{}{}{}
        \MidMv{20}{30}{3}{}{}{}
        \MidMv{30}{50}{-2}{}{}{}
        \MidMv{50}{90}{4}{}{}{}
        \MidMv{90}{110}{-.7}{}{}{}
        \MidMv{110}{130}{4}{}{}{}
        \MidMv{130}{170}{-2}{}{}{}
        \MidMv{170}{200}{1}{}{}{}
        \MidMv{200}{240}{-4}{}{}{}
        \MidMv{240}{200}{5}{}{}{}
        \MidMv{200}{300}{6}{}{}{}
        \MidMv{300}{30}{-4}{}{}{}
        \draw[red] (p5) -- (p5.11) ;
        \draw[path] plot[smooth, tension=.6] coordinates {(O) (O=5) (p5) (p5=11) (p11) (p11=20) (p20) (p20=30) (p30) (p30=50) (p50) (p50=90) (p90) (p90=110) (p110) (p110=130) (p130) (p130=170) (p170) (p170=200) (p200) (p200=240) (p240) (p240=200) (p200=300) (p300)(p300=30)};
        \path (current bounding box.south) node{$W=\{130,240,\dots\}$\qquad$E=\{5,30,50,300,\dots\}$} ;
      \end{tikzpicture}}
    \caption{Points in $E$ and $W$.}
    \label{fig:E+W}
  \end{figure}
  
  One of the left or right region of the bi-infinite path $^{\omega}\West.\Path_{\IntegerInterval{e+1}{w}}.\West^{\omega}$ must contain the entire path $\Path_{\IntegerInterval{0}{e-1}}$.
  If this is the left region, then $\Path_{\IntegerInterval{0}{w}}$ is a $0$-off-the-wall path, otherwise, it is a $0$-off-the-wall path for the system that is vertically symmetric to \TAS.
  In this way we obtain infinitely many $0$-off-the-wall paths for \TAS (or its vertically symmetric one, the definition for off-the-wall paths can be symmetrically extended to include this other case as well).

  Consider $(e,w)$ and $(e',w')$ where $e<w<e'<w'$. Then $\Path_{\IntegerInterval{0}{w'}}$ must have higher height (goes `above') than $\Path_{\IntegerInterval{0}{w}}$. 
  As indexes $(e,w)$ increase, the $0$-off-the-wall paths have to pass one `above' the other with an increasing height. Therefore the set of heights and the set of areas above for these paths is not bounded.
\end{proof}

In the rest of this section we consider that there is a path \Path with infinitely many intersections with both $\ZZOrigin.\East^{\omega}$ and $\ZZOrigin.\West^{\omega}$, that contains infinitely many off-the-wall sub-paths and that set of heights of these paths is unbounded.

For any index $i$ on a finite path \Path, we define its \emph{past}:
\begin{math}
  \PastFrom{i} = \Path_{\IntegerInterval{0}{i-1}}
\end{math}
and its \emph{future}:
\begin{math} 
  \NearFutureFrom{i} = \Path_{\IntegerInterval{i+1}{|\Path|-1}}
\end{math}.
It satisfies $\NearFutureFrom{i}\subseteq\NonCausal{i}$.

The points of interest defined below will be used in the main theorem in this section, and a similar notion is also used in the \RefSec{sec:quipu}.

\begin{definition}[PoI on an off-the-wall path]
  A \emph{point of interest} on an off the wall path \Path is any point that is east-most on any horizontal line above the wall, that is, a point $\Path_{k}$ is
  a point of interest (PoI) if \Path does not intersect $(\Path_{k}+\Vect{\East}).\East^{\omega}$.
\end{definition}

In particular, \PastFrom{k} and \NearFutureFrom{k} do not intersect $\Path_{k}.\East^{\omega}$.
We point out that PoI defined above have similar flavor as the notion of `visible glues' used in~\cite{meunier+regnault16arxiv,meunier+woods17stoc}, except, we are not concerned with the glue per se. 

\begin{theorem}
  \label{th:nice}
  A confluent tiling system \TAS either has a finite \AlphaMax or it is para-periodic.
\end{theorem}

\begin{proof}
  Suppose \AlphaMax is infinite.
  By \RefLem{lem:nice} we can assume that every infinite path in \AlphaMax is intersecting $\ZZOrigin.\East^{\omega}$ and $\ZZOrigin.\West^{\omega}$ infinitely number of times (otherwise it is para-periodic). 
  If the conditions of \RefLem{lem:OtW-unbounded-height} are satisfied, then \TAS is para-periodic. 
  Suppose that for every $\delta$ there is a height such that all $(0,\delta)$-off-the-wall paths have area above less than that height. 
  
  By \RefLem{lem:infinite-OtW} there is a $0$-off-the-wall \Path with height larger than $h$ for every $h$.
  We consider that $h$ is large enough for the existence of $k_1$ and $k_2$ below. 
  We show that either there is an ultimately periodic assembly path in \AlphaMax with the same prefix as \Path, or there is a $0$-off-the-wall path with same wall valuation as \Path and larger surface area above the wall (hence \RefLem{lem:OtW-unbounded-height} holds and \TAS is para-periodic).
  Let $\Path_{k_1}$ and $\Path_{k_2}$ (with $k_1<k_2$) be two distinct PoI on \Path such that $\AlphaMax(\Path_{k_1})=\AlphaMax(\Path_{k_2})$, i.e., the tile types match, and the edges in \Path incident to these vertices are in the same direction. 
  Let $m$ and $u$ be free paths that correspond to  $\Path_{\IntegerInterval{0}{k_1}}$ and $\Path_{\IntegerInterval{k_1}{k_2}}$ then $\Path =\ZZOrigin.mu\NearFutureFrom{k_2}$\,.
  This is illustrated in \RefFig{fig:3-PoI}.

  Consider the free paths $f_2=\NearFutureFrom{k_2}\West^{\omega}$ and $f_1=uf_2$.
  The free paths $f_1$ and $f_2$ differ because there is at least one \North in $u$.
  Let $f$ be the \CoGrow{^{\omega}\West}{f_1}{^{\omega}\West}{f_2}.
  Since the infinite free path $\ZZOrigin.\East^{\omega}$ is in the intersection of the right regions of $^{\omega}\West{f_1}$ and 
  $^{\omega}\West{f_2}$, by \RefLem{inf-co-grow}, $f$ is infinite.

  The infinite free path $f$ is either $f_1$, or $f_2$, or it is a (right most) combination of $u\NearFutureFrom{k_2}$ and $\NearFutureFrom{k_2}$ ending with $\West^{\omega}$.
  Because the right regions of $^{\omega}\West{f_1}$ and $^{\omega}\West{f_2}$ are subsets of $\NonCausal{\Path_i}$, the first vertex of \Path on the wall, by \RefLem{lem:co-grow-tile}, as a subpath of \AlphaMax the free path $f$ can start
  at both $\Path_{k_1}$ and $\Path_{k_2}$.

  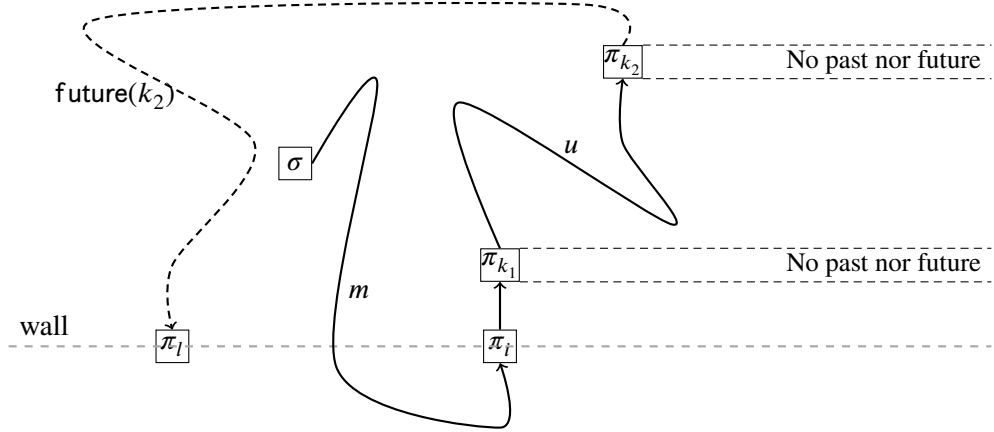
\begin{figure}[hbt]
    \centering\small\SetUnitlength{1.4em}\newcommand{\AddNpNnf}[1]{\begin{scope}[densely dashed]
        \draw (#1.north east) -- (#1.north east-|Rmax) ;
        \path (#1.east-|Rmax) node [left] {\tiny\footnotesize No past nor future} ;
        \draw (#1.south east) -- (#1.south east-|Rmax) ;
      \end{scope}} 
    \tikzset{mark position/.style args={#1(#2)}{postaction={decorate,decoration={markings,mark=at position #1 with \coordinate (#2);}}}}
    \begin{tikzpicture}
      \Tile[S]{-5,2.5}{\Seed}
      \Tile[e]{-8,-2}{\!$\Path_{l}$\!}
      \Tile[i]{0,-2}{$\Path_{i}$}
      \Tile[a2]{3,5}{$\Path_{k_2}$}
      \Tile[a1]{0,0}{$\Path_{k_1}$}
      \path (12,0) \CoorNode{Rmax} ;
      \draw[wall] (-12,-2) node [above right] {wall} -- +(24,0) ;
      \AddNpNnf{a2}
      \AddNpNnf{a1}
      \begin{scope}[path]
        \draw[mark position=0.5(a)] plot [smooth] coordinates { (S.east)
          ($(S)+(2,2)$) ($.6*(S)+.4*(a1)+(-1,-4)$) ($(i)+(0,-2)$) 
          (i.south) } ;
        \draw (i.north)-- (a1.south) ;
        \path (a) node[right] {$m$};
        \draw[mark position=0.5(a)] plot [smooth] coordinates { (a1.north)
          ($(a1)+(-1,4)$) ($.6*(a1)+.4*(a2)+(3,-1)$) ($(a2)+(0,-2)$) 
          (a2.south) } ;
        \path (a) node[left,above] {$u$};
        \draw[mark position=0.7(a),densely dashed] plot [smooth] coordinates { (a2.north) ($(a2.north)+(-1,1)$)
          ($(e)+(-2,8)$) ($(e)+(2,5)$) ($(e)+(0,2)$) 
          (e.north) } ;
        \path (a) node[left] {\NearFutureFrom{k_2}};
      \end{scope}
    \end{tikzpicture}
    \caption{Two PoI with the same tile and exit.}
    \label{fig:3-PoI}
  \end{figure}

  If $f$ is a right-most combination of both $f_1$ and $f_2$, then we consider $f$ starting from both 
  $\Path_{k_1}$ and $\Path_{k_2}$.
  At least one of these assemblies goes strictly to the right of \Path, that is, disconnects from \Path, 
  and then reconnects with it. 
  As depicted in \RefFig{fig:enlarge}, this forms an off-the-wall path that has strictly larger area above the wall without changing the wall valuation. 

  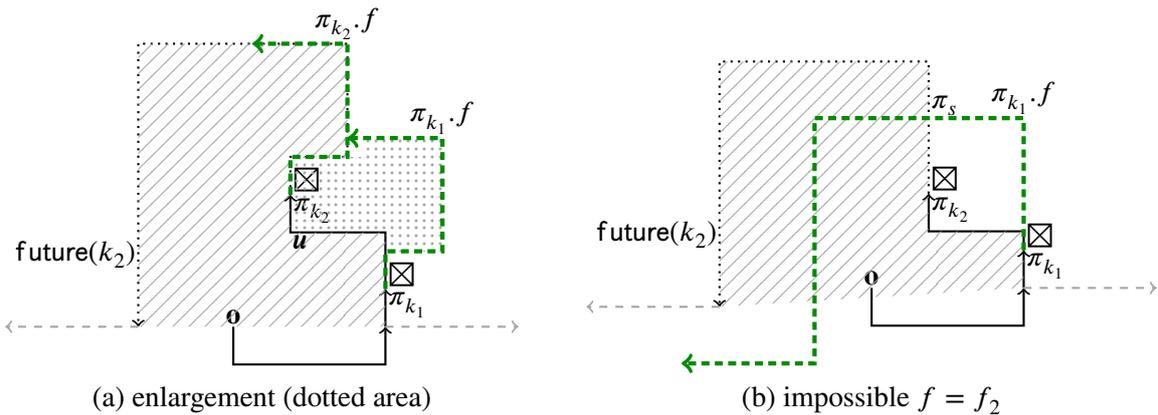
\begin{figure}[hbt]
    \centering\small\SetUnitlength{.65em}\subcaptionbox{enlargement (dotted area)\label{fig:enlarge}}{\begin{tikzpicture}[inner sep=.15ex]
        \newcommand{\VOne}{ \No{2} \Ea{3} \No{6} \VOneTag \We{5}}
        \newcommand{\ZtoO}{ \No{2} }
        \newcommand{\OtoT}{ \No{3} \We{5} \VOtoTTag \No{2} }
        \newcommand{\TtoL}{ \VOne \We{6} \So{15} }
        \newcommand{\VOneTag}{}
        \newcommand{\VOtoTTag}{}
        \path (0,0) \CoorNode{O} node[above] {\ZZOrigin} ;
        \draw[path] (O) \So{2} \Ea{8} \No{2} \CoorNode{I} ;
        \draw[path] (I) \ZtoO \CoorNode{k1} node[below right] {$\Path_{k_1}$} node[above right] {\CoGrowStart} ;
        \renewcommand{\VOtoTTag}{ node[below right] {$u$} }
        \draw[path] (k1)\OtoT \CoorNode{k2} node[below right] {$\Path_{k_2}$} node[above right] {\CoGrowStart} ;
        \draw[non-causal] (k2)\TtoL \CoorNode{L} ;
        \path (L) +(0,3) node[above left] {\NearFutureFrom{k_2}} ;
        \fill[notR1,pattern color=Grey] (I) \ZtoO \OtoT \TtoL -- cycle ;
        \fill[pattern=dots,pattern color=Grey] (k2) \No{2} \Ea{4} \No{1} \Ea{4} \So{6} \We{3} \No{1} \We{5} -- cycle ;
        \draw[wall,->] (I) \Ea{7} ;
        \draw[wall,->] (L) \We{7} ;
        \renewcommand{\VOneTag}{ node[above] {$\Path_{k_1}.f$} }
        \draw[v,ultra thick] (k1) \VOne ;
        \renewcommand{\VOneTag}{ node[above] {$\Path_{k_2}.f$} }
        \draw[v,ultra thick] (k2) \VOne ;
      \end{tikzpicture}}
    \qquad
    \subcaptionbox{impossible $f=f_2$\label{fig:impossible:f2}}{\begin{tikzpicture}[inner sep=.15ex]
        \newcommand{\VOne}{ \No{7} \VOneTag \We{5} \We{6} \So{13} }
        \newcommand{\ZtoO}{ \No{2} }
        \newcommand{\OtoT}{ \No{1} \We{5} \VOtoTTag \No{2} }
        \newcommand{\TtoL}{ \VOne }
        \newcommand{\VOneTag}{}
        \newcommand{\VOtoTTag}{}
        \path (0,0) \CoorNode{O} node[above] {\ZZOrigin} ;
        \draw[path] (O) \So{2} \Ea{8} \No{2} \CoorNode{I} ;
        \draw[path] (I) \ZtoO \CoorNode{k1} node[below right] {$\Path_{k_1}$} node[above right] {\CoGrowStart};
        \draw[path] (k1)\OtoT \CoorNode{k2} node[below right] {$\Path_{k_2}$} node[above right] {\CoGrowStart};
        \draw[non-causal] (k2) \TtoL \CoorNode{L} ;
        \fill[notR1,pattern color=Grey] (I) \ZtoO \OtoT \TtoL -- cycle ;
        \draw[wall,->] (I) \Ea{7} ;
        \draw[wall,->] (L) \We{7} ;
        \path (L) +(0,3) node[above left] {\NearFutureFrom{k_2}} ;
        \renewcommand{\VOneTag}{ node[above] {$\Path_{k_1}.f$} }
        \draw[v,ultra thick] (k1) \VOne \We{7} ;
        \renewcommand{\VOneTag}{ \We{5} node[above right] {$\Path_{s}$} }
        \path (k1) \VOne ;
      \end{tikzpicture}}
    \caption{Simple cases of co-growth.}
    \label{fig:simple-cases}
  \end{figure}

  It is not possible that $f$ is $f_2$ because $\Path_{k_1}.f_2.\West^{\omega}$ (as illustrated in \RefFig{fig:impossible:f2}) would intersect $\Path_{k_2}.f_2$ being the same segment of \Path that starts from a point, $\Path_{k_1}$, more south; thus it would have to pass north of a north shifted version of it-self.
  Let $\Path_s$ be this intersection; it cannot be in the $\West^{\omega}$ part.
  The co-growth $f$ of the two paths takes the right-most segments of the two paths $u\NearFutureFrom{k_2}$ and $\NearFutureFrom{k_2}$.
  Hence we can again take an off-the-wall path that takes the segment $f_2$ from $\Path_{k_1}$ to $\Path_s$ and then again $f_2$ from $\Path_s$ to $\Path_{l}$.
  The area above the wall for this path is larger than the area above of \Path.

  The only case left is $f=f_1$.
  Thus $\Path_{k_2}.u$ is in \AlphaMax.
  Since $\Path_{k_1}+\Vect{u}=\Path_{k_2}$, $\Path_{k_1}.u^2$ is also in \AlphaMax and by \RefLem{lem:double-pumpable} $u$ can form a forward infinite periodic path $u^\omega$.
  If $\Domain{\Path_{k_1}.u^\omega}\subseteq \NonCausal{\Path_{k_1}}$ then there is an infinite ultimately periodic path, namely $\Path_{k_1}.u^{\omega}$ in \AlphaMax and \TAS is para-periodic as depicted in \RefFig{fig:infinitely-pumpable}.

  \begin{figure}[hbt]
    \centering\small\SetUnitlength{.7em}\newcommand{\Ufp}{\No{5} \Ea{2} }
    \newcommand{\ZtoO}{ \No{2} }
    \newcommand{\OtoT}{ \Ufp }
    \newcommand{\TtoL}{ \No{1} \We{3} \No{5} \Ea{8} \No{3} \We{12} \So{16} }
    \subcaptionbox{infinitely pumpable\label{fig:infinitely-pumpable}}{\begin{tikzpicture}[inner sep=.2ex]
        \path (0,0) \CoorNode{O} node[below] {\ZZOrigin} ;
        \draw[path] (O) \No{15} \Ea{2} \So{20} \Ea{2} \No{5} \CoorNode{I} ;
        \draw[path] (I) \ZtoO \CoorNode{k1} node[below right] {$\Path_{k_1}$} node[above right] {\CoGrowStart} ; 
        \draw[path] (k1) \OtoT \CoorNode{k2} node[below right] {$\Path_{k_2}$} node[above right] {\CoGrowStart} ; 
        \draw[non-causal] (k2) \TtoL \CoorNode{L} ;
        \fill[notR1,pattern color=Grey] (I) \ZtoO \OtoT \TtoL -- cycle ;
        \draw[wall,->] (I) \Ea{6} ;
        \draw[wall,->] (L) \We{5} ;
        \path (L) +(0,3) node[above left] {\NearFutureFrom{k_2}} ;
        \draw[v,ultra thick] (k1) \Ufp \Ufp \CoorNode{k1a} ;
        \path (k1) +(0,2.5) node[right] {$u$} ;
        \path (k2) +(0,2.5) node[right] {$u$} ;
        \draw[v,ultra thick] (k1a) \Ufp node[above left] {$u^{\omega}$} ;
      \end{tikzpicture}}\quad\subcaptionbox{enlarging (dotted) by pumping\label{fig:enlarging-by-pumping}}{\begin{tikzpicture}[inner sep=.2ex]
        \path (0,0) \CoorNode{O} node[below] {\ZZOrigin} ;
        \draw[path] (O) \No{15} \Ea{2} \Ea{7} \So{1} \We{7} \So{19} \Ea{2} \No{5} \CoorNode{I} ;
        \draw[path] (I) \ZtoO \CoorNode{k1} node[below right] {$\Path_{k_1}$} node[above right] {\CoGrowStart} ;
        \draw[path] (k1) \OtoT \CoorNode{k2} node[below right] {$\Path_{k_2}$} node[above right] {\CoGrowStart} ; 
        \draw[non-causal] (k2) \TtoL \CoorNode{L} ;
        \fill[notR1,pattern color=Grey] (I) \ZtoO \OtoT \TtoL -- cycle ;
        \fill[pattern=dots,pattern color=Grey] (k2) \Ufp \No{1} \We{5} \So{5} \Ea{3} -- cycle ;
        \draw[wall,->] (I) \Ea{6} ;
        \draw[wall,->] (L) \We{5} ;
        \path (L) +(0,3) node[above left] {\NearFutureFrom{k_2}} ;
        \draw[v,ultra thick] (k1) \Ufp \Ufp \CoorNode{k1a} ;
        \path (k1) +(0,2.5) node[right] {$u$} ;
        \path (k2) +(0,2.5) node[right] {$u$} ;
        \draw[v,ultra thick] (k1a) \No{2} ;
        \path (k1a) \No{1} node[below right] {$\Path_{s}$} ;
      \end{tikzpicture}}\quad\subcaptionbox{enlarging by starting \NearFutureFrom{s} one period earlier \label{fig:one-period-back}}{\newcommand{\FutL}{ \No{1} \Ea{3} \No{3} \We{12} \So{16} }
      \renewcommand{\TtoL}{ \Ufp \FutL }
      \begin{tikzpicture}[inner sep=.2ex]
        \path (0,0) \CoorNode{O} node[below] {\ZZOrigin} ;
        \draw[path] (O) \No{15} \Ea{2} \Ea{7} \So{1} \We{7} \So{19} \Ea{2} \No{5} \CoorNode{I} ;
        \draw[path] (I) \ZtoO \CoorNode{k1} node[below right] {$\Path_{k_1}$} node[above right] {\CoGrowStart} ;
        \draw[path] (k1) \OtoT \CoorNode{k2} node[below] {$\Path_{k_2}$} node[right] {\CoGrowStart} ; 
        \draw[non-causal] (k2) \TtoL \CoorNode{L} ;
        \fill[notR1,pattern color=Grey] (I) \ZtoO \OtoT \TtoL -- cycle ;
        \fill[pattern=dots,pattern color=Grey] (k2) \No{1} \Ea{3} \No{3} \We{3} -- cycle ;
        \draw[wall,->] (I) \Ea{6} ;
        \draw[wall,->] (L) \We{5} ;
        \path (L) +(0,3) node[above left] {\NearFutureFrom{s}} ;
        \draw[v,ultra thick] (k1) \Ufp \Ufp \CoorNode{k1a} ;
        \path (k1) +(0,2.5) node[right] {$u$} ;
        \path (k2) +(0,2.5) node[right] {$u$} ;
        \path (k1a) \No{1} node[below right] {$\Path_{s}$} ; 
        \draw[v,ultra thick] (k1a) \No{2} ;
        \draw[non-causal,densely dotted,very thick,DeepPurple] (k2) \FutL ;
      \end{tikzpicture}}
    \caption{Sub-cases when $f$ starts with $u$.}
    \label{fig:sub-cases}
  \end{figure}
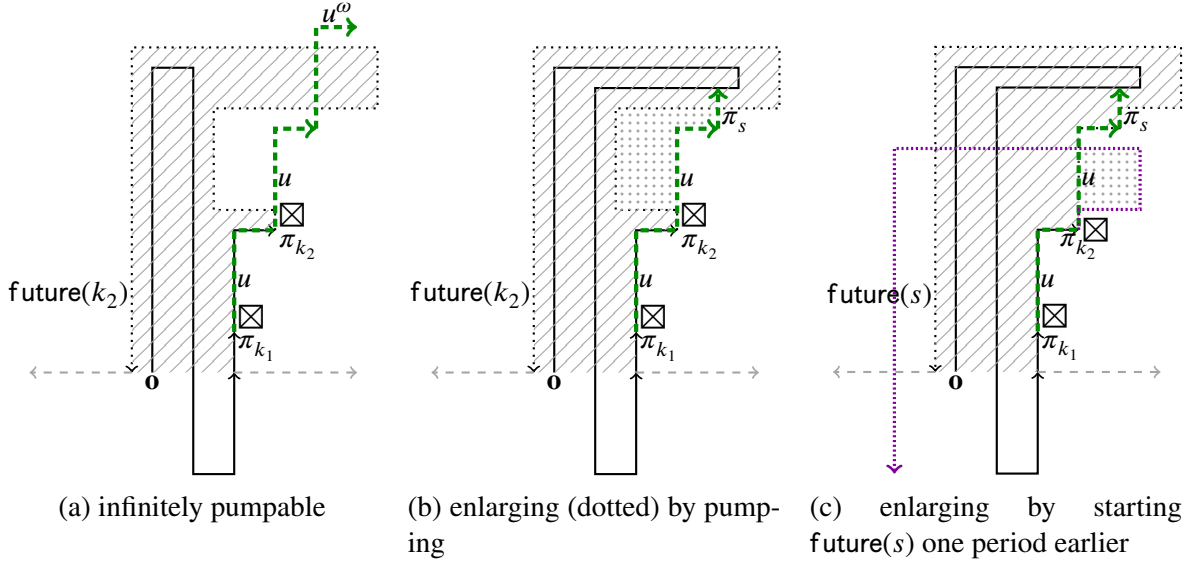

  If $\Domain{\Path_{k_1}.u^\omega}\not \subseteq \NonCausal{\Path_{k_1}}$, that is, $\Path_{k_2}.u^{\omega}$ `bumps' into \PastFrom{k_2}, it has to intersect $\Path_{\IntegerInterval{k_2}{l}}$ first, say at $\Path_s$, because \PastFrom{k_2} is in the left region of \Path.
  If $\Path_{k_2}.u$ is strictly to the right of $\Path_{\IntegerInterval{k_2}{l}}$ before reaching $\Path_s$, the area above can be enlarged by continuing with $\Path_{\IntegerInterval{s}{l}}$ as in \RefFig{fig:enlarging-by-pumping}.
  Otherwise, $\Path_{k_2}.u^{\omega}$ and $\Path_{k_2}.\NearFutureFrom{k_2}$ are the same until $\Path_{k_2}.\NearFutureFrom{k_2}$ turns right of $\Path_{k_2}.u^{\omega}$.
  But in this case let $\Path_s$ to be the vertex of \Path right before its right turn.
  By definition of co-grow, $f$ would take the right turn and $\Path_{k_1}.f$ starts with $\Path_{k_1}.\Path_{\IntegerInterval{k_2}{s}}$ and has the same right turn lower starting from $\Path_{k_1}$ as shown in \RefFig{fig:one-period-back} and enlarges the area above the wall.

  Since all cases are covered, we conclude the proof.
\end{proof}

\section{Basic Structures}
\label{sec:structures}

In this section, we define the two main structures that can be built out of ultimately periodic paths: comb and grid as well as properties useful to characterise grids.
In this section, we concentrate on (ultimately) periodic path in the domain of \AlphaMax. By 
\RefLem{path-assembly}, any periodic path of the domain of an assembly must also be an ultimately  periodic assembly.

\subsection{Comb}
\label{sub:comb}

\begin{definition}[comb]
  Let $k$ be a positive integer and $m$, $p$, $\{m_j,p_j\}_{1\leq j\leq k}$ be free paths such that for all $i$, $i'$, $j$, and $j'$, $p$ and $p_j$ are not \EmptyWord, $mp^{i}m_{j}p_{j}^{\omega}$ is a free path and
  $\left(\ZZOrigin+\Vect{mp^{i}m_{j}}\right).p_{j}^{\omega}$
  and
  $\left(\ZZOrigin+\Vect{mp^{i'}m_{j'}}\right).p_{j'}^{\omega}$
  are defined and disjoint whenever 
  $(i,j)\not=(i',j')$.
  The \emph{comb of $k$ teeth} defined by $\Comb{m}{p}{\{m_i,p_i\}_{1\leq i\leq k}}$ is the following subgraph of $\IntegerSet^2$:
  \begin{equation*}
    (\ZZOrigin+\Vect{m}).p^*
    \left(
      m_1p_1^{\omega} \vee
      m_2p_2^{\omega} \vee
      \dots \vee
      m_kp_k^{\omega} 
    \right)
    \enspace .
  \end{equation*} 
  An assembly $\Assembly$ contains a comb if the comb is a sub-graph of the assembly graph of \Assembly.
\end{definition}

\begin{figure}[hbt]
  \centering\small\footnotesize\centerline{\SetUnitlength{1.1em}\PreparePath{M}{SSEEE}
    \PreparePath{MO}{SS}
    \PreparePath{PO}{SEE}
    \PreparePath{MT}{NWWWN}
    \PreparePath{PT}{ENEE}
    \newcommand{\EEpref}{ \Ea{2} \So{2} \We{1.5} }
    \newcommand{\EE}{ \EEpref \We{.5} \So{1} \Ea{3} }
    \newcommand{\FFpref}{ \Ea{2} \No{2} \We{1.5} }
    \newcommand{\FF}{ \FFpref \We{.5} \No{1} \Ea{4} }
    \subcaptionbox{comb}{\begin{tikzpicture}[inner sep=.15ex]
        \Tile[O]{0,0}{\Seed}
        \path (O) \PathM +(0,-1) \CoorNode{s1} ;
        \Tile[s1]{s1}{$t$}
        \draw[struct-m] (O) \PathM -- (s1) ;
        \path ($.5*(O)+.5*(s1)$) node {$m$} ;
        \path (s1) \FF +(0,1) \CoorNode{n1} ;
        \Tile[n1]{n1}{$u$}
        \draw[struct-p] (s1) \FF -- (n1) ;
        \path (n1) \EE +(0,-1) \CoorNode{s2} ;
        \Tile[s2]{s2}{$t$}
        \draw[struct-p] (n1) \EE -- (s2) ;
        \path (s2) \FF +(0,1) \CoorNode{n2} ;
        \Tile[n2]{n2}{$u$}
        \draw[struct-p] (s2) \FF -- (n2) ;
        \path (n2) \EE +(0,-1) \CoorNode{s3} ;
        \Tile[s3]{s3}{$t$}
        \draw[struct-p] (n2) \EE -- (s3) ;
        \path (s3) \FF +(0,1) \CoorNode{n3} ;
        \Tile[n3]{n3}{$u$}
        \draw[struct-p] (s3) \FF -- (n3) ;
        \path (barycentric cs:s1=1,s2=1 ,n1=1) +(-.5,0) node {$p$} ;
        \path (barycentric cs:s2=1,s3=1 ,n2=1) +(-.5,0) node {$p$} ;
        \foreach \n in {1,2,3} {
          \path (s\n) \PathMO +(0,-1) \CoorNode{ss1} ;
          \Tile[ss1]{ss1}{$t_1$}
          \draw[struct-p] (s\n) \PathMO -- node [left] {$m_1$} (ss1) ;
          \path (ss1) \PathPO +(0,-1) \CoorNode{ss2} ;
          \Tile[ss2]{ss2}{$t_1$}
          \draw[struct-q] (ss1) \PathPO -- (ss2) ;
          \path (ss2) \PathPO +(0,-1) \CoorNode{ss3} ;
          \Tile[ss3]{ss3}{$t_1$}
          \draw[struct-q] (ss2) \PathPO -- (ss3) ;
          \path (barycentric cs:ss1=1,ss2=1) +(0,.5) node {$p_1$} ;
          \path (barycentric cs:ss2=1,ss3=1) +(0,.5) node {$p_1$} ;
        };
        \foreach \n in {1,2,3} {
          \path (n\n) \PathMT +(0,1) \CoorNode{nn1} ;
          \Tile[nn1]{nn1}{$t_2$}
          \draw[struct-p] (n\n) \PathMT -- (nn1) ;
          \path (barycentric cs:n\n=1,nn1=1) +(0,0.2) node {$m_2$} ;
          \path (nn1) \PathPT +(0,1) \CoorNode{nn2} ;
          \Tile[nn2]{nn2}{$t_2$}
          \draw[struct-q] (nn1) \PathPT -- (nn2) ;
          \path (nn2) \PathPT +(0,1) \CoorNode{nn3} ;
          \Tile[nn3]{nn3}{$t_2$}
          \draw[struct-q] (nn2) \PathPT -- (nn3) ;
          \path (barycentric cs:nn1=1,nn2=1) +(0,.5) node {$p_2$} ;
          \path (barycentric cs:nn2=1,nn3=1) +(0,.5) node {$p_2$} ;
        } ;
      \end{tikzpicture}}\subcaptionbox{graph representation}{\begin{tabular}{@{}c@{}}\begin{tikzpicture}
          \Tile[O]{1,0}{\Seed}
          \Tile[s]{5,0}{$t$}
          \Tile[n]{8,4}{$u$}
          \Tile[ss]{10,-5}{$t_1$}
          \Tile[nn]{10,6}{$t_2$}
          \draw[struct-m,->] (O) -- node[above] {$m$} (s) ;
          \draw[struct-p,->] (s) -- (n) ;
          \draw[struct-p,->] (n) .. controls +(-3,0) .. node[below right] {$p$} (s) ;
          \draw[struct-p,->] (s) -- node[left] {$m_1$} (ss) ;
          \draw[struct-q,->] (ss) .. controls +(5,5) and +(5,-5).. node[right] {$p_1$}(ss) ;
          \draw[struct-p,->] (s.east) -- (n.south east) -- (nn) ;
          \path (n) +(1,-1) node[right] {$m_2$} ;
          \draw[struct-q,->] (nn) .. controls +(5,5) and +(5,-5).. node[right] {$p_2$}(nn) ;
        \end{tikzpicture}\\[-1em]
        $m_2=p'm_2'$ where $p'$ is a prefix of $p$
      \end{tabular}}}
  \caption{Comb example.}
  \label{fig:struct:comb}
\end{figure}
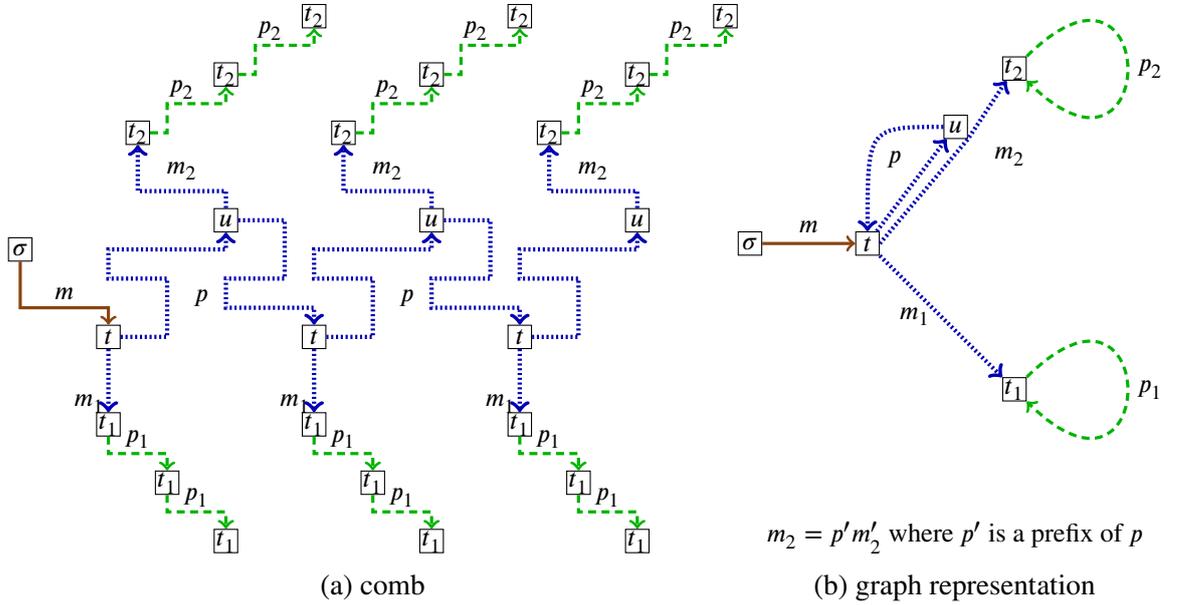

An example of a comb and its graph representation is shown in \RefFig{fig:struct:comb}.

The \emph{transient part} of the comb is the path $\ZZOrigin.{m}$, please note that only $\ZZOrigin+\Vect{m}$ is in the comb.
The \emph{period} of the comb is the free path $p$.

The \emph{backbone} of the comb is the $p$-periodic graph
\begin{math}
  (\ZZOrigin+\Vect{m}).p^*
  \left(
    m_1 \vee
    m_2 \vee
    \dots \vee
    m_k 
  \right)
  \end{math}.
The \emph{tooth $(i,j)$} for $0\leq i$ and $1\leq j\leq k$ is the graph $\left(\ZZOrigin+\Vect{mp^im_j}\right).p_j^{\omega}$.
The \emph{proto-tooth} number $j$ is the union of all the teeth $(i,j)$ for $0<i$.
It is defined by $m$, $p$, $m_{j}$, and $p_{j}$.
The tooth $(i,j)$ is an \emph{instance} of the proto-tooth $j$.
The \emph{period} of the (proto-)tooth $j$ is $p_{j}$.
Since the teeth must be disjoint, their periods are not co-linear to the period of the backbone.

To simplify our discussion, the ultimately periodic paths
that are not in a comb are considered to be the backbones of toothless combs.

Each (proto-)tooth is either on the right or the left side of the backbone.
Because the (proto-)\linebreak[2]teeth on the same side of a comb must not intersect, their periods are co-linear (otherwise there would be intersecting instances) among themselves.
And these periods are not co-linear with the period of the backbone.

\begin{lemma}[bounded teeth]
  \label{lem:struc:bounded-teeth}
  The number of different proto-teeth on each side of a comb is bounded by the length of the backbone period.
\end{lemma}

\begin{proof}
  Each tooth is at least one tile wide, and the period of the backbone provides a finite bound on the number of (disjoint) proto-teeth that could extend infinitely on each side of the backbone.
\end{proof}

\paragraph{Decorations.}
If \AlphaMax contains a comb, the subgraph of \AlphaMax that contains vertices accessible from the backbone of the comb consists of the comb (the backbone and the teeth) and finitely many finite paths starting from the backbone or from the teeth.
These finite paths are called \emph{decorations}.
These finite paths must be repeated periodically as the backbone, and the teeth they are attached to, are repeated periodically.

The comb definition could be enlarged to include the decorations as finitely many finite paths attached to its graph representation; but the definition would be uselessly complex for our analysis.
From now on, decorations on the backbone are considered to be part of the backbone and decorations on any (proto-)tooth are considered to be part of the (proto-)tooth.

\subsection{Grid}

\begin{definition}[grid]
  Let $m$, $p$ and $q$ be free paths such that, $\Vect{p}$ and $\Vect{q}$ are not co-linear and $^{\omega}p^{\omega}$ and $^{\omega}q^{\omega}$ are defined bi-infinite free periodic paths.
  The \emph{grid} \Grid{m}{p}{q} is the sub-graph of $\IntegerSet^2$ that is the union of all the paths:
  \begin{equation*}
    ^{\omega}p.
    ( \ZZOrigin + \Vect{m} +i\Vect{p}+j\Vect{q})
    .p^{\omega}
    \quad
    \text{and}
    \quad
    ^{\omega}q.
    ( \ZZOrigin + \Vect{m} +i\Vect{p}+j\Vect{q})
    .q^{\omega}
    \quad
    i,j\in\IntegerSet
    \enspace.
  \end{equation*} 
  We say that an assembly is a \emph{grid} whenever its binding graph contains a grid. 
\end{definition}

\begin{figure}[hbt]
  \centering\SetUnitlength{1em}\PreparePath{M}{EEEES}
  \PreparePath{P}{ENNN}
  \PreparePath{PP}{SSSW}
  \PreparePath{Q}{ENEEE}
  \PreparePath{QQ}{WWWSW}
  \begin{tikzpicture}
    \begin{scope}[shift={(-10,-2)}]
      \begin{scope}[shift={(0,10)}]
        \draw (-1,0) node[left] {$m$} ;
        \draw[struct-m] (0,0) \PathM ;    
      \end{scope}
      \begin{scope}[shift={(0,5)}]
        \draw (-1,0) node[left] {$p$} ;
        \draw[struct-p] (0,0) \PathP ;
      \end{scope}
      \begin{scope}[shift={(0,0)}]
        \draw (-1,0) node[left] {$q$} ;
        \draw[struct-q] (0,0) \PathQ ;
      \end{scope}
    \end{scope}
    \fill[pattern=NE lines,pattern color=DarkGrey] (5,4) \PathP \PathQ \PathPP \PathQQ ;
    \draw[struct-m] (4,3) node[left] {\ZZOrigin} \PathM ;
    \foreach \x in {-1,0,1,2,3} {
      \foreach \y in {-1,0,1,2,3} {
        \begin{scope}[shift={(\x*1+\y*4,\x*3+\y*1)}]
          \draw[struct-p] (0,0) \PathP ;
          \draw[struct-q] (0,0) \PathQ ;
        \end{scope}
      };
    };
  \end{tikzpicture}\caption{Example of \Grid{m}{p}{q} and a face (hatched).}
  \label{fig:grid}
\end{figure}
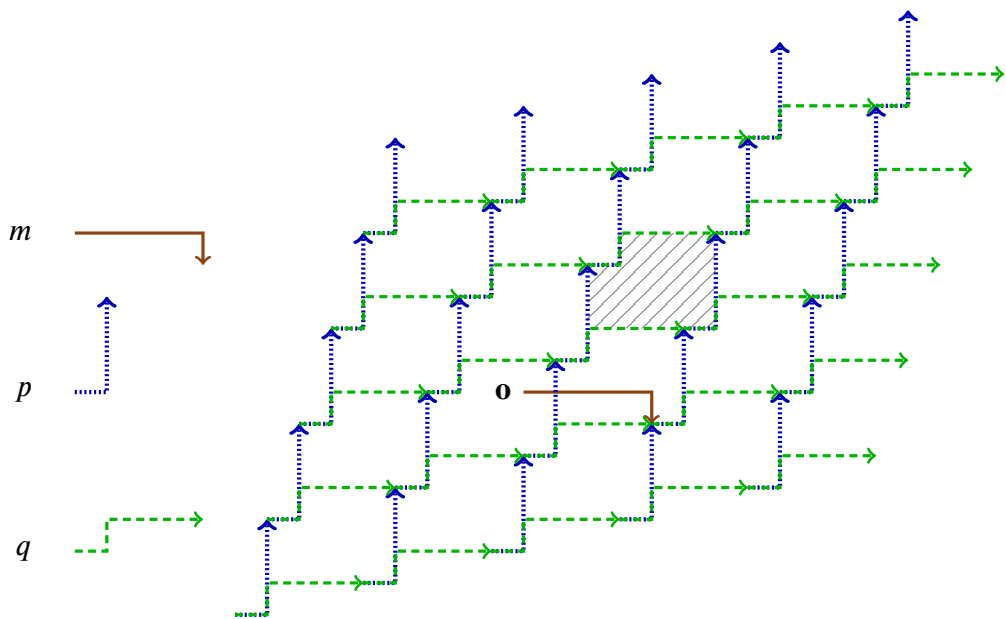

An example of a grid is represented in \RefFig{fig:grid}.

\begin{definition}[(seed) face]
  The $(i,j)$-\emph{face} of \Grid{m}{p}{q} is the finite sub-graph that is the intersection of the region between
  $^{\omega}p.
  ( \ZZOrigin + \Vect{m} +i\Vect{p}+j\Vect{q})
  .p^{\omega}$
  and   $^{\omega}p.
  ( \ZZOrigin + \Vect{m} + (i+1)\Vect{p}+j\Vect{q})
  .p^{\omega}$
  and the region between
  $^{\omega}q.
  ( \ZZOrigin + \Vect{m} +i\Vect{p}+j\Vect{q})
  .q^{\omega}$ and $^{\omega}q.
  ( \ZZOrigin + \Vect{m} +i\Vect{p}+(j+1)\Vect{q})
  .q^{\omega}$.
  
  The \emph{seed face} for the grid is the face that contains the origin vertex \ZZOrigin.
\end{definition}

Due to confluence of \TAS all faces that do not intersect $\ZZOrigin.m$ are finite and equal up to translations. Such a face is called a {\it regular} face of a grid.

If \AlphaMax is a grid, then there are infinitely many $m$, $p$ and $q$ such that \Grid{m}{p}{q} is included in \AlphaMax. 
Each of these grids have their own seed face, hence the seed face in \AlphaMax that is a grid is not unique.
If the vertex \ZZOrigin is on the border of a face, then the grid can be replaced by a coarser one such that $\ZZOrigin$ is not in the boundary of the face and the seed face is unique for this particular grid.
Similarly the {\it transient part of the grid} $m$ is finite, and there is a coarser grid for \AlphaMax such that $\ZZOrigin.m$ is entirely in the seed face.

In the rest of the paper, we always consider that the seed face is unique and that $m$ is included in the seed face. In this way, the seed face is the only face that is different than the other faces. 

The following lemma shows that in a grid there is a shifted copy of any periodic path that extends infinitely in both directions.

\begin{lemma}[extending periodic paths]
  \label{lem:struc:grid:period extension}
  Suppose \AlphaMax contains \Grid{m}{p}{q}.
  Let $r$ be a free path and $A$ be a vertex in $\IntegerSet^2$, 
  if $A.r^{\omega}\SubGraphOf\AlphaMax$, then there is a vertex $B$ such that
  $^{\omega}r.B.r^{\omega}\SubGraphOf\AlphaMax$.
\end{lemma}

\begin{proof}
  Since the vectors \Vect{p} and \Vect{q} are non co-linear they form a base of the \RationalSet-vector space $\RationalSet^2$ which contains $\IntegerSet^2$.
  So \Vect{r} can be decomposed over \Vect{p} and \Vect{q} with rational coefficients and 
  some multiple of \Vect{r} is an integer combination of \Vect{p} and \Vect{q}. Say $s\Vect{r}=t_1\Vect{p}+t_2\Vect{q}$. 
  
  At the level of faces of the grid, this multiple $s\Vect{r}$ enters the face of $A+t_1\Vect{p}+t_2\Vect{q}$ exactly as it entered the face containing $A$, so due to confluence, $r^s.A$ is also part of \AlphaMax, and therefore $r^s$ can be repeated backwards from $A$ at face level.
  This repetition can be interrupted only if the path enters an `irregular' face.
  The seed face is the only irregular face.

  If $^{\omega}r.A.r^{\omega}$ intersects the seed face, then there are integers $i$ and $j$ such that $^{\omega}{{r}}.(A+i\Vect{p}+j\Vect{q}).r^{\omega}$ does not intersect the seed face.
\end{proof}

\paragraph{Decorations.}
If \AlphaMax is a grid, the domain of \AlphaMax that contains a face may contain finite paths that start at the boundary of a face. These finite paths are repeated in all faces except maybe in the seed face.
These additional paths are called \emph{decorations}.
So when \AlphaMax is a grid, it is completely defined by the grid, the decorations of a regular face and the decorations of the seed face.
For a given grid, there are a finite number of decorations of the seed face, finite number of decorations of a regular face and all decorations are finite paths.

\subsection{Grid characterization}

\begin{lemma}[intersection of periodic paths]
  \label{lem:intersecting-period:assemble}
  Let $m$, $p$ and $q$ be free paths such that $p$ and $q$ are non co-linear.
  If \AlphaMax contains
  $\ZZOrigin.m$,
  $(\ZZOrigin+\Vect{m}).p^{\omega}$,
  $(\ZZOrigin+\Vect{m}).q^{\omega}$,
  $(\ZZOrigin+\Vect{m}+\Vect{p}).q$, and 
  $(\ZZOrigin+\Vect{m}+\Vect{q}).p$, then \AlphaMax is a grid.
\end{lemma}

\begin{figure}[hbt]
  \centering\SetUnitlength{1.2em}\newcommand{\Setting}{\TileDot[O]{3,0.3}{}
    \path (O) node[right] {\ZZOrigin} ;
    \path (3.3,2) \CoorNode{m} ;
    \draw[struct-m,transient] (O) .. controls +(-1,1) and +(-1,2.5) .. node[left] {$m$} (m) ;
  }\subcaptionbox{situation\label{fig:intersecting-period:rename}}{
    \SetUnitlength{1.3em}\begin{tikzpicture}
      \Setting
      \begin{scope}[path]
        \draw[struct-p,solid] (m) -- node[below] {$p$} ++(2,0) coordinate (p1) ;    
        \draw[struct-p] (p1) -- node[below] {$p^{\omega}$} ++(2,0) ;
        \draw[struct-q,dotted,solid] (p1) -- node[right] {$q$} ++(1.3,1.3) coordinate (p1);
        \draw[struct-q,dotted] (p1) -- node[above] {$q^{\omega}$} ++(1.3,1.3) coordinate (p1);
        \draw[struct-q,solid] (m) -- node[above] {$q$} ++(1.3,1.3) coordinate (p2) ;   
        \draw[struct-q] (p2) -- node[above] {$q^{\omega}$} ++(1.3,1.3) ;
        \draw[struct-p,dotted,solid] (p2) -- node[pos=.75,above=-.2\unitlength] {$p$} ++(2,0) coordinate (p2) ;
        \draw[struct-p,dotted] (p2) -- node[pos=.9,below] {$p^{\omega}$} ++(2,0) coordinate (p2) ;
      \end{scope}
    \end{tikzpicture}}
  \subcaptionbox{expanded\label{fig:intersecting-period:expanded}}{\newcommand{\Wid}{3.5}\begin{tikzpicture}
      \Setting\begin{scope}[path]
        \draw[struct-p] (m) -- node[below] {$p^{a}$} ++(\Wid,0) coordinate (ne) ;    
        \draw[struct-q] (ne) -- node[left] {$q^{b}$} ++(1.3,1.3) coordinate (ne) ;
        \draw[struct-p] (ne) ++(-2*\Wid,0) coordinate (nw) -- node[above] {$p^{c}$} (ne) ;    
        \draw[struct-q] (nw) ++(-\Wid,-\Wid) coordinate (sw) -- node[left] {$q^{d}$} (nw) ;    
        \draw[struct-p] (sw) -- node[below] {$p^{c}$} ++(2*\Wid,0) coordinate (se) ;    
        \draw[struct-q] (se) -- node[right] {$q^{d}$} (ne) ;    
        \draw[struct-p,dotted] (ne) -- node[below] {${p^{d}}^{\omega}$} ++(6,0) ;
        \draw[struct-p,dotted] (se) -- node[below] {${p^{d}}^{\omega}$} ++(6,0) ;
        \draw[struct-p,dotted,<-] (nw) -- node[below] {${p^{d}}^{\omega}$} ++(-6,0) ;
        \draw[struct-p,dotted,<-] (sw) -- node[below] {${p^{d}}^{\omega}$} ++(-6,0) ;
        \draw[struct-q,dotted] (ne) -- node[right] {${q^{d}}^{\omega}$} ++(2,2) ;
        \draw[struct-q,dotted] (nw) -- node[right] {${q^{d}}^{\omega}$} ++(2,2) ;
        \draw[struct-q,dotted,<-] (se) -- node[right] {${q^{d}}^{\omega}$} ++(-2,-2) ;
        \draw[struct-q,dotted,<-] (sw) -- node[right] {${q^{d}}^{\omega}$} ++(-2,-2) ;
      \end{scope}
    \end{tikzpicture}}
  \caption{Intersecting periodic parts.}
  \label{fig:intersecting-period}
\end{figure}
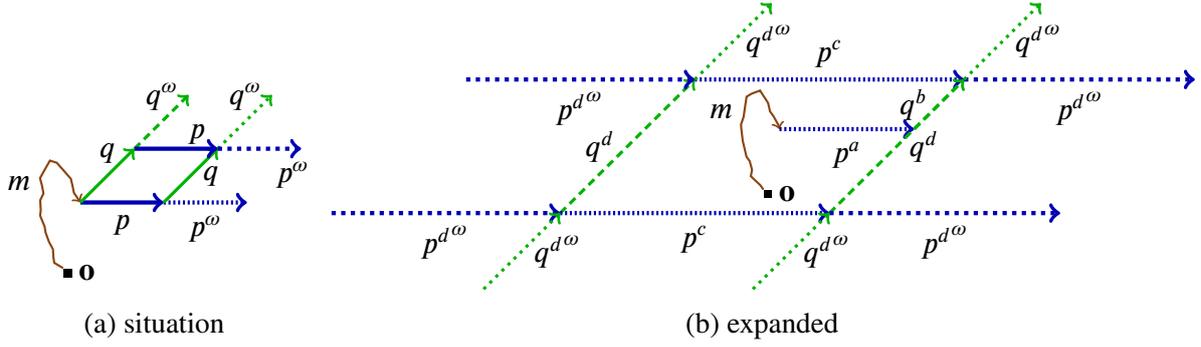

\begin{proof}
  The situation is depicted in \RefFig {fig:intersecting-period:rename} where the assumed paths are depicted with non-dashed lines.
  Note that $\ZZOrigin.p^\omega$ and $\ZZOrigin.q^\omega$ are paths in \AlphaMax and because they can be extended in $\NonCausal{\ZZOrigin+\Vect{m}+\Vect{p}+\Vect{q}}$ from the point $\ZZOrigin+\Vect{m}+\Vect{p}+\Vect{q}$ (\RefLem{non-causal-extension}), the free paths $p^{\omega}$ and $q^{\omega}$ can start at that point and are part of \AlphaMax.
  This means that for each $i,j$, the paths
  $(\ZZOrigin+\Vect{m}+i\Vect{p}+j\Vect{q}).p$
  and $(\ZZOrigin+\Vect{m}+i\Vect{p}+j\Vect{q}).q$ are part of an assembly.

  Since $\ZZOrigin.m$ is finite, by taking $a$, $b$, $c$ and $d$ large enough, the seed face of \Grid{mp^{a}q^{b}}{p^{c}}{q^{d}} contains $\ZZOrigin.m$ and this grid is in \AlphaMax (See \RefFig{fig:intersecting-period:expanded}).
\end{proof}

\begin{lemma}[intersection of periodic paths]
  \label{lem:intersecting-period:unbounded}
  Let $m$, $p$ and $q$ be non empty free paths such that $\Vect{p}$ and $\Vect{q}$ are non co-linear.
  If \AlphaMax contains $\ZZOrigin.m$, $(\ZZOrigin+\Vect{m}).p^{\omega}$, and the following corresponds to an assembly (not necessarily producible):
  $^{\omega}p.\ZZOrigin.p^{\omega} \SubGraphCup {^{\omega}q}.\ZZOrigin.q^\omega$  (with the same tiles for $p$) then \AlphaMax is a grid.
\end{lemma}

\begin{proof}
  Since $m$ is finite and $p$ and $q$ are non co-linear, with a sufficiently large $k$, the subset of $\IntegerSet^2$ ${^{\omega}q}.(\ZZOrigin+k\Vect{p}).q^\omega$ does not intersect $m$.
  So that this subgraph can be assembled.
  It can also be assembled shifted by $p$.
  Similarly $p^{\omega}$ can be assembled shifted by $q$.
  Then \AlphaMax is a grid by \RefLem{lem:intersecting-period:assemble}.
\end{proof}

The condition of above Lemma can be replaced by $^{k}p.\ZZOrigin.p^{k} \SubGraphCup {^{\omega}q}.\ZZOrigin.q^\omega$ with $k$ large enough so that both $\ZZOrigin+k\Vect{p}$ and $\ZZOrigin-k\Vect{p}$ are at Manhattan distance greater than $|p|$ of ${^{\omega}q}.\ZZOrigin.q^\omega$.
If \AlphaMax is a grid, there exists $m$, $p$ and $q$ such that the condition is true.

We observe that every $\Grid{m}{p}{q}$ contains a \Comb{m}{p}{\{q,q\}}. 
In a confluent system, the segments can be assembled to and from a comb, or between the teeth, are limited by the following lemmas (in the spirit of the observations in \citep{doty+patitz+summers11tcs}).
These lemmas show that if there is an extra path from the seed 
to a given point allowing a part of the comb and the teeth to take part in a non-causal region of that point, then, because paths are compatible, assemblies of the parts of the comb can be obtained in another way, thus revealing a grid in \AlphaMax. 

\begin{lemma}[connecting teeth]
  \label{lem:tooth-to-tooth}
  Suppose \AlphaMax has a comb.
  If there is a single edge connection in the binding graph of \AlphaMax from one tooth $T_1$ of a comb to a disjoint tooth $T_2$ of a (possibly different) comb, then \AlphaMax is a grid.
\end{lemma}

\begin{figure}[hbt]
  \centering\small\SetUnitlength{1.3em}\newcommand{\Xperiod}{5}
  \subcaptionbox{\label{fig:tooth-to-tooth:situation}situation}{
    \begin{tikzpicture}[path,inner sep=.2ex]
      \draw[struct-p] (-2,0) -- (0,0) ; 
      \draw[struct-p] (0,0) -- node[below] {$p$} (\Xperiod,0) ;  
      \draw[struct-p] (\Xperiod,0) -- node[below] {$p^{\omega}$} ++(\Xperiod,0) ;      
      \draw[struct-m,transient] (1,0) -- node[left] {$m_{j}$} +(0,2) coordinate (p1) ;
      \draw[struct-q] (p1) -- node[left] {$p_{j}$} +(0,3) coordinate (p1);
      \draw[struct-q] (p1) -- node[left] {$p_{j}$} +(0,3) coordinate (p1);
      \draw[struct-q] (p1) -- node[left] {$p_{j}^{\omega}$} +(0,3) coordinate (p1) ;
      \draw[struct-m,transient] (3,0) -- node[right] {$m_{k}$} +(0,2.5) coordinate (p2) ;
      \draw[struct-q] (p2) -- node[right] {$p_{k}$} +(0,3) coordinate (p2) ;
      \draw[struct-q] (p2) -- node[right] {$p_{k}$} +(0,3) coordinate (p2) ;
      \draw[struct-q] (p2) -- node[right] {$p_{k}^{\omega}$} +(0,3) coordinate (p2) ;
      \draw[dotted] (1,2) -- node[above] {$q$} +(2,.5) ;
      \draw (1,2) ++(0,3) -- node[above] {$q$} +(2,.5) ;
      \draw[dotted] (1,2) ++(0,3) ++(0,3) -- node[above] {$q$} +(2,.5) ;
    \end{tikzpicture}}
  \qquad
  \subcaptionbox{\label{fig:tooth-to-tooth:expanded}construction}{
    \begin{tikzpicture}[path,inner sep=.2ex]
      \draw[struct-p] (-2,0) -- (0,0) ; 
      \draw[struct-p] (0,0) -- +(1,0) ;
      \path (1,0) node[below] {$B$} ;
      \begin{scope}
        \draw[struct-p,dotted,thin] (0,0) -- node[below] {$p$} +(\Xperiod,0)coordinate (pp1) ;
        \draw[struct-p,dotted,thin] (pp1) -- node[below] {$p^{\omega}$} ++(\Xperiod,0) ; 
        \draw[struct-m,transient,thin] (1,0) -- node[left] {$m_{j}$} +(0,2) coordinate (p1) ;
        \draw[struct-q,dotted,thin] (p1) -- node[left] {$p_{j}^{l}$} +(0,6) coordinate (p1) ;
        \draw[struct-q,dotted,thin] (p1) -- node[left] {$p_{j}^{\omega}$} +(0,3) coordinate (p1) ;   
        \draw[struct-q,dotted,thin] (1,2) ++(0,3) ++(0,3) -- node[above] {$q$} +(2,.5) coordinate (a) node[above right] {$A$} ;
      \end{scope}
      \begin{scope}[densely dashed]
        \draw[struct-p,thin] (0,6) -- node[below] {$p$} ++(\Xperiod,0) coordinate (pp1) ;
        \draw[struct-p,thin] (pp1) -- node[below] {$p^{\omega}$} ++(\Xperiod,0) ;   
        \draw[struct-m,transient,->,thin] (a) -- node[right] {$\RevBack{{m_{k}}}$} +(0,-2.5) coordinate (p2) ;
      \end{scope}
    \end{tikzpicture}}
  \caption{Path $q$ between teeth $i$ and $j$.}
  \label{fig:tooth-to-tooth}
\end{figure}
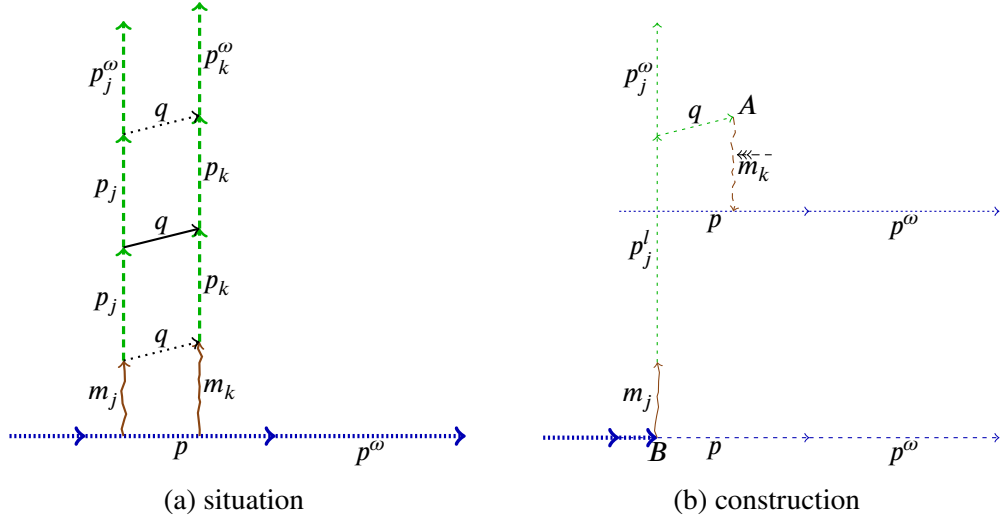

\begin{proof}
  We first suppose that both teeth $T_1$ and $T_2$ belong to the same comb and let $d$ be the direction of the edge connecting a vertex in $T_1$ with a vertex in $T_2$. This implies that the two vertices in $\IntegerSet^2$ are mapped by \AlphaMax to tiles whose sides $d$ and, respectively $-d$, have matching glues. 
  Consider the path that starts from 
  a vertex of the periodic portion of $T_1$, to the endpoint of
  the connecting edge  in $T_1$, following direction $d$ (this is some decoration) ending with the other endpoint of the edge that leads to $T_2$. Let $q$ be the associated free path. 
  
  If there is such a path between teeth $(\ZZOrigin+\Vect{mp^i m_j}).p_j^{\omega}$ and $(\ZZOrigin+\Vect{m p^i m_k}).p_{k}^{\omega}$, then by periodicity, and the confluence of \TAS, there is such a path for any $i$ 
  (starting from some value), and the path $q$ repeats infintely often between $T_1$ and $T_2$. Moreover, the periods between consecutive vertices in $T_1$ that are starts of these paths must have equal displacement vector to the one between consecutive vertices in $T_2$. 
  Let $q$ be the associated free path.
  Up to some rotations of $p_{j}$ and $p_{k}$, we can assume that $q$ goes from the beginning of some $p_{j}$ to the beginning of some $p_{k}$.
  The situation is as depicted in \RefFig{fig:tooth-to-tooth:situation}.

  Let $A=(\ZZOrigin+\Vect{mp^im_jp_j^l q})$ for $l$ sufficiently large be a vertex in $T_2$.
  Since there are two paths in \AlphaMax that lead to 
  $A$ from the `last' backbone vertex leading to $T_1$ and $T_2$ 
  (point $B$ in \RefFig{fig:tooth-to-tooth:expanded}), all vertices of paths starting at $B$ leading to vertex $A$ are in \NonCausal{A} (thin paths).
  Therefore there is a path $A.\RevBack{{m_{k}}}$ forming a cycle $(\ZZOrigin+\Vect{mp^im_jp_j^l}).q\RevBack{{m_{k}}}\RevBack{{p'}}$ for some factor $p'$ of $p$.
  So $(\ZZOrigin+\Vect{mp^im_jp_j^l}).p^{\omega}$ exists in \AlphaMax
  arbitrarily away from the backbone (that is, $l>(|p_j|+|p|)$).
  By \RefLem{lem:intersecting-period:assemble}, \AlphaMax is a grid.

  If the two teeth $T_1$ and $T_2$ belong to different combs, the analysis follows the same argument except this time the period of the backbone of the second comb is assembled at point $\ZZOrigin+\Vect{mp^im_jp_j^l}$.
\end{proof}

\begin{lemma}[disjoint comb connection]
  \label{lem:comb-connections}
  Let $C_1$ and $C_2$ be two disjoint combs in \AlphaMax that have infinitely many edges in the binding graph of \AlphaMax with one endpoint in $C_1$ and the other in $C_2$. 
  Then either \AlphaMax is a grid or the connecting edges connect vertices of the comb backbones.
\end{lemma}

\begin{proof}
  Since the combs are disjoint, each comb is in the non-causal region for every vertex of the other comb.
  Sufficiently away from the seed, due to the connecting edges, any path that starts at one comb and continues with a path in the other comb exists in \AlphaMax. 
  If the connecting edges between the combs appear between the teeth, then \AlphaMax is a grid by \RefLem{lem:tooth-to-tooth}.
  
  Suppose one tooth $T$ of $C_1$ has infinite number of edges connecting it to the backbone of $C_2$ in \AlphaMax, then (sufficiently away from the seed) there is a copy of $C_2$ at any instance of the proto-tooth containing $T$.
  In particular, the backbone of $C_2$ (similarly as in \RefLem{lem:tooth-to-tooth}) can be infinitely extended 
  from a vertex in a far copy of $T$ in $C_1$.
  Therefore, \AlphaMax is a grid.
\end{proof}

\section{Quipu  Definition and Properties}
\label{sec:quipu}

In this section, we define what quipus are and some primitive operation on them.

\begin{definition}[Quipu]
  \label{def:quipu}
  A \emph{quipu associated to \TAS} is $\Quipu=(V,E,\Root,\VertexLabel,\ArcLabel)$ such that:
  $(V,E)$ is a finite directed graph, $ \Root \in V$ is called the {\it root}, 
  $\VertexLabel:V\to \TileSet\cup\{\Seed\}$ is labelling of vertices with tile types, and
  $\ArcLabel:E\to \DirectionSet$ is a labelling of arcs with directions
  that satisfies:
  \begin{enumerate}
  \item the in-degree of the root (\Root) is zero, $\VertexLabel(\Root)=\Seed$ and no other vertex is labelled with \Seed unless $\Seed\in \TileSet$,
  \item for every arc $e=(x,y)$ with $\lambda(e)=d$, we have $\eta(x)_d=\eta(y)_{\Backward{d}}$, that is, the glue of tile type $\VertexLabel(x)$ in direction $\ArcLabel(e)=d$ matches the one of $\VertexLabel(y)$ in direction $\Backward{\ArcLabel(e)}$,
  \item there is a unique path in $(V,E)$ from the root $r$ to every vertex, and the $\lambda$-labels are deterministic,
  \item\label{def:quipu:one-arc-to-cycle}
    no two cycles share a vertex, and there is exactly one arc incoming to any cycle,
  \item any path meets at most two cycles,
  \item\label{def:quipu:free-path}
    the sequence of edge labels (\ArcLabel-labels) of any walk from \Root (finite or infinite) is a free path, in particular, the sequence of \ArcLabel-labels of any cycle has a non-null displacement $\IntegerSet^2$-vector, and
  \item\label{def:quipu:unique-walk}
    the sequences of \ArcLabel-labels of any two distinct rooted walks, $w_1$ and $w_2$ correspond to different $\IntegerSet^2$-vectors ($\Vect{w_1}\neq\Vect{w_2}$).
  \end{enumerate}
\end{definition}

The shape of the graph of a quipu is illustrated in \RefFig{fig:zones}. Observe that \RefCndEnum{def:quipu:free-path} implies that no two consecutive arcs in \Quipu have a label $d\Backward{d}$.
Hence each vertex that is not a \Root has at most three outgoing arcs. 

Directly from the definition of quipu we have the lemma below.

\begin{lemma}
  Let \Quipu be any quipu and $p$ be the free path formed by a sequence of the \ArcLabel-labels of any rooted (finite or infinite) walk.
  The path $\ZZOrigin.p$ is in \AlphaMax, and for every vertex $x$ visited by $\ZZOrigin.p$, $\VertexLabel(x)=\AlphaMax(x)$.
\end{lemma}

The \emph{assembly corresponding to a quipu \Quipu}, \AlphaQuipu is the union of $\lambda$-labels of all rooted paths in \Quipu.
By the above Lemma, $\AlphaQuipu\SubGraphOf\AlphaMax$ and \AlphaQuipu is a producible assembly.
As with \AlphaMax, we refer to \AlphaQuipu as to an assembly, as well as to a subgraph of $\IntegerSet^2$.
Whenever there might be ambiguities, we use \Quipu-paths to address paths in the quipu and similarly $\AlphaMax$-path, $\IntegerSet^2$-path, \Quipu-vertex, etc.
In the rest of the paper we show that for every \TAS there is a quipu such that $\AlphaQuipu=\AlphaMax$.

\begin{figure}[hbt]
  \centering
  \small\newcommand{\DisplayThree}[4][]{\begin{scope}[shift={(#4)}]
      \TileDot[#31]{0,0}{}
      \TileDot[#32]{1.45,.9}{}
      \TileDot[#33]{1.45,-.9}{}
      \draw[#2] (#31) to [bend left] node [above] {#1} (#32) ;
      \draw[#2] (#32) to [bend left] (#33) ;
      \draw[#2] (#33) to [bend left] (#31) ;
    \end{scope}
  }
  \newcommand{\DisplayFour}[3]{\begin{scope}[shift={(#3)}]
      \TileDot[#21]{0,0}{}
      \TileDot[#22]{1,1}{}
      \TileDot[#23]{2,0}{}
      \TileDot[#24]{1,-1}{}
      \draw[#1] (#21) to [bend left] (#22) ;
      \draw[#1] (#22) to [bend left] (#23) ;
      \draw[#1] (#23) to [bend left] (#24) ;
      \draw[#1] (#24) to [bend left] (#21) ;
    \end{scope}
  }
  \subcaptionbox{the three zones\label{fig:zones}}{\newcommand{\EndZZ}{5}
    \newcommand{\EndZO}{{2*\EndZZ}}
    \SetUnitlength{1.4em}
    \begin{tikzpicture}
      \TileDot[O]{0,1}{}
      \TileDot[O1]{[shift={(1.5,1)}]O}{}
      \TileDot[O11]{[shift={(1,2.5)}]O1}{}
      \path (O) node [left] {\Root} ;
      \DisplayThree{struct-z1}{a}{\EndZZ,0}
      \TileDot[a21]{[shift={(1.5,-.25)}]a2}{}
      \DisplayFour{struct-z1}{b}{\EndZZ,4.5}
      \TileDot[b21]{[shift={(2,.5)}]b2}{}
      \TileDot[b41]{[shift={(1,-1)}]b4}{}
      \DisplayThree{struct-q}{c}{\EndZO,-.5}
      \DisplayFour{struct-q}{d}{\EndZO,3.5}
      \DisplayThree{struct-q}{e}{\EndZO,6.5}
      \TileDot[d21]{[shift={(2,0)}]d2}{}
      \TileDot[e21]{[shift={(1.5,.25)}]e2}{}
      \draw[transient,struct-z0] (O) -- +(3,-1.5) ;
      \draw[transient,struct-z0] (O) -- (O1) ;
      \draw[transient,struct-z0] (O1) -- (a1) ;
      \draw[transient,struct-z0] (O1) -- (b1) ;
      \draw[transient,struct-z0] (O1) -- +(2.5,.5) ;
      \draw[transient,struct-z0] (O1) -- (O11) ;
      \draw[transient,struct-z0] (O11) -- +(1,1.5) ;
      \draw[transient,struct-z0] (O11) -- +(1.5,1) ;
      \draw[transient,struct-z1] (a2) -- (a21) ;
      \draw[transient,struct-z1] (a21) -- (c1) ;
      \draw[transient,struct-z1] (a21) -- +(1.5,.5) ;
      \draw[transient,struct-z1] (a3) -- +(2,-.5) ;
      \draw[transient,struct-z1] (b2) -- (b21) ;
      \draw[transient,struct-z1] (b21) -- (e1) ;
      \draw[transient,struct-z1] (b3) -- (d1) ;
      \draw[transient,struct-z1] (b21) -- +(.5,1.5) ;
      \draw[transient,struct-z1] (b4) -- (b41) ;
      \draw[transient,struct-z1] (b41) -- +(2,.5) ;
      \draw[transient,struct-z1] (b41) -- +(1.5,-1) ;
      \draw[transient,struct-z2] (c2) -- +(3,.25) ;
      \draw[transient,struct-z2] (c3) -- +(2,.5) ;
      \draw[transient,struct-z2] (d2) -- (d21) ;
      \draw[transient,struct-z2] (d21) -- +(2,.5) ;
      \draw[transient,struct-z2] (d21) -- +(2,-.5) ;
      \draw[transient,struct-z2] (d3) -- +(3,0) ;
      \draw[transient,struct-z2] (d4) -- +(3.5,-.1) ;
      \draw[transient,struct-z2] (e2) -- (e21) ;
      \draw[transient,struct-z2] (e21) -- +(1,0) ;
      \draw[transient,struct-z2] (e21) -- +(1.5,.5) ;
      \draw[transient,struct-z2] (e3) -- +(3,.5) ;
      \path (\EndZZ,-2) +(-.1,0) \CoorNode{Zz} ;
      \path (\EndZO,-2) +(-.1,0) \CoorNode{Zo} ;
      \begin{scope}[loosely dashed]
        \draw (Zz) -- +(0.,10) ;
        \draw (Zo) -- +(0.,10) ;
      \end{scope}
      \draw[decorate,decoration=brace] (Zz) -- node[below=.5ex] {\ZoneZero} (current bounding box.west |- 0, -2 ) ;
      \draw[decorate,decoration=brace] (Zo) -- node[below=.5ex] {\ZoneOne} (Zz) ;
      \draw[decorate,decoration=brace] (current bounding box.east |- 0, -2 ) -- node[below=.5ex] {\ZoneTwo} (Zo) ;
    \end{tikzpicture}}\quad\newcommand{\SetArm}[4]{\TileDot[a11]{[shift={(.25,-2)}]#1}{}
    \draw[transient,struct-z#3] (#1) -- (a11) ;
    \DisplayThree{struct-z#4}{c}{[shift={(2.4,.8)}]a11}
    \draw[transient,struct-z#3] (a11) -- (c1) ;
    \TileDot[c11]{[shift={(1.5,0)}]a11}{}
    \draw[transient,struct-z#3] (a11) -- (c11) ;
    \path (a11) node [above right] {#2} ;
  }
  \newcommand{\SetUnroll}[1]{\begin{tikzpicture}
      \TileDot[O]{2.5,2}{}
      \path (O) node [left] {\Root} ;
      \TileDot[OO]{[shift={(1.5,0)}]O}{}
      \DisplayThree[$d'$]{struct-p}{a}{[shift={(.5,-1.5)}]OO}
      \path (barycentric cs:a1=1,a2=1,a3=1) node {$c$} ;
      \draw[transient,struct-z0] (O) -- (OO) ;
      \path (a1) node [below left] {$x_c$} ;
      \SetArm{a1}{$G$}{1}{2}
      #1
    \end{tikzpicture}
  }
  \SetUnitlength{1.6em}
  \begin{tabular}[b]{@{}c@{}}
    \subcaptionbox{Quipu \Quipu before one-step unrolling $c$\label{fig:unroll:before}}{\qquad
    \SetUnroll{\draw[struct-z0] (OO) -- node [left] {$d$} (a1) ;
    }\qquad
    }
    \\
    \subcaptionbox{Quipu $Q'$ after one-step unrolling $c$\label{fig:unroll:after}}{\SetUnroll{\TileDot[OO1]{[shift={(2.5,.75)}]OO}{}
    \path (OO1) node [above right] {$x$} ;
    \draw[struct-z0] (OO) -- node [above] {$d$} (OO1) ;
    \draw[struct-z0] (OO1) -- node [left] {$d'$\!} (a2) ;
    \SetArm{OO1}{$G'$}{0}{1}
    }
    }
  \end{tabular}
  \caption{Quipu.}
  \label{fig:quipu}
\end{figure}

\subsection{Quipu Zones}

A quipu represents the union of paths, which by definition of quipu can visit at most two cycles. 
Therefore, the vertices of a quipu are partitioned in three zones (illustrated in \RefFig{fig:zones}):
\begin{itemize}
\item \ZoneZero: all vertices that can be reached from the root without entering a vertex of a cycle,
\item \ZoneOne: all vertices of cycles that can be reached 
  from $r$ without visiting another cycle and all vertices that can be accessed from those vertices without entering another cycle, and
\item \ZoneTwo: all vertices of cycles that are reached after visiting another cycle and all vertices that can be accessed from them.
\end{itemize}
Every walk initiated from the root is composed of a part in $\ZoneZero$, then an optional part in $\ZoneOne$, then finally an optional part in $\ZoneTwo$ if there is a part in $\ZoneOne$.

\begin{fact}
  We observe that the union of all paths that visit two cycles, a cycle $c$ in \ZoneOne and another cycle in \ZoneTwo connected by a path from $c$, forms a comb. 
  Since all paths in \Quipu visit at most two cycles, and there is a unique path from $r$ to any of the vertices in \Quipu, the assembly corresponding to \Quipu is a union of combs. The paths in \ZoneZero correspond to the transient part, those that visit vertices in 
  \ZoneOne but not \ZoneTwo are backbones, and the paths that visit cycles in \ZoneTwo are the teeth.
  The \RefCndEnum{def:quipu:unique-walk} in \RefDef{def:quipu} implies that the combs are mutually disjoint.
\end{fact}

In the rest of this section, \Quipu and \QuipuOther designate quipus.

\medskip

Let $x$ be any \Quipu-vertex.
We denote by \CoverBy[\Quipu]{x} (or \CoverBy{x} if \Quipu is understood from the context) the subset of $\IntegerSet^2$ that consists of all displacement vectors associated to rooted walks ending in $x$.
These subsets of $\IntegerSet^2$ are semi-linear and disjoint.
For a vertex $x$ in $\ZoneZero$, $\CoverBy{x}=\{\Vect{m}\}$ where $m$ is the label of a path in \Quipu from $r$ to $x$. 
For a vertex $x$ in $\ZoneOne$, $\CoverBy{x}=\Vect{a}+\NaturalSet\Vect{b}$ for some \Vect{a} and non-null \Vect{b} in $\IntegerSet^2$.
For a vertex $x$ in $\ZoneTwo$, $\CoverBy{x}=\Vect{a}+\NaturalSet\Vect{b}+\NaturalSet\Vect{c}$ for some \Vect{a} and non-null and non co-linear \Vect{b} and \Vect{c} in $\IntegerSet^2$.
The cover of a \Quipu-vertex set $S$ is defined as the (disjoint) union of the covers of $x$ for $x\in S$.
In particular, $\CoverBy[\Quipu]{\Quipu}=\Domain{\AlphaQuipu}$.

\subsection{Cycle Unrollings}

For every circle $c$ there is only one vertex with an incoming arc from a vertex outside the cycle, by the \RefCndEnum{def:quipu:one-arc-to-cycle} of \RefDef{def:quipu}. We refer to this vertex as `entry' to $c$. 

\begin{definition}[one-step unrolling] Let $c$ be a cycle in a quipu \Quipu and let 
  $x_c$ be the entry vertex of $c$.
  The \emph{(one-step) unrolling} of $c$ modifies \Quipu to $\Quipu'$ in the following way:
  \begin{itemize}
  \item add a new vertex $x$ with $\VertexLabel(x)=\VertexLabel(x_c)$,
  \item if $x_c$ is a root of a subgraph $G$ of \Quipu that has no intersection with $c$, add
    a copy $G'$ of $G$ (with the same labels) rooted at $x$,
  \item for each arc $(y,x_c)$ in \Quipu where $y$ is not in $c$, erase the arc $(y,x_c)$ and add an arc $(y,x)$ with the same label, and 
  \item for the arc $(x_c,z)$ in $c$, add an arc $(x,z)$ with the same label.
  \end{itemize}
\end{definition}

The figures \ref{fig:unroll:before} and \ref{fig:unroll:after} show a quipu before and after one-step unrolling.
The result of the one-step unrolling, \QuipuOther, is still a quipu.
The sets of sequences of \ArcLabel-labels of any walk in \Quipu and \QuipuOther are equal; $\AlphaQuipu=\AlphaQuipuOther$.
The cycle $c$ is still in the quipu but its entry vertex is the vertex following $x_c$ in $c$.
The new vertex $x$ can be seen as a copy of $x_c$ because the set of labels of paths that visit $x_c$ in \Quipu is the same as the set of labels of paths that visit $Q'$.
In addition, the copy of vertex $x_c$, the vertex $x$, has moved one zone up (i.e., from \ZoneOne to \ZoneZero or from \ZoneTwo to \ZoneOne) and the copies of any subgraph rooted at $x_c$ in $Q'$ is a subgraph rooted at $x$ also one zone up. 
Moreover, $\CoverBy[Q']{x_c}=\CoverBy[Q']{x}\CupDis \CoverBy[\Quipu]{x_c}$.
The shortest path to $x_c$ in $\QuipuOther$ is longer than that in \Quipu; in some sense, $x_c$ has `moved away from the seed'.
In \RefSec{sec:quipu-extension} we use unrolling to `move' a vertex $x$ to add new vertices adjacent to it.

The lemma below follows directly from the definitions.

\begin{Lemma}
  \label{fact:unrolling-cycle}
  The one-step unrolling $Q'$ has the property that every vertex belonging to a cycle of \Quipu or on the \Quipu-path to a cycle is still on the $\Quipu'$-path to the same cycle or to a copy of this cycle. 
\end{Lemma}

The {\it unrolling} of a cycle $c$ is the quipu obtained from \Quipu by a sequence of $n$, ($1\le n$) one-step unrollings of $c$. 
The \emph{full unrolling} of a cycle is the unrolling of all its vertices, i.e., $n$ consecutive unrolling where $n$ is the length of $p$.
Let $z$ be in \ZoneTwo, then $\CoverBy[\Quipu]{z}=\Vect{a_z}+\NaturalSet\Vect{b_b}+\NaturalSet\Vect{c_t}$ where $b_b$ is the period of the backbone and $c_t$ of the tooth.
If the first cycle leading to $z$ is fully unrolled, then $\CoverBy[\QuipuOther]{z}=\CoverBy[\Quipu]{z}+\Vect{b_b}$.
If the second cycle leading to $z$ is fully unrolled, then $\CoverBy[\QuipuOther]{z}=\CoverBy[\Quipu]{z}+\Vect{c_t}$.

We introduce another operation on quipu \Quipu that is used to extend \Quipu.

\begin{definition}
  Let $c=x_0,\ldots,x_{s-1}$ be a cycle in a quipu \Quipu and let 
  $x_c=x_0$ be the entry vertex of $c$. The \emph{$k$-multiple of $c$} modifies
  \Quipu to a graph $\Quipu'$ in the following way:
  \begin{itemize}
  \item add new vertices $x_0^1,x_1^1,\ldots,x_{s-1}^1,\ldots,x_0^{k-1},x_1^{k-1},\ldots,x_{s-1}^{k-1}$ with $\VertexLabel(x_i)=\VertexLabel(x_i^j)$; and add arcs $(x_i^j,x_{i+1}^j)$
    with arc label 
    $\ArcLabel{(x_i^j,x_{i+1}^j)}=\ArcLabel{(x_i,x_{i+1})}$
    for all $0\le i<s-1$ and $1\le j\le k-1$,
  \item remove the arc $(x_{s-1},x_0)$ and add arcs $(x_{s-1},x_0^1)$, $(x_{s-1}^{k-1},x_0)$ and 
    $(x_{s-1}^j,x_0^{j+1})$ for $1\le j< s-1$ with the same arc label as 
    $(x_{s-1},x_0)$, and
  \item for all $i=0,\ldots,s-1$, if $x_i$ is a root of a subgraph $G$ of \Quipu that has no intersection with $c$, add
    a copy $G'$ of $G$ (with the same labels) rooted at $x_i^j$. 
  \end{itemize}
\end{definition}

The $k$-multiple of $c$ essentially expands the cycle $c$ and forms a cycle $c^k=cc\cdots c$.
Similarly as with unrolling, $\AlphaQuipuOther=\AlphaQuipu$, while the new cycle $c^k$ remains in the same zone where $c$ was. As with unrolling we have that $\CoverBy{\QuipuOther}=\CoverBy{\Quipu}$, and we consider the vertices $x_i^j$ are copies of $x_i$.

\subsection{Relations between the sets covered by the \Quipu-vertices}

From here to the end of the section, we have the following set up: $x$ is a \Quipu-vertex and $d\in D$ is such that there is $t$, a tile type in \TileSet, whose glue on $-d$ side equals the glue $\VertexLabel(x)_d$.

Before proceeding with quipu construction in the next section we investigate the intersection of the set $\CoverBy{x}+d$ 
with the covers of the other vertices in \Quipu.

\begin{fact}
  The vertices of $\CoverBy{x} + \Vect{d}$ must belong to $\Domain{\AlphaMax}$; either those vertices are mapped to $t$ by \AlphaMax, or those vertices are in causal portion of vertices in $\CoverBy{x} + \Vect{d}$ and the producible assembly has already placed other tiles in those locations that are obstacles for $t$ and \Quipu already account for them.
\end{fact}

Let $y$ be any \Quipu-vertex.

\begin{lemma}[$\ZoneZero$]
  \label{lem:finite-intersection:Z0}
  If $x$ or $y$ belongs to \ZoneZero then
  $\left(\CoverBy{x} + \Vect{d}\right)
  \cap
  \CoverBy{y}\subseteq\{z\}$
  for some $z\in \IntegerSet^2$.
\end{lemma}

\begin{proof}
  The lemma follows directly from the fact that the cover of every vertex in \ZoneZero is a singleton. 
\end{proof}

\begin{lemma}[\ZoneOne and \ZoneOne]
  \label{lem:finite-intersection:Z1+Z1}
  If $x$ and $y$ belong to \ZoneOne then:
  \begin{equation}
    \left( \CoverBy{x} + \Vect{d} \right)
    \cap
    \CoverBy{y}
    =
    F_0
    \cup
    \left(
      F_1+\NaturalSet\Vect{b}
    \right)\label{Z_1}
  \end{equation}
  where $F_0$ and $F_1$ are finite sets.
  If $F_1$ is not empty, then $x$ and $y$ belong to combs with co-linear backbones in the same direction and $b$ is the $\lcm$ of their periods.
\end{lemma}

\begin{proof}Since $x$ and $y$ belong to \ZoneOne, $\CoverBy{x} = \Vect{a_x}+\NaturalSet\Vect{b_x}$ and $\CoverBy{y} = \Vect{a_y}+\NaturalSet\Vect{b_y}$ where 
  $b_x$ and $b_y$ are periods of the backbone of $x$ and $y$ respectively. Then $\CoverBy{x}+\Vect{d} = \Vect{a_x}+\Vect{d}+\NaturalSet\Vect{b_x}$.

  If  $\Vect{b_x}$ and $\Vect{b_y}$ are not co-linear or $\Vect{b_x}=c\Vect{b_y}$ for some negative rational number $c$, then the intersection in \RefEq{Z_1} is finite.
  If the backbones are different but co-linear and the displacement vectors of their periods are in the same direction, then we take the $\Vect{b}=\text{lcm}(\Vect{b_x},\Vect{b_y})$ 
  which is also in the same direction.
  Then $\CoverBy{x} = A_x+\NaturalSet\Vect{b}$ and $\CoverBy{y} = A_y+\NaturalSet\Vect{b}$ where $A_x$ and $A_y$ are finite subsets of $\IntegerSet^2$.
  The lemma follows.
\end{proof}

\begin{lemma}[\ZoneOne and \ZoneTwo]
  \label{lem:finite-intersection:Z1+Z2}
  If one of vertices $x, y$ belongs to \ZoneOne and the other to \ZoneTwo then one of the three conditions hold:
  \begin{enumerate}
  \item $\left|
      \left( \CoverBy{x} + \Vect{d} \right)
      \cap 
      \CoverBy{y}
    \right| < \infty$,
  \item $\left( \CoverBy{x} + \Vect{d} \right)
    \cap 
    \CoverBy{y}
    =
    \Vect{a}+\NaturalSet\Vect{b}$, $x$ and $y$ belong to the same comb and \Vect{b} is the period of the backbone, or
    \enspace 
  \item \AlphaMax is a grid.
  \end{enumerate}
\end{lemma}

\begin{proof} Consider the case when $\left( \CoverBy{x} + \Vect{d} \right)\cap\CoverBy{y}$ is infinite.
  If $x$ and $y$ belong to different combs, because 
  these are combs of \Quipu, they are disjoint.
  Since (sufficiently away from the seed) points of the two combs are connected by an edge (corresponding to the connection with glue $d$) 
  the two combs are connected in \AlphaMax by infinite number of edges, from a tooth of one comb to the backbone of the other comb. 
  By \RefLem{lem:comb-connections}, \AlphaMax is a grid.

  In the case when $x$ and $y$ belong to the same comb, let $b$ be the period of the comb backbone and $c$ be the period of the tooth. 
  Without loss of generality, we can assume $x$ to be in \ZoneOne and $y$ in \ZoneTwo. 
  Then
  $\CoverBy{x} = \Vect{a_x}+\NaturalSet\Vect{b}$ and
  $\CoverBy{y} = \Vect{a_y}+\NaturalSet\Vect{b}+\NaturalSet\Vect{c}$ where $b$ is the period of the backbone and $c$ the tooth of $y$. 
  Since $c$ is not co-linear with $b$, the lemma follows.
\end{proof}

\begin{lemma}[\ZoneTwo and \ZoneTwo] \label{lem:finite-intersection:Z2}
  If both $x$ and $y$ belong to \ZoneTwo then one of the three conditions hold,
  \begin{enumerate}
  \item \label{lem:finite-intersection:Z2:1}
    $\left( \CoverBy{x} + \Vect{d} \right)
    \cap \CoverBy{y}
    =\emptyset$,
  \item\label{cnd:finite-intersection:Z2:same-comb}
    $\left( \CoverBy{x} + \Vect{d} \right)=\CoverBy{y}$
    \ or \
    $\left( \CoverBy{x} + \Vect{d} \right)
    \CupSym\CoverBy{y}
    =
    \Vect{a}+\NaturalSet\Vect{b}$ (where $\CupSym$ denotes the symmetric difference),
    $x$ and $y$ belong to the same comb and \Vect{b} is the period of the backbone, or
  \item\label{lem:finite-intersection:Z2:3}
    \AlphaMax is a grid.
  \end{enumerate}
\end{lemma}

\begin{proof}
  Suppose $\left( \CoverBy{x} + \Vect{d} \right)\cap \CoverBy{y}$ is not empty.
  If $x$ and $y$ belong to different proto-teeth, due to periodicity of the teeth, there are infinite number of vertices in the intersection $\CoverBy{y} \cap \left( \CoverBy{x} + \Vect{d} \right)$ and hence infinite number of edges with endpoints on both teeth.
  By \RefLem{lem:tooth-to-tooth}, \AlphaMax is a grid.

  Hence $x$ and $y$ must belong to the same proto-tooth, and
  $\CoverBy{x} = \Vect{a_x}+\NaturalSet\Vect{b}+\NaturalSet\Vect{c}$ and
  $\CoverBy{y} = \Vect{a_y}+\NaturalSet\Vect{b}+\NaturalSet\Vect{c}$
  where $b$ is the period of its backbone and $c$ is the period of the tooth. 
  Since we assume \RefEnum{lem:finite-intersection:Z2:1} and \RefEnum{lem:finite-intersection:Z2:3} do not hold, we have that:
  \begin{equation*}
    \left( \CoverBy{x} + \Vect{d} \right)
    \CupSym \CoverBy{y}
    =
    \left(F+\NaturalSet\Vect{b}\right)
    \cup
    \left(G+\NaturalSet\Vect{c}\right)
    \enspace 
  \end{equation*}
  where $F$ and $G$ are finite subsets of $\IntegerSet^2$.
  Because $x$ and $y$ are part of the same proto-tooth, due to periodicity, if there is a connection between a point in $\IntegerSet^2$ in \CoverBy{x} and in \CoverBy{y}, this connection appears throughout the tooth and the set $G+\NaturalSet\Vect{c}$ must be empty, except possibly close to the backbone, which repeats with every period of the backbone.
  But the vertices $x$ and $y$ can be either part of a decoration of a tooth, or be part of the periodic portion of the tooth, and within one period of $c$, $\CoverBy{x} + \Vect{d}$ and $\CoverBy{y}$ can intersect at most once. 
  Hence, if $2\le|F|$ then there are two periods of $c$ where $\CoverBy{x} + \Vect{d}$ and $\CoverBy{y}$ miss their intersection.
  Using a similar argument as in the proof of \RefLem{lem:double-pumpable}, it follows that $\CoverBy{x} + \Vect{d}$ and $\CoverBy{y}$ do not intersect through out the tooth, implying \RefCndEnum{cnd:finite-intersection:Z2:same-comb}.
  Therefore, $\left( \CoverBy{x} + \Vect{d} \right)\CupSym \CoverBy{y}$ is either empty, or $\Vect{a}+\NaturalSet\Vect{b}$.
\end{proof}

\section{Quipu Extension}
\label{sec:quipu-extension}

In this section we show that a quipu \Quipu associated with \TAS can be extended to $\QuipuOther$ if $\Domain{\AlphaQuipu}\subsetneq\Domain{\AlphaMax}$ such that $\Domain{\AlphaQuipu}\subsetneq\Domain{\AlphaQuipuOther}\subseteq\Domain{\AlphaMax}$.
We use a formal notation $\Quipu[\text{grid}]$ to indicate that \AlphaMax is a grid. 

\begin{definition}[quipu extension]
  \label{hyp:quipu:induction}
  A quipu \QuipuOther associated to \TAS is an \emph{extension} of \Quipu if the following hold:
  \begin{enumerate}
  \item
    \label{hyp:quipu:induction:keep-paths}the sequence of $\ArcLabel$-labels of any rooted \Quipu-walk is also the sequence of $\ArcLabel'$-labels of some rooted \QuipuOther-walk ($\AlphaQuipu\SubGraphOf\AlphaQuipuOther$),
  \item
    \label{hyp:quipu:induction:cycles}every cycle $c$ of \Quipu is (possibly unrolled) in \QuipuOther or there is a $k$-multiple of $c$ in \QuipuOther.
  \end{enumerate}
  If \AlphaMax is a grid, then for every quipu \Quipu associated to \TAS we consider $\Quipu[\text{grid}]$ to be \emph{an extension} of \Quipu. 
\end{definition}

Suppose $\Domain{\AlphaQuipu}\subsetneq\Domain{\AlphaMax}$ for a quipu \Quipu associated with \TAS.
Because the subgraphs of $\IntegerSet^2$ induced by $\Domain{\AlphaQuipu}$ and $\Domain{\AlphaMax}$ are connected and intersect, there is a vertex $y$ in $\Domain{\AlphaMax}\setminus\Domain{\AlphaQuipu}$ that is incident to a vertex $x$ in \AlphaQuipu in some direction $d$.
Suppose $\AlphaQuipu(x)=\AlphaMax(x)=t$ and $\AlphaMax(y)=t'$ such that $t_d=t'_{\Backward{d}}$.
The following Lemma shows that there is an extension $\QuipuOther$ of a \Quipu such that $x+\Vect{d}\in \Domain{\AlphaQuipuOther}$.
The extension may increase the domain of \AlphaQuipuOther by more than one point of the lattice $\IntegerSet^2$.

\begin{Lemma}[add a vertex to \Quipu]
  \label{lem:finite-union}
  Let $X$ be a set of \Quipu-vertices with label $t\in T$ and $d$ a direction such that there is tile type $t'$ in \TileSet with $t_d=t'_{\Backward{d}}$.
  Then either \AlphaMax is a grid or there is an extension \QuipuOther of quipu \Quipu  satisfying 
  \begin{equation*}
    \CoverBy[\Quipu]{X}+\Vect{d}
    \subseteq
    \CoverBy{\QuipuOther}
    \enspace .
  \end{equation*}
\end{Lemma}

\begin{proof}
  Recall that unrolling of cycles or forming $k$-multiples of cycles may add in $X$ copies of vertices already in $X$.
  The copies are added to $X$ immediately so that $\CoverBy{X}$ does not change with these operations.
  Let $x\in X$ such that $\CoverBy{x}+\Vect{d}\not\subseteq \Domain{\AlphaQuipu}$ (but $\CoverBy{x}+\Vect{d} \subseteq\Domain{\AlphaMax}$).

  If $x\in X$ belongs to \ZoneZero, then $\CoverBy[\Quipu]{x}+\Vect{d}$ is either empty or a singleton.
  In former case $\QuipuOther=\Quipu$, while in the latter case \QuipuOther is obtained by adding a vertex $x'$ with $\VertexLabel{(x')}=t'$ and an edge $(x,x')$ with $\ArcLabel{(x,x')}=d$.
  
  Suppose now $x$ belongs to $\ZoneOne\cup\ZoneTwo$.
  
  If $\left|\left(\CoverBy[\Quipu]{x} + \Vect{d}\right)\cap\CoverBy{\Quipu}\right| =\emptyset$, the desired quipu \QuipuOther is obtained such that we add a vertex $x'$ in $ \Quipu$ and an arc $(x,x')$ with $\VertexLabel(x')=t'$ and $\ArcLabel(x,x')=d$.
  If $\left|\left(\CoverBy[\Quipu]{x} + \Vect{d}\right)\cap\CoverBy{\Quipu}\right| < \infty$, then there are a finite number of vertices $y$ such that 
  $\left|\left( \CoverBy[\Quipu]{x} + \Vect{d}\right)\cap\CoverBy[\Quipu]{y}\right| < \infty$ and for each such $y$ we do the following.
  We `move' $x$ away from the seed by unrolling the closest to $x$ cycle in \Quipu leading to $x$ finite number of times.
  With this we obtain a new quipu $\hat \Quipu$ with copies $\hat x$ of $x$ such that $\left(\CoverBy[\hat\Quipu]{\hat x} + \Vect{d}\right)\subseteq\Domain{\AlphaQuipu}$ and $\CoverBy[\hat\Quipu]{y}\cap \left(\CoverBy[\hat\Quipu]{x} + \Vect{d}\right)=\emptyset$ for every $\hat \Quipu$-vertex $y$.
  Then the desired quipu \QuipuOther is obtained such that we add a vertex $x'$ in $\hat \Quipu$ and an arc $(x,x')$ with $\VertexLabel(x')=t'$ and $\ArcLabel(x,x')=d$.
  Please note that the copies of $x$ are one zone below.
  
  Consider the case when $\left(\CoverBy[\Quipu]{x} + \Vect{d}\right)\cap\CoverBy{\Quipu}$ is infinite.
  Let $y$ be a \Quipu-vertex (necessarily in $\ZoneOne\cup\ZoneTwo$) such that $\CoverBy{x} + \Vect{d}\cap\CoverBy{y}$ is infinite.
  When $x$ and $y$ both belong to \ZoneTwo, by \RefLem{lem:finite-intersection:Z2}, if \AlphaMax is not a grid, it must be 
  either (a) $\CoverBy{y}=( \CoverBy{x} + \Vect{d})$ and in that case there is no extension of \Quipu, i.e., $\QuipuOther=\Quipu$, or, (b) 
  $\CoverBy{y}\CupSym( \CoverBy{x} + \Vect{d})=\Vect{a}+\NaturalSet\Vect{b}$ and $y$ and $x$ belong to the same proto-tooth and $\Vect{b}$ is the period of the backbone.
  A full unrolling of the cycle corresponding to the proto-tooth forms quipu $\hat \Quipu$ by adding 
  copies $x'$ to $x$ and $y'$ to $y$ in \ZoneOne and $\CoverBy[\Quipu]{x} = \CoverBy[\hat\Quipu]{x}\CupDis\CoverBy[\hat\Quipu]{x'}$ (and similarly for $y'$).
  It follows that $\Vect{a}+\NaturalSet\Vect{b}$ (being in \ZoneOne) is either $\CoverBy[\hat\Quipu]{x'}+\Vect{d}$ or $\CoverBy[\hat \Quipu]{y'}$.
  Since $\CoverBy{\hat\Quipu}$ does not contain $\CoverBy[\Quipu]{x}+\Vect{d}$, it is the former case.
  Then the desired quipu \QuipuOther is obtained by adding a vertex $x''$ in $\hat \Quipu$ and an arc $(x',x'')$ with 
  $\VertexLabel(x'')=t'$ and $\ArcLabel(x',x'')=d$.
  The process is repeated for every such $x$ and $y$ in \ZoneTwo.
  
  When $x$ belongs to \ZoneTwo and $y$ belongs to \ZoneOne, or $x$ belongs to \ZoneOne and $y$ belongs to \ZoneTwo, by \RefLem{lem:finite-intersection:Z1+Z2}, if \AlphaMax is not a grid, it must be $\CoverBy{y}\cap\left( \CoverBy{x} + \Vect{d}\right)=\Vect{a}+\NaturalSet\Vect{b}$ and $y$ and $x$ belong to the same comb. By fully unrolling the cycle in \ZoneTwo the intersection in question becomes an intersection between two vertices in \ZoneOne.

  When both $x$ and $y$ belong to \ZoneOne, by \RefLem{lem:finite-intersection:Z1+Z1}, if \AlphaMax is not a grid, it must be 
  $\CoverBy{y}
  \cap
  \left(\CoverBy{x} + \Vect{d}\right)
  = F_0\cup \left(F_1+\NaturalSet\Vect{b}\right)$
  where $F_0$ and $F_1$ are finite sets, and when 
  $F_1$ is not empty,
  then $x$ and $y$ belong to combs with co-linear backbones with displacement in the same direction and $\Vect{b}$ is the $\lcm$ of their cycles.
  The vertices in $F_0$ are at the beginning of the cycles corresponding to the back-bone cycles.
  Both cycles can be unrolled until the cover of the entry vertex for both has a vertex in $(F_1+\NaturalSet\Vect{b})$.
  New vertices and appropriate arcs to copies of $x$ are added similarly as above.
  We perform a $|b|/|b_x|$-multiple of $b_x$, the cycle with vertex $x$.
  The new cycle contains $x$, and $x_1,\ldots,x_{k-1}$ copies of $x$ and $\CoverBy{y}\cap (\CoverBy{x_i} + \Vect{d}) = \emptyset$ for $i=1,\ldots,k-1$, and $\left(\CoverBy{x} + \Vect{d}\right)\subseteq\CoverBy{y}$.
  Then we obtain $\QuipuOther$ by adding vertices $x_i'$ with $\VertexLabel(x_i')=t'$ and arcs $(x_i,x_i')$ with $\ArcLabel(x_i,x_i')=d$ when $\left(\CoverBy{x_i} + \Vect{d}\right)\subseteq\CoverBy{y}=\emptyset$.

  In all cases, it follows directly from the construction, the conditions \textit{\ref{hyp:quipu:induction:keep-paths}.} and \textit{\ref{hyp:quipu:induction:cycles}.} of \RefDef{hyp:quipu:induction} are satisfied. 
\end{proof}

The new quipu $\QuipuOther$ obtained with extension in direction $d$ as in the construction in the proof of \RefLem{lem:finite-union} is denoted $\Quipu[X,t,d]$.

The following lemma is an essential observation in the construction of a quipu that corresponds to \AlphaMax.
It means that we can concentrate on finite and ultimately periodic paths.
It (and its proof) is similar in flavor to \RefLem{lem:nice}. 

\begin{lemma}[Ultimately periodic path and quipu]
  \label{lem:quipu:utp}
  If there is an infinite path $\Path$ in \AlphaMax such that $\Path\cap\AlphaQuipu$ is finite, then there exists an infinite ultimately periodic path $\Path'=\ZZOrigin.mp^{\omega}$ 
  such that $\Path'\cap\AlphaQuipu$ is finite.
\end{lemma}

\begin{proof}
  If the quipu has no cycles then \AlphaQuipu is finite.
  The existence of \Path, by \RefTh{th:nice}, implies that there is an ultimately periodic path in \AlphaMax.
  This path can intersect \AlphaQuipu at only finite number of points.

  So we suppose that \Quipu has a cycle $C$. 
  Let $\ZZOrigin.mp^{\omega}$ be the path in \AlphaQuipu corresponding to the cycle $C$ in the quipu.
  Since the number of intersections between \Path and \AlphaQuipu is finite, by replacing $m$ by $mp\cdots p$ if necessary, we can assume that there is no intersection between \Path and $(\ZZOrigin+\Vect{m}).p^{\omega}$.
  Let $i$ be the largest index of \Path of any intersection with $\ZZOrigin.m$ and let $m'$ be the suffix of $m$ starting at $\Path_{i}$.
  Consider the bi-infinite path $^{\omega}\RevBack{p}\RevBack{m'}.\Path_{\IntegerInterval{i}{+\infty}}$ and its (infinite) right region $R$. 
  Because any path in \AlphaQuipu generated by a cycle $C'$ in \Quipu has a finite intersection 
  with both \Path and $\ZZOrigin.mp^{\omega}$, and the number of cycles in \Quipu is finite, we can choose $\ZZOrigin.mp^\omega$ such that the intersection of $R$ and \CoverBy{C'} for any cycle $C'\not = C$ is finite.

  Then the statement of the lemma follows directly from \RefLem{lem:nice} and \RefCor{cor:new-periodic-path}.
\end{proof}

\subsection{Transient Step: extending quipu \Quipu with a finite path $\ZZOrigin.m$  in \AlphaMax}

The quipu is modified to cover each vertex of the path, inductively one displacement at the time.
Through the traversal of the displacements in $m$, we record a quipu, \QuipuOther together with a set of \QuipuOther-vertices, $X$ that have the same label and contains a vertex covering the \AlphaMax-vector $\ZZOrigin+\Vect{l}$ where $l$ is the current prefix of $m$.
The induction starts at $l=\EmptyWord$ with $(\Quipu,\{\Root\})$.

Let $l'=ld$ where $d\in\DirectionSet$ and $l'$ is a prefix of $m$.
Suppose $(\Quipu,X)$ is generated with by the traversal of the path $\ZZOrigin.l$.
Using \RefLem{lem:finite-union}, we generate the quipu $\QuipuOther=\Quipu[X,\AlphaMax(\ZZOrigin+l\Vect{d}),d]$ and a set of \QuipuOther-vertices $Y$ such that $\CoverBy[\Quipu]{X}+\Vect{d}\subseteq\CoverBy[\QuipuOther]{Y}$.
We set $X'$ to be the subset of $Y$ of vertices labelled by the tile type at $\ZZOrigin+\Vect{l'}$ in \AlphaMax.
If the extension step detects that \AlphaMax is a grid, the extension stops and output $\Quipu[{\text{grid}}]$.

At each step, by ~\RefLem{lem:finite-union}, quipu \QuipuOther is an extension of \Quipu.
In the construction some cycles may be unrolled, as in the proof of \RefLem{lem:finite-union}.

Since $\ZZOrigin.m$ belongs to \AlphaMax, all necessary vertices traversing $\ZZOrigin.m$ in \AlphaMax are part of $\Quipu[\ZZOrigin.m]$, (but not necessarily all the directions as edges, which are added only with new vertices). 
The set $\CoverBy[\Quipu']{X'}$ where $X'$ is the last set of quipu-vectors contains the point $\ZZOrigin+\Vect{m}$, plus possibly others.

The result of the extension is the last generated quipu.
It is denoted $\Quipu[m]$.

\subsection{Periodic Step: extending quipu \Quipu with an ultimately periodic path $A.mp^{\omega}$  in \AlphaMax such that $\Domain{A.mp^{\omega}}\cap\Domain{\AlphaQuipu}=\{A\}$}

Denote with $T_m$ and $C_p$ the path and the cycle of tile types populating vertices in $\IntegerSet^2$ corresponding to $A.mp^\omega$ in \AlphaMax, i.e. $\AlphaMax((A+\Vect{m}).p^\omega)=
T_mC_p^\omega$ (by \RefLem{path-assembly}, $C_p$ and $T_m$ exist). If $T_m$ is empty, set $T_m=C_p$ and consider extension $C_pC_p^\omega$. We have that $|C_p|=k|p|$ for some $k\ge1$.
Because $A.mp^\omega$ is in domain of $\AlphaMax$ and intersects \AlphaQuipu only at point $A$,
addition of a path $T_m$ and a cycle $C_p$ to the quipu \Quipu is possible, with possibly unrolling one or two cycles.
Let $x_A$ be the vertex in \Quipu such that $A\in\CoverBy{x_A}$.

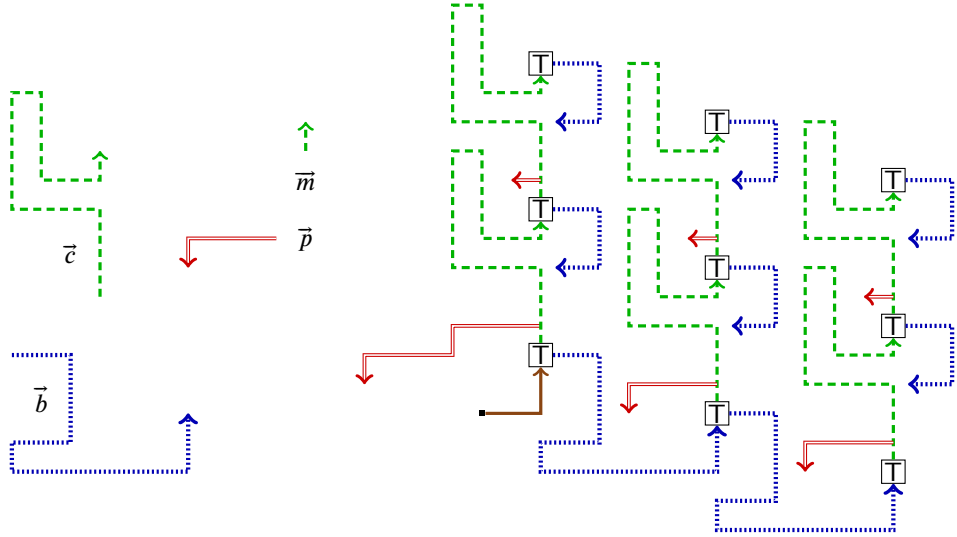
\begin{figure}[hbt]
  \centering\small\footnotesize\SetUnitlength{1.1em}\PreparePath{M}{EEN}
  \PreparePath{N}{NNNWWWNNNNESSSEE}
  \PreparePath{E}{EESSSWWSEEEENNNESSSEN}
  \PreparePath{W}{WWWSWWWS}
  \PreparePath{Wbase}{WWWS}
  \PreparePath{Wmid}{WWWS}
  \PreparePath{Wshort}{W}
  \newcommand{\EEpref}{ \Ea{2} \So{2} \We{1.5} }
  \newcommand{\EE}{ \EEpref \We{.5} \So{1} \Ea{3} }
  \newcommand{\FFpref}{ \Ea{2} \No{2} \We{1.5} }
  \newcommand{\FF}{ \FFpref \We{.5} \No{1} \Ea{3} }
  \PreparePath{E}{EESSSWWSEEEEEEN}
  \begin{tikzpicture}[inner sep=.15ex]
    \begin{scope}[shift={(-6,2)}]
      \draw[struct-q] (0,6) node {\Vect{m}} ++(0,1) \No{1};
      \draw[struct-add-once] (0,4) node {\Vect{p}} ++(-1,0) \PathWbase ;
    \end{scope}
    \begin{scope}[shift={(-14,4)}]
      \path (0,1.5) node {\Vect{c}} ;
      \draw[struct-q] (1,0) \PathN \No{1} ;
    \end{scope}
    \begin{scope}[shift={(-16,2)}]
      \path (1,-1.5) node {\Vect{b}} ;
      \draw[struct-p] (0,0) \PathE \No{1} ; 
    \end{scope}
    \TileDot[O]{0,0}{T}
    \path (O) \PathM +(0,1) \CoorNode{n1} ;
    \Tile[n1]{n1}{T}
    \draw[struct-m] (O) \PathM -- (n1) ;
    \path (n1) \PathE +(0,1) \CoorNode{n2} ;
    \Tile[n2]{n2}{T}
    \draw[struct-p] (n1) \PathE -- (n2) ;
    \draw[struct-add-once] (n1) ++(0,1) \PathW ;
    \path (n2) \PathE +(0,1) \CoorNode{n3} ;
    \Tile[n3]{n3}{T}
    \draw[struct-p] (n2) \PathE -- (n3) ;
    \foreach \n in {1,2,3} {
      \path (n\n) \PathN +(0,1) \CoorNode{nn1} ;
      \Tile[nn1]{nn1}{T}
      \draw[struct-q] (n\n) \PathN -- (nn1) ;
      \draw[struct-p] (nn1) \EEpref ;
      \path (nn1) \PathN +(0,1) \CoorNode{nn2} ;
      \Tile[nn2]{nn2}{T}
      \draw[struct-q] (nn1) \PathN -- (nn2) ;
      \draw[struct-p] (nn2) \EEpref ;
      \draw[struct-add-once] (nn1) ++(0,1) \PathWshort ;
    } ;
    \draw[struct-add-once] (n2) ++(0,1) \PathWmid ;
    \draw[struct-add-once] (n3) ++(0,1) \PathWmid ;
  \end{tikzpicture}
  \caption{Adding an ultimately periodic path to \ZoneTwo.}
  \label{fig:adding-to-Z2}
\end{figure}

\Subpar{$x_A\in \ZoneTwo$, move $x_A$ to \ZoneOne}
Consider the case $x_A$ in \ZoneTwo and let $c$ be the cycle in \ZoneTwo with a path leading to $x_A$.
If $\Vect{p}$ is colinear to $\Vect{c}$ then $A.mp^\omega$ is possibly a new pseudo-tooth and we unroll $c$ such that $x_A$ ends in \ZoneOne.
If $\Vect{p}$ is not colinear to $\Vect{c}$ then since $A$ is the only intersection with \AlphaQuipu, the situation has to be as in \RefFig{fig:adding-to-Z2} (double line paths). We unroll $c$ such that $x_A$ ends in \ZoneOne.
All other repetitions of free paths $mp^\omega$ starting at periodic appearances of points in $\CoverBy{x_A}$ have to be `blocked', as shown in the figure. Otherwise, if $mp^\omega$ can start at a point in $\CoverBy{x_A}$ other than $A$, (from the backbone, or from a tooth), it can be repeated infinitely many times and would intersect another tooth implying that \AlphaMax is a grid by \RefLem{lem:intersecting-period:unbounded}.

\Subpar{$x_A\in \ZoneOne$, adding a backbone or a proto-tooth}
Consider the case $x_A$ in \ZoneOne.
We first unroll $b$ until $\CoverBy{x_A}=A+\Vect{b}\NaturalSet$ where $b$ is a backbone cycle.
If $\Vect{p}$ is co-linear to $\Vect{b}$, then $A.mp^\omega$ is an independent backbone and we unroll $b$ such that $x_A$ ends in \ZoneZero.

Otherwise, if for all points $B\in \CoverBy{x_A}=A+\Vect{b}\NaturalSet$ we have that $\Domain{B.mp^{\omega}}\cap\Domain{\AlphaQuipu}=\{B\}$ then we add to \Quipu at $x_A$ a path $T_m$ and add a new cycle with arc labels $p^k$ having $|C_p|$ new vertices and $|C_p|$ arcs.
This new cycle is a proto-tooth for the backbone $b$.
Otherwise, if for some $\{B\}\subsetneq\Domain{B.mp^{\omega}}\cap\Domain{\AlphaQuipu}$ we unroll $b$ such that $A$ is in \ZoneZero.

\Subpar{$x_A\in \ZoneZero$, adding a backbone}
Consider $x_A$ in \ZoneZero.
Then $\CoverBy{x_A}=\{A\}$.
Let $r_A$ be the path rooted at $\Root$ to $x_A$ and $\Domain{A.mp^\omega}$ intersects $\Domain{r_A}$ at $x_A$.
We add to \Quipu a path $T_m$ and add a new cycle with $|C_p|$ new vertices and $|C_p|$ arcs as above.
All new vertices have covers that do not intersect with \Domain{\AlphaQuipu}.

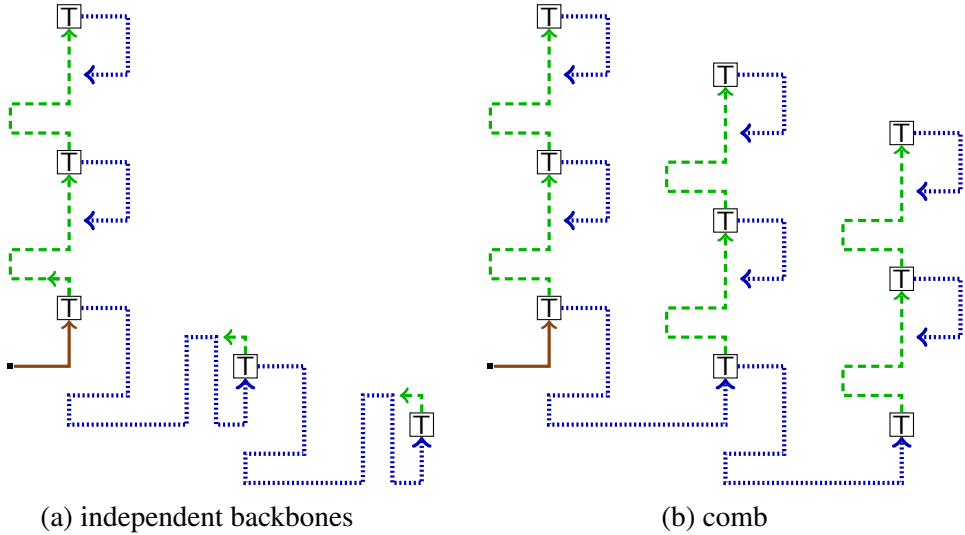
\begin{figure}[hbt]
  \centering\small\footnotesize\SetUnitlength{1.1em}\PreparePath{M}{EEN}
  \PreparePath{N}{NWWNEENN}
  \subcaptionbox{independent backbones
    \label{fig:quipu:share:2-backbone}}{\PreparePath{E}{EESSSWWSEEEENNNESSSEN}
    \newcommand{\EEpref}{ \Ea{2} \So{2} \We{1.5} }
    \newcommand{\EE}{ \EEpref \We{.5} \So{1} \Ea{3} }
    \newcommand{\FFpref}{ \Ea{2} \No{2} \We{1.5} }
    \newcommand{\FF}{ \FFpref \We{.5} \No{1} \Ea{3} }
    \begin{tikzpicture}[inner sep=.15ex]
      \TileDot[O]{0,0}{T}
      \path (O) \PathM +(0,1) \CoorNode{n1} ;
      \Tile[n1]{n1}{T}
      \draw[struct-m] (O) \PathM -- (n1) ;
      \path (n1) \PathE +(0,1) \CoorNode{n2} ;
      \Tile[n2]{n2}{T}
      \draw[struct-p] (n1) \PathE -- (n2) ;
      \path (n2) \PathE +(0,1) \CoorNode{n3} ;
      \Tile[n3]{n3}{T}
      \path (n1) \PathN +(0,1) \CoorNode{nn1} ;
      \Tile[nn1]{nn1}{T}
      \draw[struct-q] (n1) \PathN -- (nn1) ;
      \draw[struct-p] (nn1) \EEpref ;
      \path (nn1) \PathN +(0,1) \CoorNode{nn2} ;
      \Tile[nn2]{nn2}{T}
      \draw[struct-q] (nn1) \PathN -- (nn2) ;
      \draw[struct-p] (nn2) \EEpref ;
      \draw[struct-p] (n2) \PathE -- (n3) ;
      \foreach \n in {1,2,3} {
        \draw[struct-q] (n\n) \No{1} -- ++(-0.75,0) ;
      } ;
    \end{tikzpicture}
  }
  \newcommand{\EEpref}{ \Ea{2} \So{2} \We{1.5} }
  \newcommand{\EE}{ \EEpref \We{.5} \So{1} \Ea{3} }
  \newcommand{\FFpref}{ \Ea{2} \No{2} \We{1.5} }
  \newcommand{\FF}{ \FFpref \We{.5} \No{1} \Ea{3} }
  \subcaptionbox{comb
    \label{fig:quipu:share:comb}}{\PreparePath{E}{EESSSWWSEEEEEEN}
    \begin{tikzpicture}[inner sep=.15ex]
      \TileDot[O]{0,0}{T}
      \path (O) \PathM +(0,1) \CoorNode{n1} ;
      \Tile[n1]{n1}{T}
      \draw[struct-m] (O) \PathM -- (n1) ;
      \path (n1) \PathE +(0,1) \CoorNode{n2} ;
      \Tile[n2]{n2}{T}
      \draw[struct-p] (n1) \PathE -- (n2) ;
      \path (n2) \PathE +(0,1) \CoorNode{n3} ;
      \Tile[n3]{n3}{T}
      \draw[struct-p] (n2) \PathE -- (n3) ;
      \foreach \n in {1,2,3} {
        \path (n\n) \PathN +(0,1) \CoorNode{nn1} ;
        \Tile[nn1]{nn1}{T}
        \draw[struct-q] (n\n) \PathN -- (nn1) ;
        \draw[struct-p] (nn1) \EEpref ;
        \path (nn1) \PathN +(0,1) \CoorNode{nn2} ;
        \Tile[nn2]{nn2}{T}
        \draw[struct-q] (nn1) \PathN -- (nn2) ;
        \draw[struct-p] (nn2) \EEpref ;
      } ;
    \end{tikzpicture}
  }
  \caption{Two cycles sharing a vertex.}
  \label{fig:teeh-on-backbone}
\end{figure}

The example in \RefFig{fig:teeh-on-backbone} illustrates the situation when $x_A$ is in \ZoneOne and $m=\EmptyWord$. Unless \AlphaMax is a grid, there are two possibilities: independent backbones or a comb.
If these are two independent backbones as in \RefFig{fig:quipu:share:2-backbone} then both cycles are unrolled to that they do not intersect.
Otherwise it is the situation in \RefFig{fig:quipu:share:comb}, the proto-tooth cycle has to be unrolled.

The result of this extension is denoted $\Quipu[A.mp^{\omega}]$.

\subsection{Quipu Build-up Step: extending quipu \Quipu with an ultimately periodic path $\ZZOrigin.mp^{\omega}$  in \AlphaMax such that $\Domain{\ZZOrigin.mp^{\omega}}\cap\Domain{\AlphaQuipu}$ is finite}

Let $A\in \IntegerSet^2$ and $m'$ and $m''$ be free paths such that $m'.A.m''p^\omega=\ZZOrigin.mp^{\omega}$ and $\Domain{A.m''p^\omega}\cap \Domain{\AlphaQuipu}=\{A\}$.
The quipu extension \QuipuOther is defined to be the quipu obtained from \Quipu by first performing the Periodic extension for $A.m''p^\omega$ followed by the Transient extension for the path $\ZZOrigin.m''$.
In that case we write $\QuipuOther=\Quipu[mp^{\omega}]=\left(\Quipu[A.m'p^{\omega}]\right)[m'']$.

By the construction of the Periodic and the Transient Extension Steps, it might happen that the Quipu Build-up Step yields $\Quipu[\text{grid}]$.
If it is not the case, \QuipuOther is an extension of \Quipu such that $\Domain{\ZZOrigin.mp^\omega}\subseteq\Domain{\AlphaQuipuOther}$ as explained below.

\begin{lemma}\label{lem:add-cycle}
  Let $\Domain{\ZZOrigin.mp^\omega}\cap\Domain{\AlphaQuipu}$ be finite for a path $\ZZOrigin.mp^\omega$ in \AlphaMax, and let $\QuipuOther$ be obtained from \Quipu by a Quipu Build-up Step.
  If $\QuipuOther \neq \Quipu[\text{grid}]$ then the number of cycles in $\QuipuOther=\Quipu[mp^\omega]$ is larger than the number of cycles in \Quipu and $\Domain{\ZZOrigin.mp^\omega}\subseteq\Domain{\alpha_{\QuipuOther}}$.
\end{lemma}

\begin{proof}
  Let $A\in \IntegerSet^2$ is such that $\Domain{A.m''p^\omega}\cap \Domain{\AlphaQuipu}=\{A\}$ and $m'.A.m''p^\omega=\ZZOrigin.mp^{\omega}$.
  So, there is a vertex in \Quipu whose cover contains $A$. 
  The lemma follows directly from the construction of the Periodic Extension step.
  Because $\ZZOrigin.mp^\omega$ is in \AlphaMax, $A.m''p^\omega$ must be in the $\NonCausal{A}$.
  Unless \AlphaMax is a grid, the Periodic Extension Step adds a cycle.
  Let $\Quipu'$ be the quipu obtained from \Quipu at the end of the Periodic Extension Step.
  By the single direction extensions within the Transient Extension Step, for each prefix $l$ of $m'$, $\CoverBy{\Quipu'}$ contains $\ZZOrigin +\Vect{l}$. 
\end{proof}

If there is no ultimately periodic path in \AlphaMax that intersects \AlphaQuipu a finite number of times, by \RefLem{lem:quipu:utp}, there are no infinite paths in \AlphaMax that have a finite intersection with \AlphaQuipu.
If $\Domain{\AlphaQuipu}\subsetneq\Domain{\AlphaMax}$, then there are vertices in \AlphaMax adjacent to \Domain{\AlphaQuipu} but not in \AlphaQuipu.
Let $d$ be a direction of one such edge and $t$ the label of the corresponding vertex in \AlphaQuipu incident to the edge. 
Then we set $\QuipuOther=\Quipu[X,d,t]$ to be an extension of \Quipu obtained by addition of a direction $d$ with tile $t$ as in \RefLem{lem:finite-union}.

If $\Domain{\AlphaQuipu}=\Domain{\AlphaMax}$ then we, set $\QuipuOther=\Quipu[\text{halt}]$.

\medskip

The theorem below shows that if the Quipu Build-up Step cannot extend \Quipu with an ultimately periodic path, then there are only decorations that can be added to its structure such that its paths cover whole domain of \AlphaMax, and those decorations can be added in a finite number of steps. 

\begin{theorem}
  \label{th:finite-decorations}
  Let \Quipu be such that $\Domain{\AlphaQuipu}\subsetneq\Domain{\AlphaMax}$.
  Suppose for all infinite paths $\pi$ in \AlphaMax such that $\Domain{\pi}\not\subseteq\Domain{\AlphaQuipu}$ the intersection $\Domain{\pi}\cap \Domain{\AlphaQuipu}$ is infinite. 
  Then there are finite number of Quipu Build-up 
  Steps of the form $\Quipu[X,d,t]$ that extend \Quipu to \QuipuOther such that $\Domain{\AlphaQuipuOther}=\Domain{\AlphaMax}$.
\end{theorem}

\begin{proof}
  Because both \AlphaMax and \AlphaQuipu are connected and have a non-empty intersection, there is a vertex $y$ in \AlphaMax and $x$ in \Quipu such that $y\not\in\CoverBy{\Quipu}$ and $y\in\CoverBy{x}+d$ for some $d\in\DirectionSet$.
  Then \Quipu is extended to $\Quipu[\{x\},d,\eta(x)]$ whose cover contains $y$.
  An extension of the form $\Quipu[\{x\},d,t]$ is performed step by step as long as there is a vertex in \AlphaMax that is not in \AlphaQuipu. When such a vertex does not exist, $\Domain{\AlphaMax}=\Domain{\AlphaQuipu}$ and the extension becomes $\Quipu[\text{halt}]$.
  In order to cover the whole domain of \AlphaMax with only finite number of such extensions, we first extend with directions that are 
  incident to covers of \Quipu-vertices in \ZoneTwo, then in \ZoneOne, and finally in \ZoneZero.
  
  If $x$ belongs to \ZoneTwo, then the extension $\Quipu[\{x\},d,t]$ adds vertices to some proto-tooth $P$.
  The set of vertices covered by $P$ is of the form $F+\NaturalSet\Vect{b}+\NaturalSet\Vect{c}$ where $F$ is a finite set $b$ is the period of its backbone and $c$ is the period of the tooth.
  If a vertex $y$ is added to \Quipu at proto-tooth $P$, then $\CoverBy[Q']{y}=F'+\NaturalSet\Vect{b}+\NaturalSet\Vect{c}$ for some finite $F'$.
  Since the covers of distinct vertices of \Quipu are disjoint, the number of vertices that can be added are bounded between two consecutive teeth of $P$ that are 
  \Vect{b} apart and the periods of the teeth $\Vect{c}$. Hence, the number of additions is bounded by the number of vertices in the parallelogram of sides \Vect{b} and \Vect{c} and there are only finitely many such vertices, and therefore, finitely many additions to teeth in \ZoneTwo.
  
  If $x$ belongs to \ZoneOne, it belongs to some backbone $b$.
  Any extension with a vertex incident to a vertex in \ZoneOne adds to the cover of quipu a set ${F}+\NaturalSet\Vect{b}$ where $F$ is a finite set and $\Vect{b}$ is the period of its backbone.
  With every addition, the vertices that are being added are at an eventually increasing distance from the periodic portion $b$ of the backbone.
  Hence, by continuous addition, after some possible unrollings of $b$, there is a branch of arbitrary length attached to $b$ in the quipu. 
  If infinitely many vertex additions are made to $b$, by K\"onig's lemma there exist an infinite path
  $\pi$ in \ZoneOne starting at a vertex $z$ in $b$. By construction, the infinite path $\ZZOrigin.m\pi$ where $m$ is the label of the shortest path from the root to $z$ intersects \Quipu only finite number of times.
  We also observe that $\ZZOrigin.m\pi$ must be an infinite path in \AlphaMax.
  Let $X=\CoverBy{\ZoneZero}$.
  Then $X$ is finite and remains unchanged after new vertices are added to \Quipu.
  Because the added vertices are at increased distance from the backbone $b$, there is $k$ such that the cover of the $j$th added vertex in the quipu does not intersect any $\pm k\cdot\Vect{b}$ translation of $X$. This means that sufficiently away from \ZoneZero, every addition of a new vertex in \Quipu appears in its non-causal part, and no unrolling of $b$ is performed, that is the number of times $b$ is unrolled is finite. 
  Thus $\ZZOrigin.m\pi$ belongs to \AlphaMax, a contradiction with assumption that there is no infinite path in \AlphaMax that has only a finite number of intersections with \AlphaQuipu.
  Therefore the number of vertex additions in \ZoneOne is finite.

  If infinitely many vertex additions on \ZoneZero were possible, then using similar arguments as above, by K\"onig's lemma there would be an infinite branch, with an infinite path in \AlphaMax finitely intersecting \AlphaQuipu. This is again contradiction with our assumption that no such path exists. 
  
  Therefore, in a finite number of steps, an extension $\QuipuOther$ of \Quipu is obtained satisfying $\Domain{\QuipuOther}=\Domain{\AlphaMax}$.
\end{proof}

\section{Quipu Construction in Finite Steps}
\label{sec:quipu-finite}

In this section, we prove that it is possible to build quipu that covers \AlphaMax.
We assume that \AlphaMax is infinite, otherwise, \AlphaMax is covered by a quipu identical to a spanning tree of \AlphaMax.

\begin{definition}
  A \emph{quipu filtration} corresponding to TAS \TAS is a sequence $\Quipu_{\TAS}=(\Quipu_0,\Quipu_1,\ldots,\QuipuI,\ldots)$ of quipus associated to \TAS such that $\Quipu_0=\{r\}$, and for each $i>0$, $\QuipuIpO$ is generated from $\QuipuI$ by some extension.
\end{definition}

A filtration $\Quipu_{\TAS}$ is finite if there is $i$ such that $\QuipuI=\Quipu[\text{grid}]$ or $\QuipuI=\Quipu[\text{halt}]$.
The aim of this section is to show that for each TAS \TAS there is a finite quipu filtration.
If the filtration ends with $\Quipu[\text{halt}]$, then by definition of $\Quipu[\text{halt}]$, the domain of the last quipu in the filtration coincides with \Domain{\AlphaMax}. 

For a chosen order of directions $D$, we order ultimately periodic producible assembly paths $\ZZOrigin.mp^\omega$ for $m$ and $p$ not empty as follows:
\begin{equation*}
  m_1p_1^\omega < m_2p_2^\omega \text{ if } \left\{
    \begin{array}{ll}
      |m_1|{+}|p_1|<|m_2|{+}|p_2| & \\
      |m_1|{+}|p_1|=|m_2|{+}|p_2| \wedge |m_1|<|m_2| \\
      |m_1|{+}|p_1|=|m_2|{+}|p_2| \wedge |m_1|=|m_2| \wedge m_1p_1<m_2p_2 \text{ in alphabetical order}.
    \end{array}
  \right.
\end{equation*}
This is a well ordering and such sequences can be enumerated by a breath first search.

In the rest of the section $\Quipu_{\TAS}$ refers to a filtration for a TAS \TAS that is defined inductively as follows. 
The induction starts with the quipu $\QuipuZero=\{ \Root\}$, then $\AlphaQuipuZero$ is $\{\ZZOrigin\}$.
Let $m_0p_0^{\omega}$ be the least ultimately periodic path according to the above order such that $\ZZOrigin.m_0p_0^\omega$ is in \AlphaMax and finitely intersect \AlphaQuipu.
The ultimately periodic path $\ZZOrigin.m_0p_0^\omega$ exists in \AlphaMax because if \AlphaMax is not finite it is para-periodic by \RefTh{th:nice}.
Then we set $\Quipu_1=\QuipuZero[m_0p_0^{\omega}]$ as a result of a Quipu Build-up Step.
It follows that every $\QuipuI\in \Quipu_{\TAS}$ has a cycle for positive $i$.

If $\QuipuI=\Quipu[\text{grid}]$, or $\QuipuI=\Quipu[\text{halt}]$, the filtration stops and it is finite.
Otherwise, let $P_i$ be the ordered list of ultimately periodic paths in \AlphaMax that finitely intersect $\AlphaQuipuI$. 

If $P_i=\emptyset$, i.e., for each paths $mp^{\omega}$  it holds that either it is not in $\AlphaMax$, or its intersection with $\Domain{\alpha_{\QuipuI}}$ is not finite, then by \RefLem{lem:quipu:utp}, there are no infinite paths in \AlphaMax that have a finite intersection with \AlphaQuipu.
Let $d$ be a direction for which there is a vertex $x$ in $\alpha_{\QuipuI}$ whose cover is adjacent to a vertex in $\Domain{\AlphaMax}\setminus \Domain{\alpha_{\QuipuI}}$ in direction $d$.
Let $t=\eta(x)$.
Then we replace \QuipuI by $\QuipuI[\{x\},d,t]$ as long as there is such an $x$, first we consider vertices $x$ in \ZoneTwo, then in \ZoneOne, and at the end vertices in \ZoneZero.
By \RefTh{th:finite-decorations}, there is a finite number of extensions of this type, and eventually vertices like $x$ do not exist, the obtained quipu is \QuipuIpO hence $\Domain{\AlphaMax}=\Domain{\alpha_{\QuipuIpO}}$ and $\QuipuIpO=\Quipu[\text{halt}]$.

\smallskip

Suppose $P_i\not =\emptyset$ and let $mp^\omega$ be the least element in $P_i$.
We perform the following steps:
\begin{enumerate}
\item for every cycle $c$ in $\QuipuI$ (in \ZoneOne or in \ZoneTwo) such that $p$ and $c$ are not co-linear, for any $\hat p\in\Rotate{p}$ and $\hat c\in\Rotate{c}$ test whether the following is an assembly in \TAS starting from some tile $t$ (not necessarily producible):
  \begin{equation}
    \label{eq:grid-detection}
    ^{\omega}\hat p
    .\ZZOrigin.
    \hat p^{\omega}
    \SubGraphCup
    {^{\omega}\hat c}
    .\ZZOrigin.
    \hat c^{\omega}
    \enspace .
  \end{equation}
  If any such exists, then \AlphaMax is a grid by \RefLem{lem:intersecting-period:unbounded} and we set  $\QuipuIpO=\Quipu[\text{grid}]$, otherwise
\item set $\QuipuIpO$ to be $\QuipuI[mp^{\omega}]$ as a result of a Quipu Build-up Step.
\end{enumerate}

In order to show that the filtration $\Quipu_{\TAS}$ is finite, it only remains to show that the induction uses step 2. above only finite number of times. 
In the rest of this section, we show that there is an $i$ such that either $\QuipuI=\Quipu[\text{grid}]$ or $P_i$ is empty.

\subsection{Grids are detected within a finite number of steps}

We observe that if \AlphaMax is a grid, the sequence of quipu extensions in the filtration described above stops in a finite number of steps.

\begin{lemma}[Grid detection]
  \label{lem:grid-detection}
  If \AlphaMax is a grid, then there the filtration $\Quipu_{\TAS}$ is finite and ends with $\Quipu[\text{grid}]$.
\end{lemma}

\begin{proof}
  Suppose $\AlphaMax=\GridFunction(m,p,q)$.
  Let $p_0$ be the first cycle added to the filtration.
  Consider the least $ns^{\omega}$ in the ordered list of ultimately periodic paths that is in \AlphaMax such that $s$ is non colinear to $p_0$.
  By \RefLem{lem:struc:grid:period extension}, since there are assembly periodic paths with cycles $p_0$ and $s$, there are also bi-infinite assembly periodic paths with the same cycles (not necessarily producible).
  Because the cycles are not colinear, and \AlphaMax is a grid, there is an assembly (not necessarily producible) where these bi-infinite paths intersect at a vertex so that the condition \RefEq{eq:grid-detection} is satisfied.

  The list of ultimately periodic free paths is well-ordered so the filtration process will reach the free path $ns^\omega$.
  Let $\QuipuI$ be the quipu constructed thus far. 
  At this point, all the cycles in $\QuipuI$ are colinear to $p_0$; and therefore there are no proto-teeth added to these backbones.
  The ultimately periodic path $\ZZOrigin.ns^{\omega}$ has only finitely many intersections with the finitely many non-colinear toothless backbones in $\QuipuI$.
  Hence, the corresponding $P_i$ has $ns^{\omega}$ as its least element and the extension $\Quipu[ns^{\omega}]$ is considered.
  The first step in this extension process is checking condition \RefEq{eq:grid-detection} which uncovers the grid. 
\end{proof}

In the rest of this section, it is assumed that \AlphaMax is not a grid. Suppose for each $i$, $P_i\not =\emptyset$. We have that $P_{i+1}\subsetneq P_i$.
We obtain a contradiction in the construction of $\Quipu_{\TAS}$. 

\subsection{Comb representatives}

We consider a special set of combs, which we call a set of comb representatives.

\begin{definition}
  A set of combs with non-empty teeth \SetCombRepresentative is said to be a {\it set of comb representatives} for \AlphaMax if 
  \begin{enumerate}
  \item the backbone of every comb in \AlphaMax with non-empty teeth intersects infinitely often a backbone of a comb in \SetCombRepresentative, 
  \item backbones of any two combs in \SetCombRepresentative do not intersect.
  \end{enumerate}
\end{definition}

The first condition implies that any backbone of every comb in \AlphaMax with non-empty teeth is co-linear with a backbone of a comb in \SetCombRepresentative.

\begin{lemma}[finite comb representatives]
  If \AlphaMax has a finite set of comb representatives, then all comb representatives are finite.
\end{lemma}

\begin{proof}
  Suppose there is a set of comb representatives \SetCombRepresentative that is infinite. If a backbone intersects another backbone infinite number of times, then they have colinear displacement vectors. Hence, 
  if two combs in \SetCombRepresentative have parallel backbones, they to not intersect. Let $p_1^\omega$ and $p_2^\omega$ be such backbones with $\Vect{p_1}=a\Vect{p_2}$. Suppose $a$ is positive. 
  Because these are combs with teeth, if \AlphaMax is not a grid, the teeth cannot intersect, nor they can intersect with the backbone of the other comb.
  So either the right, or the left region of the bi-infinite path $^\omega\RevBack{p_1}\RevBack{m_1}.\ZZOrigin.m_2p_2^\omega$ cannot have any teeth ($m_1$ and $m_2$ are the corresponding transient parts).
  So there cannot be a third backbone parallel to $p_1$ and $p_2$ in the same direction.
  Therefore, \SetCombRepresentative can have at most four combs with co-linear backbones.
  Hence, the set of backbone displacement vectors of combs in \SetCombRepresentative is infinite and there cannot be any finite set of comb representatives (with a finite set of displacement vectors).
\end{proof}

This proves that if \SetCombRepresentative is finite, then the number of non-colinear backbones with teeth in \AlphaMax, $e$ satisfies:
$e \leq |\SetCombRepresentative| \leq 4e$.

\begin{lemma}[quipu with finite comb representatives]
  \label{lem:comb:incorporation}
  If \AlphaMax has a finite set of comb representatives, then there is $j$ and $\Quipu_j\in \Quipu_{\TAS}$ such that 
  the set of non-empty teeth combs in $\Quipu_j$ forms a set of comb representatives. 
\end{lemma}

\begin{proof}
  The proof follows directly from the extensions in Quipu Build-up steps.
  If a path $mp^\omega$ is the least element in $P_i$ then, $\QuipuIpO=\QuipuI[mp^{\omega}]$ and $\Domain{\ZZOrigin.mp^\omega}\subseteq\Domain{\QuipuIpO}$ (\RefLem{lem:add-cycle}). 
  Since there is a finite set of comb representatives for \AlphaMax, there is a bound $B$ on the backbone transient and period lengths for combs in this set of representatives.
  So, there is $j$ such that all ultimately periodic paths $mp^\omega$ such that $p$ is a backbone of a comb with teeth and $|p|>B$ intersect $\Quipu_j$ infinitely often.
  Hence, the set of cycles in \ZoneOne of $\Quipu_j$ contains backbone periods of all combs in a set of representatives.
  The teeth of these combs (if their respective cycles are not already in $\Quipu_j$) intersect $\Quipu_j$ finitely often (otherwise, there is an intersection between two combs and by \RefLem{lem:tooth-to-tooth}, \AlphaMax is a grid).
  Hence $P_j$ contains ultimately periodic paths that correspond to all these teeth.
  Observe that if two backbones are co-linear and they intersect infinitely often, far from the seed, after both transient parts are passed, a tooth for one of the backbone is also a tooth for the other (differing by a segment close to the backbone). 
  There are only finite number of teeth (close to the seed) that might be teeth for one of the backbone, but not the other.
  Hence, after finite Quipu Build-up extension steps a quipu is extended with a tooth of a comb backbone, which will show as a proto-tooth cycle in \ZoneTwo attached to one of the backbones of comb representatives.
  Because there are finite number of proto-teeth for each comb (\RefLem{lem:struc:bounded-teeth}) in finite number of steps, at least one proto-tooth for all backbones of comb representatives will appear in an extension of $\Quipu_j$. 
\end{proof}

\subsection{Pseudo-combs and infinite hits}

\begin{lemma}\label{lem:path-with-cycles}
  Suppose the filtration $\Quipu_{\TAS}$ is infinite.
  For every $k\ge 1$ there is an $i$ and a direction $d$ such that $\QuipuI$ has a rooted path $\pi$ in $\ZoneZero$ and $k$ cycles in \ZoneOne corresponding to ultimately periodic paths $\ZZOrigin.m_j'dm_j''p_j^\omega\in \AlphaQuipuI$ ($j=1,\ldots,k$) such that $m_j'$ are distinct prefixes of $\pi$ and no $m_j'd$ is a prefix of $\pi$. 
\end{lemma}

\begin{proof} 
  If the filtration is infinite, by \RefTh{th:finite-decorations} there are infinite number of steps $\QuipuIpO=\QuipuI[mp^{\omega}]$ and by \RefLem{lem:add-cycle} with each extension step the quipu is extended by at least a cycle.
  Hence, the extension steps construct a quipu with arbitrarily large number of cycles.
  If there is a bound $B$ on the number of cycles that can appear in \ZoneOne, then there is a bound on the number of backbones that can appear in \AlphaQuipuI.
  By \RefLem{lem:struc:bounded-teeth}, each backbone can have only finite number of proto-teeth, and therefore from each cycle in \ZoneOne there are only finite number of cycles in \ZoneTwo.
  This implies that the total number of cycles in $\QuipuI$ is bounded, which is contradictory to the assumption that the number of steps in the filtration that add cycles to the quipu is not bounded. Hence, the number of cycles that can appear in \ZoneOne of a quipu in $\Quipu_{\TAS}$ is not bounded. 
  By \RefDef{def:quipu}, and the construction steps, each quipu is deterministic, hence from each vertex in $\QuipuI$ there are at most four outgoing arcs (each arc per direction). The vertices of \ZoneZero in $\QuipuI$ form a rooted tree with at most four branches per vertex that can lead to arbitrarily number of leafs (cycles in \ZoneOne).
  Hence there must be $\QuipuI$ with a path $\pi$ in \ZoneZero that is long enough, and a direction $d$ with arbitrarily large number of branches leading out of $\pi$ in direction $d$. Because $\Domain{\alpha_{Q_{i}}}\subset\Domain{\alpha_{Q_{i+1}}}$, the lemma follows.
\end{proof}

The situation described in ~\RefLem{lem:path-with-cycles} implies that in case of an infinite filtration, there is a path $\pi$ in the sequence of quipus that is part of \ZoneZero, that can have arbitrarily many backbones starting on one of its sides. 
In order to facilitate our discussion is the lemmas that follow we call such situation a {\it pseudo-comb} and investigate its properties. The proofs use the techniques developed in~\RefSec{sec:nice}, in particular co-grow and points of interest in an infinite path relative a periodic path.
As observed with the main theorem, these properties imply that a confluent \TAS must have a finite filtration. 

\begin{definition}[pseudo-comb associated with a filtration]
  A \emph{pseudo-comb associated with the filtration $\Quipu_{\TAS}$}, \PseudoComb, is a pair $(\Path,\{\beta_i,m_ip_i\}_{i\in\NaturalSet})$ such that:
  \begin{enumerate}
  \item \Path is an infinite path in \AlphaMax starting at \ZZOrigin, $\beta_i$ is an increasing sequence of integers and $m_ip_i^\omega$ are ultimately periodic free paths ($0\leq i$),
  \item $\Path\cap \Path_{\beta_i}.m_ip_i^{\omega}=\{\Path_{\beta_i}\}$, 
  \item for all $i$ there is some $k$ such that the ultimately periodic path $\Path_{\IntegerInterval{0}{\beta_i}}.m_ip_i^{\omega}$ belongs to $\alpha_{\Quipu_k}$, and $\Domain{(\Path_{\beta_i}+\Vect{m_i}).p_i^{\omega}}$ belongs to \CoverBy[\Quipu_k]{\ZoneOne}.
  \end{enumerate}
  A pseudo-comb is \emph{two-sided} if there are infinitely many $\Path_{\beta_i}.m_ip_i^{\omega}$ in each of the left and the right regions delimited by $^{\omega}\RevBack{p_0}\RevBack{m_0}.\Path_{\IntegerInterval{\beta_0}{+\infty}}$.
  Otherwise it is {\it left-sided} or {\it right-sided} depending on which region contains infinitely many $\Path_{\beta_i}.m_ip_i^{\omega}$. A pair of pseudo-combs one being left-sided and the other right-sided are called {\em complementary}.
\end{definition}

\begin{figure}[hbt]
  \small\SetUnitlength{1.6em}\newcommand{\VStep}{3}\newcommand{\VInc}{.4}\newcommand{\HLong}{3.5}\newcommand{\HShort}{2}\centering\subcaptionbox{right-sided\label{quipu:pseudo-comb:basic}}{\ \renewcommand{\VStep}{2}\begin{tikzpicture}[x=.7\unitlength,inner sep=.25ex]
      \TileDot[O]{\HShort,\VStep/3+.5}{\Seed} ;
      \path (O) node [below=.2em] {\ZZOrigin\ } ;
      \TileDot[g1]{\HShort,\VStep}{} ;
      \TileDot[g2]{\HShort,\VStep*2}{} ;
      \TileDot[g3]{\HShort,\VStep*3}{} ;
      \TileDot[g4]{\HShort,\VStep*4}{} ;
      \TileDot[g5]{\HShort,\VStep*5}{} ;
      \begin{scope}[shift={(0,\VInc)}]
        \TileDot[a1]{\HLong,\VStep}{} ;
        \TileDot[a2]{\HLong+1.25,\VStep*2+.25}{} ;
        \TileDot[a3]{\HLong+1.5,\VStep*3+.5}{} ;
        \TileDot[a4]{\HLong+1,\VStep*4+.25}{} ;
        \TileDot[a5]{\HLong+1.5,\VStep*5+.5}{} ;
        \path (a1) +(\HLong,-\VStep/2) \CoorNode{aa1} ;
        \path (a1) +(-0.5,-\VStep) \CoorNode{aaa1} ;
        \path ($(a2)+(aaa1)-(a1)+(.5,.25)$) \CoorNode{aa2} ;
        \path ($(a3)+(aaa1)-(a1)+(1.5,.25)$) \CoorNode{aa3} ;
        \path ($(a4)+(aaa1)-(a1)+(2.5,.75)$) \CoorNode{aa4} ;
        \path ($(a5)+(aaa1)-(a1)+(2.5,1)$) \CoorNode{aa5} ;
        \path (g5) +(0,\VStep) node[below left] (g6) {\Path} ;
      \end{scope}
      \begin{scope}
        \TransientPathStyleAdd{ultra thick}
        \TransientPath{O}{}{g1}
        \TransientPath{g1}{}{g2}
        \TransientPath{g2}{}{g3}
        \TransientPath{g3}{}{g4}
        \TransientPath{g4}{}{g5}
        \TransientPath{g5}{}{g6}
      \end{scope}
      \TransientPath{g1}{}{a1}
      \TransientPath{g2}{}{a2}
      \TransientPath{g3}{}{a3}
      \TransientPath{g4}{}{a4}
      \TransientPath{g5}{}{a5}
      \PeriodicPath{a1}{}{aaa1}
      \PeriodicPath{a2}{}{aa2}
      \PeriodicPath{a3}{}{aa3}
      \PeriodicPath{a4}{}{aa4}
      \PeriodicPath{a5}{}{aa5}
    \end{tikzpicture}\ }\qquad\subcaptionbox{two-sided\label{quipu:pseudo-comb:two-sided}}{\renewcommand{\HShort}{.5}\begin{tikzpicture}[x=.7\unitlength,inner sep=.25ex]
      \TileDot[O]{0,\VStep/3+.5}{\Seed} ;
      \path (O) node [below=.2em] {\ZZOrigin\ } ;
      \TileDot[g1]{.5*\HShort,\VStep}{} ;
      \TileDot[g2]{\HShort,\VStep*2}{} ;
      \TileDot[g3]{\HShort,\VStep*3}{} ;
      \TileDot[h1]{-.5*\HShort,\VStep*1.5}{} ;
      \TileDot[h2]{-\HShort,\VStep*2.25}{} ;
      \TileDot[h3]{-\HShort,\VStep*3.3}{} ;
      \begin{scope}[shift={(0,\VInc)}]
        \TileDot[a1]{\HLong,\VStep}{} ;
        \TileDot[a2]{\HLong-.25,\VStep*2+.25}{} ;
        \TileDot[a3]{\HLong,\VStep*3+.5}{} ;
        \TileDot[b1]{-\HLong,\VStep}{} ;
        \TileDot[b2]{-\HLong-.25,\VStep*1.75+.5}{} ;
        \TileDot[b3]{-\HLong,\VStep*2.75+1}{} ;
        \path (a1) +(\HLong,-\VStep*.8) \CoorNode{aa1} ;
        \path (a1) +(-0.5,-\VStep) \CoorNode{aaa1} ;
        \path ($(a2)+(aa1)-(a1)+(0,1)$) \CoorNode{aa2} ;
        \path ($(a3)+(aa1)-(a1)+(0,2)$) \CoorNode{aa3} ;
        \path (b1) +(-\HLong*.5,-\VStep*.6) \CoorNode{bb1} ;
        \path ($(b2)+(bb1)-(b1)+(0,1)$) \CoorNode{bb2} ;
        \path ($(b3)+(bb1)-(b1)+(0,2)$) \CoorNode{bb3} ;
        \path (g3) +(0,\VStep) node[below left] (g4) {\Path} ;
        \path (h3) +(0,\VStep) \CoorNode{h4} ;
      \end{scope}
      \begin{scope}
        \TransientPathStyleAdd{ultra thick}
        \TransientPath{O}{}{g1}
        \TransientPath{g1}{}{h1}
        \TransientPath{g2}{}{h2}
        \TransientPath{g3}{}{h3}
        \TransientPath{h1}{}{g2}
        \TransientPath{h2}{}{g3}
        \TransientPath{h3}{}{g4}
      \end{scope}
      \TransientPath{g1}{}{a1}
      \TransientPath{g2}{}{a2}
      \TransientPath{g3}{}{a3}
      \PeriodicPath{a1}{}{aaa1}
      \PeriodicPath{a2}{}{aa2}
      \PeriodicPath{a3}{}{aa3}
      \TransientPath{h1}{}{b1}
      \TransientPath{h2}{}{b2}
      \TransientPath{h3}{}{b3}
      \PeriodicPath{b1}{}{bb1}
      \PeriodicPath{b2}{}{bb2}
      \PeriodicPath{b3}{}{bb3}
    \end{tikzpicture}}\qquad\subcaptionbox{complementary\label{quipu:pseudo-comb:bi-infinite}}{\begin{tikzpicture}[x=.7\unitlength,inner sep=.25ex]
      \TileDot[O]{0,\VStep/3+.5}{\Seed} ;
      \path (O) node [below=.2em] {\ZZOrigin\ } ;
      \TileDot[g1]{.5*\HShort,\VStep}{} ;
      \TileDot[g2]{\HShort,\VStep*2}{} ;
      \TileDot[g3]{\HShort,\VStep*3}{} ;
      \TileDot[h1]{-.5*\HShort,\VStep*1.5}{} ;
      \TileDot[h2]{-\HShort,\VStep*2.25}{} ;
      \TileDot[h3]{-\HShort,\VStep*3.1}{} ;
      \begin{scope}[shift={(0,\VInc)}]
        \TileDot[a1]{\HLong,\VStep}{} ;
        \TileDot[a2]{\HLong-.25,\VStep*2+.25}{} ;
        \TileDot[a3]{\HLong,\VStep*3+.5}{} ;
        \TileDot[b1]{-\HLong,\VStep}{} ;
        \TileDot[b2]{-\HLong-.25,\VStep*1.75+.5}{} ;
        \TileDot[b3]{-\HLong,\VStep*2.75+1}{} ;
        \path (a1) +(\HLong,-\VStep*.7) \CoorNode{aa1} ;
        \path (a1) +(-0.5,-\VStep) \CoorNode{aaa1} ;
        \path ($(a2)+(aa1)-(a1)+(0,1)$) \CoorNode{aa2} ;
        \path ($(a3)+(aa1)-(a1)+(0,2)$) \CoorNode{aa3} ;
        \path (b1) +(-\HLong*.5,-\VStep*.6) \CoorNode{bb1} ;
        \path ($(b2)+(bb1)-(b1)+(0,1)$) \CoorNode{bb2} ;
        \path ($(b3)+(bb1)-(b1)+(0,2)$) \CoorNode{bb3} ;
        \path (g3) +(0,\VStep) node[below left] (g4) {\Path} ;
        \path (h3) +(0,\VStep) node[below left] (h4) {$\Path'$} ;
      \end{scope}
      \begin{scope}
        \TransientPathStyleAdd{ultra thick}
        \TransientPath{O}{}{g1}
        \TransientPath{g1}{}{g2}
        \TransientPath{g2}{}{g3}
        \TransientPath{g3}{}{g4}
        \TransientPath{g1}{}{h1}
        \TransientPath{h1}{}{h2}
        \TransientPath{h2}{}{h3}
        \TransientPath{h3}{}{h4}
      \end{scope}
      \TransientPath{g1}{}{a1}
      \TransientPath{g2}{}{a2}
      \TransientPath{g3}{}{a3}
      \PeriodicPath{a1}{}{aaa1}
      \PeriodicPath{a2}{}{aa2}
      \PeriodicPath{a3}{}{aa3}
      \TransientPath{h1}{}{b1}
      \TransientPath{h2}{}{b2}
      \TransientPath{h3}{}{b3}
      \PeriodicPath{b1}{}{bb1}
      \PeriodicPath{b2}{}{bb2}
      \PeriodicPath{b3}{}{bb3}
    \end{tikzpicture}}
  \caption{Pseudo-combs.}
  \label{quipu:pseudo-comb}
\end{figure}
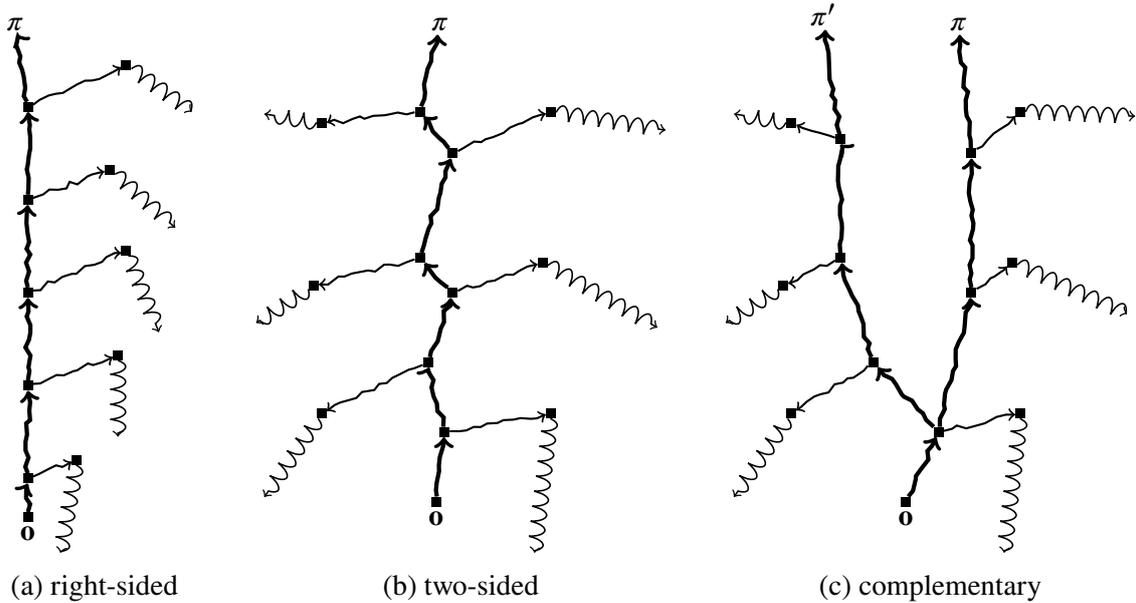

\RefFigure{quipu:pseudo-comb} depicts the general shape of (right-sided) pseudo-combs.
A two-sided pseudo-comb is depicted in \RefFig{quipu:pseudo-comb:two-sided}.
Two complementary pseudo-combs are depicted in \RefFig{quipu:pseudo-comb:bi-infinite}.

In the remaining discussion we consider all pseudo-combs to be associated with the fixed filtration $\Quipu_{\TAS}$, and we will refer simply to pseudo-combs.

We note that a comb can be a pseudo-comb if we set that $\Path$ is its back-bone and $m_ip_i^\omega$ are its teeth, however, a comb in $\Quipu_k$ whose domain includes points from $\CoverBy[\Quipu_k]{Z_2}$ does not satisfy the condition 3 above and hence, is not a pseudo-comb.

\begin{corollary}
  \label{cor:path-with-cycles}
  If the filtration $\Quipu_{\TAS}$ is infinite, then there is a pseudo-comb that is not a comb for any quipu in the filtration.
\end{corollary}

\begin{proof} 
  The proof follows directly from \RefLem{lem:path-with-cycles} and the definition for pseudo-comb.
  The path \Path cannot become ultimately periodic during this construction, because it would appear in the ordered list of ultimately periodic paths and hence the quipu extension step would add \Path as a cycle outside zone \ZoneZero. 
\end{proof}

\begin{lemma}
  \label{lem:bi-infinite-pseudo-comb}
  If the filtration $\Quipu_{\TAS}$ is infinite, either there is a two-sided pseudo-comb or there are two complementary pseudo-combs.\end{lemma}

\begin{proof}
  Suppose $\Quipu_{\TAS}$ is infinite. Then the path \Path of a pseudo-comb $\PseudoComb$ constructed in \RefLem{lem:path-with-cycles} extends to an infinite path in \AlphaMax by Koenig's lemma. We observe that \Path cannot be ultimately periodic because at certain step it would appear as the least element in some $P_k$, and the extension step for $\Quipu_k$ would add \Path as a cycle outside \ZoneZero. Suppose that the pseudo-comb $\PseudoComb$
  is left sided and $\beta_j$ is maximal such that the ultimately periodic path $\Path_{\beta_j}.m_jp_j^\omega$ is in \AlphaMax and the interior of the right region of 
  $^\omega\RevBack{p_j}\RevBack{m_j}.\Path_{\beta_j}.\Path_{\IntegerInterval{\beta_j}{+\infty}}$
  has no other paths from the pseudo-comb $\PseudoComb$. If there are no paths on the right side of $\Path$, we take $\beta_j=\beta_0$ where $\Path_{\beta_0}.m_0p_0^\omega$ is the first path on the left side of the pseudo-comb $\PseudoComb$. 
  There is $\QuipuI$ in the filtration $\Quipu_{\TAS}$ such that the domain of $\AlphaQuipuI$ contains $\Path_{\beta_j}.m_jp_j^\omega$. 
  Then the right region of $^\omega\RevBack{p_j}\RevBack{m_j}.\Path_{\beta_j}.\Path_{\IntegerInterval{\beta_j}{+\infty}}$ and $\Path_{\beta_j}.m_jp_j^\omega$ satisfy the conditions of~\RefCor{cor:new-periodic-path} and hence, there is an ultimately periodic path in $\AlphaMax$ in this region starting from a
  vertex in $\pi$ that finitely intersects $\QuipuI$. Because the filtration is infinite, either $\PseudoComb$ cannot have finite number of ultimately periodic paths on the right of $\pi$, and it must be two sided, or (in case $\beta_0=\beta_j$)
  $\PseudoComb$ can be extended to a right sided pseudo-comb. 
\end{proof}

By definition of pseudo-comb, all $\Path_{\beta_i}.m_ip_i^{\omega}$ paths are disjoint (each such path is a backbone in \ZoneOne of some quipu in the filtration, and they are distinct cycles because they differ at $\Path_{\beta_i}$). 
Similarly as in~\RefSec{sec:nice}, we define points of interest on the path $\pi$ of a given pseudo-comb, except in this case we choose one of the periods of the backbones to play the role of the horizontal lines in \RefFig{fig:3-PoI}.
One assumes that a bi-infinite periodic path $^{\omega}p.A.p^\omega$ slides in one direction $d$, and if \Path extends in a direction non-colinear to $d$, then there is an `end-point' intersection between \Path and a shifted bi-infinite path.
This end-point of intersection becomes a point of interest for that bi-infinite path.
These points are chosen on \Path such that every vertex in \Path that follows from a point of interest is in the non-causal part of that point.
We define these points of interest more precisely below. 

\begin{definition}[infinitely hit]\label{def:infinitely-hit}
  Let $\PseudoComb=(\Path,\{\beta_i,m_i,p_i\}_{i\in\NaturalSet})$ be a pseudo-comb.
  For a period $p=p_{i_0}$ ($0<i_0$) and a direction $d$ in \DirectionSet that is non colinear to $p$, and for each $k\in \NaturalSet$ let $\Path_{d,p}(k)$ be the bi-infinite periodic path $^{\omega}p.(\Path_{i_0}+\Vect{m_{i_0}}+kd).p^{\omega}$.
  The pseudo-comb $\PseudoComb$ is \emph{$(p,d)$-infinitely hit} if there is some $k_0$ such that, for all $k>k_0$ there is $j_k$ and some $q\in\Rotate{p}$ such that:
  \begin{enumerate}
  \item $\Path_{j_k}.q^{\omega}\subset \Path_{d,p}(k)$\,,
  \item $\Path\cap \Path_{j_k}.q^{\omega}=\{\Path_{j_k}\}$, and
  \item $\Path_{j_k}.q^{\omega}$
    and $\Path_{\beta_{i_0}}.m_{i_0}p_{i_0}^{\omega}$ belong to the same side of $^{\omega}\RevBack{p_0}\RevBack{m_0}.\Path_{\IntegerInterval{\beta_0}{+\infty}}$ and there are infinitely many $\Path_{\beta_i}.m_ip_i^{\omega}$ on this side.
  \end{enumerate}
  The path vertex $\Path_{j_k}$ is called a {\em $(p,d)$-point of interest at shift $k$}.
\end{definition}

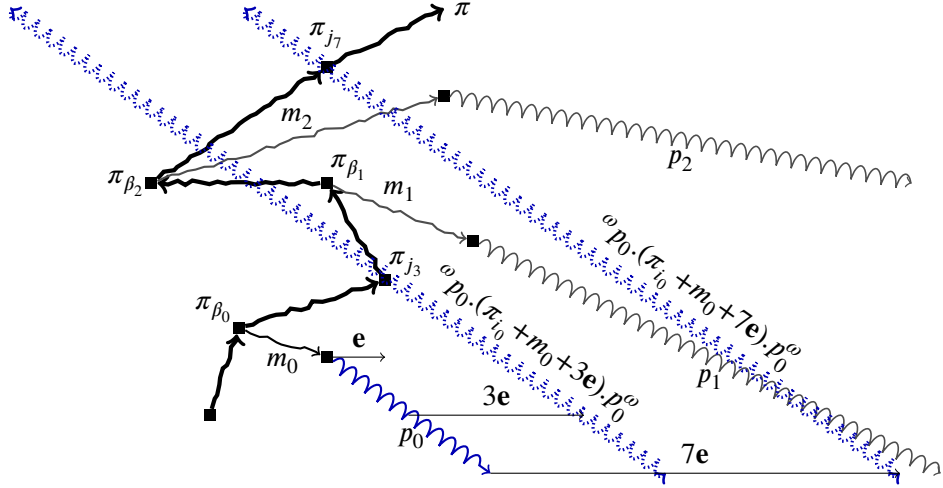
\begin{figure}[hbt]
  \small\centering\SetUnitlength{2em}\newcommand{\Slope}{1.4}
  \begin{tikzpicture}
    \TileDot[O]{0,2}{}
    \TileDot[p0]{0.5,3.5}{}
    \TileDot[m0]{2,3}{}
    \TileDot[m2]{4.5,5}{}
    \TileDot[m3]{4,7.5}{}
    \TileDot[p1]{3,4.33}{}
    \TileDot[p2]{2,6}{}
    \TileDot[p3]{-1,6}{}
    \TileDot[p4]{2,8}{}
    \path (4,9) \CoorNode{p5} node [right] {\Path};
    \path (m0) +(2*\Slope,-2) \CoorNode{a0} ;
    \path (a0) ++(3,0) \CoorNode{b1} ++(-8*\Slope,8) \CoorNode{c1} ;
    \path (a0) ++(7,0) \CoorNode{b2} ++(-8*\Slope,8) \CoorNode{c2} ;
    \path (m2) +(8,-4) \CoorNode{a2} ;
    \path (m3) +(8,-1.5) \CoorNode{a3} ;
    \begin{scope}
      \TransientPathStyleAdd{ultra thick}
      \TransientPath{O}{}{p0}
      \TransientPath{p0}{}{p1}
      \TransientPath{p1}{}{p2}
      \TransientPath{p2}{}{p3}
      \TransientPath{p3}{}{p4}
      \TransientPath{p4}{}{p5}
    \end{scope}
    \TransientPath[below]{p0}{m_0}{m0}
    \begin{scope}
      \PeriodicPathStyleAdd{draw=DarkBlue,thick}
      \PeriodicPath[below]{m0}{p_0}{a0}
    \end{scope}
    \begin{scope}
      \TransientPathStyleAdd{draw=VeryDarkGrey}
      \TransientPath[above]{p2}{m_1}{m2}
      \TransientPath[above]{p3}{m_2}{m3}
      \PeriodicPathStyleAdd{draw=VeryDarkGrey}
      \PeriodicPath[below]{m2}{p_1}{a2}
      \PeriodicPath[below]{m3}{p_2}{a3}
    \end{scope}
    {\PeriodicPathStyleAdd{densely dotted,very thick,draw=DarkBlue,<->}
      \PeriodicPath[above right,sloped]{b1}{\quad\qquad^{\omega}p_0.(\Path_{i_0}{+}m_0{+}3\East).p_0^{\omega}}{c1}
      \PeriodicPath[above right,sloped]{b2}{^{\omega}p_0.(\Path_{i_0}{+}m_0{+}7\East).p_0^{\omega}}{c2}
    }
    \path (p0) node [above left=0.1em,inner sep=.1em] {$\Path_{\beta_0}$} ;
    \path (p1) node [above right=0em,inner sep=.1em] {$\Path_{j_3}$} ;
    \path (p2) node [above right=0em,inner sep=.1em] {$\Path_{\beta_1}$} ;
    \path (p3) node [left=.1em,inner sep=.1em] {$\Path_{\beta_2}$} ;
    \path (p4) node [above=.3em] {$\Path_{j_7}$} ;
    \draw[->] (m0) -- node[above]{\East} +(1,0) ;
    \draw[->] ($.5*(m0)+.5*(a0)$) -- node[above]{3\East} +(3,0) ;
    \draw[->] (a0) -- node[above]{7\East} +(7,0) ;
  \end{tikzpicture}
  \caption{A ($p_0$,\East)-infinitely hit right-sided pseudo-comb. The periodic paths parallel to $p_0$ that hit the pseudo-comb are dotted, the other backbones with periods $p_1$ and $p_2$ are in grey.}
  \label{quipu:pseudo-comb:inf-hit}
\end{figure}

\RefFigure{quipu:pseudo-comb:inf-hit} depicts a situation of a path \Path being ($p_0$,\East)-infinitely hit.
The other backbones are drawn in dark grey.
There is no intersection between the backbones, but they might intersect any of the bi-infinite periodic shift (dotted) of $p_0$.

With the following lemmas we observe that in an infinite filtration, there must be a pseudo-comb that is infinitely hit. 

\begin{lemma}\label{lem:infinite-hit:non-colinear:two-sided}
  If a two-sided pseudo-comb $\PseudoComb$ has a backbone $m_ip_i^\omega$ on the left side and $m_jp_j^\omega$ on the right side such that $p_i$ and $p_j$ are not colinear, then $\PseudoComb$ is infinitely hit. 
\end{lemma}

\begin{proof}
  Let $A=\Path_{\beta_i}+\Vect{m_i}$, $B=\Path_{\beta_j}+\Vect{m_j}$ and let $d$ and $d'$ be directions that are not colinear to $p_i$, $p_j$, respectively. Consider the bi-infinite shifts $\delta_k={}^{\omega}p_i.(A+kd).p_i^\omega$ and $\gamma_k={}^{\omega}p_j.(B+kd').p_j^\omega$ for all $k\in \IntegerSet$.
  Observe that the interior of the left region of $^\omega\RevBack{p_i}\RevBack{m_i}.\Path_{\IntegerInterval{\beta_i}{\beta_{i'}}}.m_{i'}p_{i'}^\omega$ where $\Path_{\beta{i'}}.m_{i'}p_{i'}^\omega$ is a left backbone and $i<i'$, contains only finite number of vertices from $\Path$.
  Hence for all $k\in \IntegerInterval{0}{\beta_{i'}}$, the shifted bi-infinite path $\delta_k$ either intersects $\Path$ at a `left most' point satisfying the conditions 1. and 2. of \RefDef{def:infinitely-hit} or there is no intersection with $\Path$.
  The situation is symmetric on the right with $j$ and $j'$, for $j<j'$. 
  If the `left most' (or resp. `right most') intersections always exist with $\delta_k$ (or resp. $\gamma_k$) for all $k>0$, then because $\PseudoComb$ is two sided, $\PseudoComb$ is $(p_i,d)$-infinitely hit, (or resp. $(p_j,d')$-infinitely hit). 
  Otherwise there are $k_1$ and $k_2$ such that $\Path$ has no intersection with $\delta_{k_1}$ and $\gamma_{k_2}$. Since $p_i$ and $p_j$ are not colinear, $\delta_{k_1}$ and $\gamma_{k_2}$ intersect; call that point $C$, and \Path must be completely in one of the regions bounded by $\delta_{k_1}$ and $\gamma_{k_2}$. 
  Now we can continue and consider the shifts for $k\le 0$ (which is equivalent to taking positive shifts in directions $\Backward{d}$ and $\Backward{d'}$). 
  By the symmetry of the argument, either $\PseudoComb$ is infinitely hit or there are $k_1',k_2'<0$ such that $\delta_{k_1'}$ and $\gamma_{k_2'}$ intersect at some point $D$ and bound a region that contain \Path. But the latter is impossible because $^\omega\RevBack{p_i}.C.p_j^\omega$ and $^\omega\RevBack{p_i}.D.p_j^\omega$ bound a finite parallelogram region which cannot contain the infinite path \Path. 
\end{proof}

\begin{corollary}\label{cor:infinite-hit:non-colinear:complementary}
  If a two complementary pseudo-combs have one a backbone 
  $m_ip_i^\omega$ on the left side, and the other a backbone $m_jp_j^\omega$ on the right side such that $p_i$ and $p_j$ are not colinear, then at least one of the pseudo-combs is infinitely hit. 
\end{corollary}

\begin{proof}
  Suppose the backbones of the two complementary pseudo-combs are attached to the infinite paths $\Path$ and $\Path'$ with $\Path$ carying backbones on the left side and $\Path'$ on the right side. 
  Because these pseudo-combs are associated with the same filtration, from the tile where \Path and $\Path'$ become disjoint, they do not intersect further, nor do their backbones. The proof follows similar lines to the one of
  \RefLem{lem:infinite-hit:non-colinear:two-sided}.
\end{proof}

\begin{lemma}\label{lem:infinite-hit:two-sided}
  Every two sided pseudo-comb is infinitely hit.
\end{lemma}

\begin{figure}[hbt]
  \small\centering\SetUnitlength{1.75em}\newcommand{\Slope}{.4}
  \subcaptionbox{beginning\label{fig:quipu:co-linear:opposed:base}}{\SetUnitlength{1.75em}\begin{tikzpicture}
      \TileDot[O]{.5,0}{}
      \TileDot[p0]{2,-.5}{}
      \TileDot[p1]{3,.75}{}
      \TileDot[m0]{2.75,-1.5}{}
      \TileDot[m1]{2.5,2.25}{}
      \path (6,.75) \CoorNode{p2} node [right] {\Path};
      \path (m0) +(3*\Slope,-3) \CoorNode{a0} ;
      \path (m1) +(-3*\Slope,3) \CoorNode{a1} ;
      \begin{scope}
        \TransientPathStyleAdd{ultra thick}
        \TransientPath{O}{}{p0}
        \TransientPath{p0}{}{p1}
        \TransientPath{p1}{}{p2}
      \end{scope}
      \TransientPath[right]{p0}{m_0}{m0}
      \TransientPath[right]{p1}{m_1}{m1}
      \PeriodicPath[right]{m0}{p_0}{a0}
      \PeriodicPath[right]{m1}{p_1}{a1}
    \end{tikzpicture}}
  \subcaptionbox{impossible switch\label{fig:quipu:co-linear:opposed:impossible}}{\begin{tikzpicture}
      \TileDot[O]{.5,0}{}
      \TileDot[p0]{2,-.5}{}
      \TileDot[p1]{3,.75}{}
      \TileDot[p2]{4.5,.75}{}
      \TileDot[m0]{2.75,-1.5}{}
      \TileDot[m1]{2.75,3}{}
      \TileDot[m2]{6,2.5}{}
      \path (3.5,5) \CoorNode{pxx} node [right] {\Path};
      \path (m0) +(3*\Slope,-3) \CoorNode{a0} ;
      \path (m1) +(-3*\Slope,3) \CoorNode{a1} ;
      \path (m2) +(-3*\Slope,3) \CoorNode{a2} ;
      \begin{scope}
        \TransientPathStyleAdd{ultra thick}
        \TransientPath{O}{}{p0}
        \TransientPath{p0}{}{p1}
        \TransientPath{p1}{}{p2}
        \TransientPath{p2}{}{pxx}
      \end{scope}
      \TransientPath[right]{p0}{m_0}{m0}
      \TransientPath[right]{p1}{m_1}{m1}
      \TransientPath[right]{p2}{m_2}{m2}
      \PeriodicPath[right]{m0}{p_0}{a0}
      \PeriodicPath[right]{m1}{p_1}{a1}
      \PeriodicPath[right]{m2}{p_2}{a2}
    \end{tikzpicture}}
  \subcaptionbox{necessary scheme\label{fig:quipu:co-linear:opposed:scheme}}{\begin{tikzpicture}
      \TileDot[O]{.5,0}{}
      \TileDot[p0]{2,-.5}{}
      \TileDot[p1]{3,.75}{}
      \TileDot[p2]{4.5,-.25}{}
      \TileDot[p3]{5.4,.5}{}
      \TileDot[m0]{2.75,-1.5}{}
      \TileDot[m1]{2.75,3}{}
      \TileDot[m2]{4.75,-.75}{}
      \TileDot[m3]{5.25,2.25}{}
      \path (7,1) \CoorNode{pxx} node [right] {\Path};
      \path (m0) +(3*\Slope,-3) \CoorNode{a0} ;
      \path (m1) +(-3*\Slope,3) \CoorNode{a1} ;
      \path (m2) +(3*\Slope,-3) \CoorNode{a2} ;
      \path (m3) +(-3*\Slope,3) \CoorNode{a3} ;
      \begin{scope}
        \TransientPathStyleAdd{ultra thick}
        \TransientPath{O}{}{p0}
        \TransientPath{p0}{}{p1}
        \TransientPath{p1}{}{p2}
        \TransientPath{p2}{}{p3}
        \TransientPath{p3}{}{pxx}
      \end{scope}
      \TransientPath[right]{p0}{m_0}{m0}
      \TransientPath[right]{p1}{m_1}{m1}
      \TransientPath[right]{p2}{m_2}{m2}
      \TransientPath[right]{p3}{m_3}{m3}
      \PeriodicPath[right]{m0}{p_0}{a0}
      \PeriodicPath[right]{m1}{p_1}{a1}
      \PeriodicPath[right]{m2}{p_2}{a2}
      \PeriodicPath[right]{m3}{p_3}{a3}
      \path ($.5*(m3)+.5*(a3)+(1.75,0)$) node {\dots} ;
      \path ($.5*(m2)+.5*(a2)+(1.75,0)$) node {\dots} ;
    \end{tikzpicture}
  }
  \caption{Co-linear but opposite directions on left and right.}
  \label{fig:quipu:co-linear:opposed}
\end{figure}
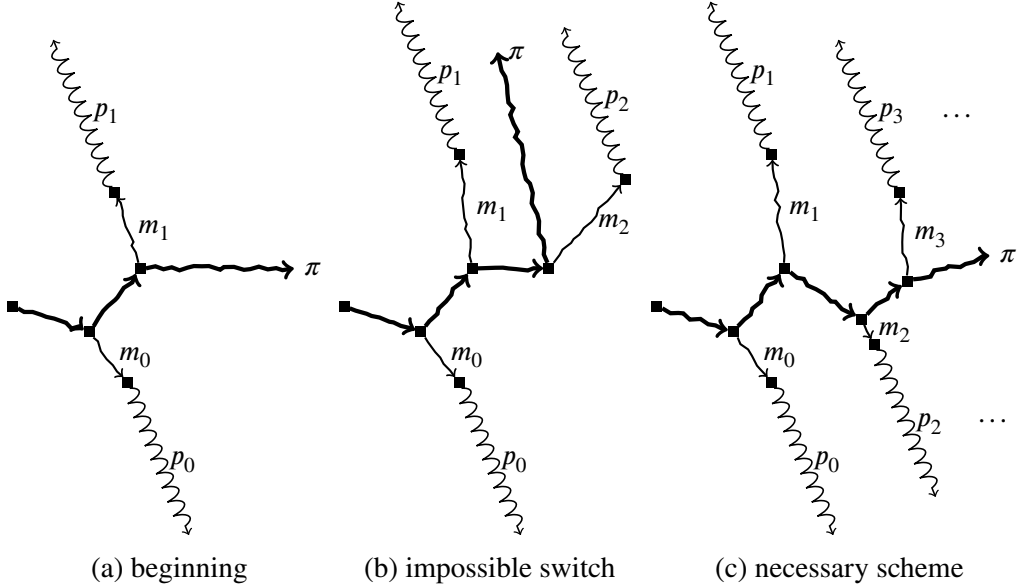

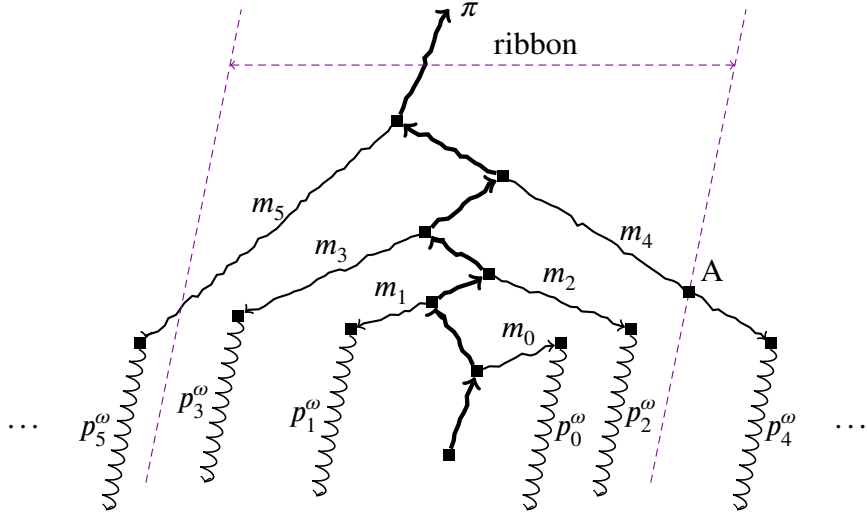
\begin{figure}
  \centering\SetUnitlength{1.75em}\newcommand{\Slope}{.2}
  \begin{tikzpicture}
    \TileDot[O]{0,-1.5}{}
    \TileDot[p0]{.5,0}{}
    \TileDot[p1]{-1.75/6,1.25}{}
    \TileDot[p2]{4.25/6,1.75}{}
    \TileDot[p3]{-2.5/6,2.5}{}
    \TileDot[p4]{5.75/6,3.5}{}
    \TileDot[p5]{-5.5/6,4.5}{}
    \path (0,6.5) \CoorNode{pxx} node [right] {\Path};
    \TileDot[m0]{2,.5}{}
    \TileDot[m1]{-1.75,.75}{}
    \TileDot[m2]{3.25,.75}{}
    \TileDot[m3]{-3.75,1}{}
    \TileDot[m4]{5.75,.5}{}
    \TileDot[m5]{-5.5,.5}{}
    \path (m0) +(-3*\Slope,-3) \CoorNode{a0} ;
    \path (m1) +(-3*\Slope,-3) \CoorNode{a1} ;
    \path (m2) +(-3*\Slope,-3) \CoorNode{a2} ;
    \path (m3) +(-3*\Slope,-3) \CoorNode{a3} ;
    \path (m4) +(-3*\Slope,-3) \CoorNode{a4} ;
    \path (m5) +(-3*\Slope,-3) \CoorNode{a5} ;
    \begin{scope}
      \TransientPathStyleAdd{ultra thick}
      \TransientPath{O}{}{p0}
      \TransientPath{p0}{}{p1}
      \TransientPath{p1}{}{p2}
      \TransientPath{p2}{}{p3}
      \TransientPath{p3}{}{p4}
      \TransientPath{p4}{}{p5}
      \TransientPath{p5}{}{pxx}
    \end{scope}
    \TransientPath{p0}{m_0}{m0}
    \TransientPath{p1}{m_1}{m1}
    \TransientPath{p2}{m_2}{m2}
    \TransientPath{p3}{m_3}{m3}
    \TransientPath{p4}{m_4}{m4}
    \TransientPath{p5}{m_5}{m5}
    \PeriodicPath[right]{m0}{p_0^{\omega}}{a0}
    \PeriodicPath[left]{m1}{p_1^{\omega}}{a1}
    \PeriodicPath[right]{m2}{p_2^{\omega}}{a2}
    \PeriodicPath[left]{m3}{p_3^{\omega}}{a3}
    \PeriodicPath[right]{m4}{p_4^{\omega}}{a4}
    \PeriodicPath[left]{m5}{p_5^{\omega}}{a5}
    \path ($.5*(m4)+.5*(a4)+(1.75,0)$) node {\dots} ;
    \path ($.5*(m5)+.5*(a5)-(1.75,0)$) node {\dots} ;
    \begin{scope}[densely dashed,draw=DeepPurple]
      \draw (-5,0) +(-2*\Slope,-2) -- +(6.5*\Slope,6.5) \CoorNode{tl} ;
      \draw[name path=rib] (4,0) +(-2*\Slope,-2) -- +(6.5*\Slope,6.5) \CoorNode{tr} ;
      \draw[<->] ([shift={(-1*\Slope,-1)}]tl) -- node [above right]{ribbon} ([shift={(-1*\Slope,-1)}]tr) ;
      \path[name path=M4] (p4) -- (m4);
      \SetIntersect{M4}{rib}{A}
      \TileDot[Aa]{A}{}
      \path (Aa) node[above right] {A} ; 
    \end{scope}
    \end{tikzpicture}
  \caption{Co-linear and same directions.}
  \label{fig:quipu:co-linear:same-dir}
\end{figure}

\begin{proof}
  Consider a two-sided pseudo-comb $\PseudoComb$.
  By \RefLem{lem:infinite-hit:non-colinear:two-sided}, If there are non-colinear backbones, then $\PseudoComb$ is infinitely hit. So we suppose that all backbones have colinear periods.
  Without loss of generality we enumerate the backbones such that the periodic path $p_{k}^\omega$ is on the right side of \Path for even values of $k$ and it is on the left side of \Path for odd values of $k$.

  The situation when $\Vect{p_0}$ and $\Vect{p_1}$ are colinear in opposite directions is shown in \RefFig{fig:quipu:co-linear:opposed:base}.
  If there is a backbone $p_{2k}^\omega$ on the right side whose period displacement vector $\Vect{p_{2k}}$ is in opposite direction of $\Vect{p_0}$ (like $p_2^\omega$ in \RefFig{fig:quipu:co-linear:opposed:impossible}), then \Path is bounded within a finite-width region delimited by $p_{2k-1}^\omega$ and $p_{2k}^\omega$ (as depicted in \RefFig{fig:quipu:co-linear:opposed:impossible} with $p_1^\omega$ and $p_2^\omega$). 
  This situation is impossible because every vertex in $\IntegerSet^2$ is covered by at most a single quipu vertex, but the finite-width region containing \Path must also be covered by arbitrarily high number of backbone cycles of the pseudo-comb. 
  Therefore every backbone on the right extends in the same direction as $p_0^\omega$ and every backbone on the left extends in the same direction as $p_1^\omega$. The situation is as in \RefFig{fig:quipu:co-linear:opposed:scheme}.
  In this case, \Path is infinitely hit by both $(p_0,d)$ and $(p_1,d)$
  for a direction $d$ not colinear to $\Vect{p_0}$ and $\Vect{p_1}$.

  Now consider the situation when all period vectors $\Vect{p_k}$ are colinear and in the same direction as depicted in \RefFig{fig:quipu:co-linear:same-dir}.
  If there is a direction $d$ not colinear to $p_0$ such that \Path crosses $^{\omega}p_0.(\Path_{i_0}+m_{i_0}+kd).p_0^{\omega}$ for any $k$ sufficiently large (say $d=\East$ in \RefFig{fig:quipu:co-linear:same-dir}), then the pseudo-comb is $(p_0,d)$-infinitely hit. 
  
  Finally we consider the case when \Path extends infinitely between  $^{\omega}p_0.(\Path_{i_0}+m_{i_0}+k_1d).p_0^{\omega}$ and $^{\omega}p_0.(\Path_{i_0}+m_{i_0}+k_2d).p_0^{\omega}$ for a direction $d$ not colinear to $p_0$; that is in some finite-width ribbon as depicted in \RefFig{fig:quipu:co-linear:same-dir}.
  If the extension of \Path is in the same direction as $\Vect{p_0}$, the situation is similar to the one depicted in \RefFig{fig:quipu:co-linear:opposed:impossible}, and we have already observed that it cannot occur in an infinite filtration.
  Hence \Path extends in direction colinear but opposite to $\Vect{p_0}$.
  
  Starting from some $j$ the periodic parts of all backbones $m_{j'}p_{j'}^\omega$ for $j'\ge j$ are `outside' the ribbon.
  Let $A$ be in the first intersection of $\Path_{\beta_j}.m_j$ with  $^{\omega}p_0.(\Path_{i_0}+m_{i_0}+k_2d).p_0^{\omega}$.
  Let $m'$ be the prefix of  $\Path_{\beta_j}.m_j$ joining $A$.
  Now \Path, and $\Path_{\beta_j}.m'\RevBack{p_0}^\omega$ satisfy the condition of \RefCor{cor:new-periodic-path} and there is an ultimately periodic path in \AlphaMax inside the right region ${}^\omega\RevBack{p_0}\RevBack{m'}.\Path_{\beta_j}.\Path_{\IntegerInterval{\beta_j}{\infty}}$.
  Let $\ZZOrigin. nr^\omega$ be this path, and because this path is inside the ribbon, it must be that $\Vect{r}$ is colinear oppositely oriented to $\Vect{p_0}$. 
  Moreover, since every quipu extension $\QuipuI$ in the filtration $\Quipu_{\TAS}$ contains only finite number of backbones of the pseudo-comb $\PseudoComb$, there is $l$ such that $\ZZOrigin.nr^\omega$ appears as the minimal element of $P_l$, and we have $\Quipu_l=\Quipu[nr^{\omega}]$. 
  But that implies that a backbone cycle with period colinear to $\Vect{p_0}$ is added to \Path in direction opposite to $\Vect{p_0}$, obtaining a situation as in \RefFig{fig:quipu:co-linear:opposed:impossible} which we already observed cannot appear. 
\end{proof}

\begin{corollary}\label{cor:infinite-hit}
  Every pseudo-comb is infinitely hit or has a complementary pseudo-comb infinitely hit.
\end{corollary}

\begin{proof}
  Similarly to the proof of \RefLem{lem:infinite-hit:two-sided}
  can be shown that at least one pseudo-comb of a complementary pair of pseudo-combs is infinitely hit.
\end{proof}

\subsection{The filtration $\Quipu_{\TAS}$ is always finite}

In the following lemma, we consider a pseudo-comb $\PseudoComb=(\Path,\{\beta_i,m_i,p_i\}_{i\in\NaturalSet})$ that is $(p_i,d)$-infinitely hit, and without loss of generality we take that it is hit on the right side. 
We denote $\Path_{\beta_j}$ with $C_j$. 
We concentrate on points of interest of \Path that are in shifts between backbones, and we set $(p_i,d)$-points of interest at shifts $i_j$ to be
$\Path_{i_j}$, where $C_j$ is between $\Path_{i_j}$ and $\Path_{i_{j+1}}$, possibly $C_j$ equal to $\Path_{i_{j+1}}$.
We denote with $\rho_{i_j}=\Path_{i_j}.q^\omega$ the path that satisfies conditions 1. and 2. in \RefDef{def:infinitely-hit}. We will use the technique of superimposing paths through co-growth as developed in \RefSec{sec:nice}. 

\begin{lemma}
  \label{lem:extra-comb}Unless \AlphaMax is a grid, every $p_j^\omega$ ($j>i$) that is a backbone period corresponding to $C_j.m_jp_j^\omega$ in the pseudo-comb $\PseudoComb$ is also a tooth of a comb in \AlphaMax.
\end{lemma}

\begin{figure}[hbt]
  \centering\small\SetUnitlength{1.6em}\newcommand{\VStep}{3}\newcommand{\VInc}{.4}\newcommand{\HLong}{3.5}\newcommand{\HShort}{1}\subcaptionbox{context of $\Path$
    \label{quipu:infinite-path}}{\begin{tikzpicture}[x=.7\unitlength,inner sep=.25ex]
      \TileDot[O]{\HShort,\VStep/3+.5}{\Seed} ;
      \path (O) node [above left] {\ZZOrigin\ } ;
      \TileDot[g1]{\HShort,\VStep}{$G_1$} ;
      \TileDot[g2]{\HShort,\VStep*2}{$G_2$} ;
      \TileDot[g3]{\HShort,\VStep*3}{$G_3$} ;
      \begin{scope}[shift={(0,\VInc)}]
        \TileDot[a1]{\HLong,\VStep}{$A_1$} ;
        \TileDot[a2]{\HLong-.25,\VStep*2+.25}{$A_2$} ;
        \TileDot[a3]{\HLong,\VStep*3+.5}{$A_3$} ;
        \begin{scope}[shift={(0,\VInc*3)}]
          \path (a1) +(\HLong,\VStep/6) \CoorNode{aa1} ;
          \path ($(a2)+(aa1)-(a1)+(0,.5)$) \CoorNode{aa2} ;
          \path ($(a3)+(aa1)-(a1)+(0,1)$) \CoorNode{aa3} ;
          \path (g3) +(0,\VStep) node[below left] (g4) {\Path} ;
        \end{scope}
      \end{scope}
      \begin{scope}
        \TransientPathStyleAdd{ultra thick}
        \TransientPath{O}{}{g1}
        \TransientPath{g1}{}{g2}
        \TransientPath{g2}{}{g3}
        \TransientPath{g3}{}{g4}
      \end{scope}
      \TransientPath{g1}{}{a1}
      \TransientPath{g2}{}{a2}
      \TransientPath{g3}{}{a3}
      \PeriodicPath{a1}{}{aa1}
      \PeriodicPath{a2}{}{aa2}
      \PeriodicPath{a3}{}{aa3}
    \end{tikzpicture}}\quad\SetUnitlength{1.7em}\subcaptionbox{first co-growth\label{fig:quipu:PoI}}{\begin{tikzpicture}[x=.8\unitlength,inner sep=.25ex]
      \TileDot[O]{\HShort,\VStep*.6}{\Seed} ;
      \path (O) node [above left] {\ZZOrigin\ } ;
      \Tile[g1]{\HShort,\VStep*1.3}{$C_i$} ;
      \Tile[i1]{\HShort,\VStep*2}{$\Path_{i_{j}}$} ;
      \Tile[i3]{\HShort,\VStep*3.5}{$\Path_{i_{j+1}}$} ;
      \begin{scope}[shift={(0,\VInc)}]
        \TileDot[a1]{\HLong,\VStep*1.25}{$A_1$} ;
        \TileDot[a2]{\HLong*1.25,\VStep*3}{$A_2$} ;
        \begin{scope}[shift={(0,\VInc*3)}]
          \path (a1) +(\HLong+1.75,\VStep/6) \CoorNode{aa1} ;
          \path (a2) +($(aa1)-(a1)+(-1,.3)$) \CoorNode{aa2} ;
          \path (0,\VStep*4+2) \CoorNode{e} node[above left] {\Path} ;
        \end{scope}
      \end{scope}
      \begin{scope}[shift={(0,.45*\VStep)}]
        \Tile[i2]{0,\VStep*2.25}{$C_j$} ;
        \TileDot[i2m]{2*\HShort,\VStep*2.4}{} ;
        \TileDot[i3m]{$(i3)+(i2m)-(i1)$}{} ;
      \end{scope}
      \begin{scope}
        \TransientPathStyleAdd{ultra thick}
        \TransientPath{O}{}{g1}
        \TransientPath{g1}{}{i1}
        \TransientPath{i1}{}{i2}
        \TransientPath{i2}{r\ \ }{i3}
        \TransientPath{i3}{}{e} 
      \end{scope}
      \TransientPath{g1}{m_i}{a1}
      \draw[transient,-] (i2) -- (i2m) ;
      \TransientPath[pos=.4,above]{i2m}{l}{a2}
      \TileDot[a3]{$(i3m.center)+(a2.center)-(i2m.center)$}{} ; 
      \PeriodicPathRight{a1}{p_i^{\omega}}{aa1}
      \path ($(a3)+(aa2)-(a2)$) \CoorNode{aa3} ;
      \PeriodicPathRight{a2}{p_j^{\omega}}{aa2}
      \begin{scope}[densely dashed]
        \TransientPath{i3m}{l}{a3}
        \PeriodicPathRight{a3}{p_j^{\omega}}{aa3}
        \TransientPath[right]{i1}{f_0}{i2m}
        \TransientPath[right]{i3}{f_0}{i3m}
        \path (i1) ++ ($1.4*(aa1)-1.4*(a1)$) \CoorNode{ii1} ;
        \path (i3) ++($(ii1)-(i1)$) \CoorNode{ii3} ;
        \begin{scope}[draw=DarkGray]
          \draw[periodic] (i1) -- 
          node[sloped,below] {\DarkGray no intersection with \Path}
          node[sloped,above=.7em] {$\rho_{i_j}$} (ii1) ;
          \draw[periodic] (i3) -- node[sloped,below] {\DarkGray no intersection with \Path}
          node[sloped,above=.7em] {$\rho_{i_{j{+}1}}$} (ii3) ;
        \end{scope}
      \end{scope}
      \path (i2m) node[above left] {$A_1$} ;
      \path (i3m) node[above left] {$A_2$} ;
      \path[b1f1,->] ([shift={(0,.4)}]ii1) -- (i1.north east) ;
      \path[b1f1] (i1.north) -- (i2.east) -- (i3.south) ;
      \path[b1f1] (i2.north east) -- (i3.south) ;
      \path[b1f1,->] (i3.north) -- ([shift={(.25,0)}]e) ;
      \path[b2f2,->] ([shift={(0,.3)}]ii3) -- (i3.north east) ;
      \path[b2f2,->] ([shift={(.25,0)}]i3.north) -- ([shift={(.5,0)}]e) ; 
      \path (i1.north east) node[above] {\CoGrowStart} ;
      \path (i3.north east) node[above] {\CoGrowStart} ;
    \end{tikzpicture}}\quad\SetUnitlength{1.9em}\subcaptionbox{second co-growth\label{fig:quipu:cg2}}{\begin{tikzpicture}[x=.8\unitlength,inner sep=.25ex]
      \TileDot[O]{\HShort-1,\VStep*1.6}{\Seed} ;
      \path (O) node [above left] {\ZZOrigin\ } ;
      \Tile[i1]{\HShort,\VStep*2}{$\Path_{i_{j}}$} ;
      \Tile[i3]{\HShort,\VStep*3.5}{$\Path_{i_{j+1}}$} ;
      \begin{scope}[shift={(0,\VInc)}]
        \TileDot[a2]{\HLong*1.25,\VStep*3}{$A_2$} ;
        \begin{scope}[shift={(0,\VInc*3)}]
          \path (a2) +(\HLong+1,\VStep/6+.3) \CoorNode{aa2} ;
          \path (0,\VStep*5.6) \CoorNode{e} node[above left] {\Path} ;
        \end{scope}
      \end{scope}
      \begin{scope}[shift={(0,.45*\VStep)}]
        \Tile[i2]{0,\VStep*2.25}{$C_j$} ;
        \TileDot[i2m]{2*\HShort,\VStep*2.4}{} ;
        \TileDot[i3m]{$(i3)+(i2m)-(i1)$}{} ;
      \end{scope}
      \TileDot[a3]{$(i3m.center)+(a2.center)-(i2m.center)$}{} ; 
      \TileDot[i2ma]{$.5*(i2m.center)+.5*(a2.center)$}{}
      \TileDot[i3ma]{$.5*(i3m.center)+.5*(a3.center)$}{}
      \begin{scope}
        \TransientPathStyleAdd{ultra thick}
        \TransientPath{O}{}{i1}
        \TransientPath{i1}{}{i2}
        \TransientPath{i2}{r\ \ }{i3}
        \TransientPath{i3}{}{e} 
      \end{scope}
      \TransientPath[below]{i2}{l'}{i2m}
      \draw [transient,-] (i2m) -- (i2ma) ;
      \TransientPath{i2ma}{}{a2}
      \path (i2ma) node [above=3.5] {$l$} ;
      \TransientPath{i3m}{n'}{i3ma}
      \TransientPath{i3ma}{n''}{a3}
      \PeriodicPathRight{a2}{p_j^{\omega}}{aa2}
      \path ($(a3)+(aa2)-(a2)$) \CoorNode{aa3} ;
      \PeriodicPathRight{a3}{p_j^{\omega}}{aa3}
      \TransientPath[right]{i1}{f_0}{i2m}
      \TransientPath[right]{i3}{\!f_0}{i3m}
      \path (i2m) node [below right] {$A_1$} ;
      \path (i3m) node[left=2.0] {$A_2$} ;
      \TileDot[a4]{$(a3)+(a3)-(a2)$}{}
      \path ($(aa3)+(a3)-(a2)$) \CoorNode{aa4} ;
      \TileDot[i4m]{$(i3m)+(a3)-(a2)$}{}
      \TileDot[i4ma]{$(i3ma)+(a3)-(a2)$}{}
      \path (i4m) node[above left] {$A_3$} ;
      \path (i2ma) node[below right] {\ $B_1$} ;
      \path ([shift={(.1,-.2)}]i3ma) node[below right] {$B_2$} ;
      \path (i4ma) node[below right] {$B_3$} ;
      \begin{scope}[dashed]
        \path ($.5*(i2.north east)+.5*(i2m)+(0,.3)$) \CoorNode{i2mg0} ;
        \path ($(i3m)+(i2mg0)-(i2m)$) \CoorNode{i3mg0} ;
        \TransientPath[-]{i2m}{}{i2mg0}
        \TransientPath[below right]{i2mg0}{g_0}{i3ma}
        \TransientPath[-]{i3m}{}{i3mg0}
        \TransientPath[below right]{i3mg0}{g_0}{i4ma}
        \TransientPath{i4m}{n'}{i4ma}
        \TransientPath{i4ma}{n''}{a4}
        \PeriodicPathRight{a4}{p_j^{\omega}}{aa4}
      \end{scope}
      \path[b1f1,->] ([shift={(0,.3)}]aa2) -- ([shift={(0,.2)}]a2.north) -- (i2m.north east) ;
      \path[b1f1,->] (i2m.north west) -- (i2.north east) -- (i3.south east) -- ([shift={(.1,-.1)}]i3m.south east) -- ([shift={(0,-.2)}]a3.south east) -- ([shift={(0,-.3)}]aa3) ;
      \path[b2f2,->] ([shift={(0,.3)}]aa3) -- ([shift={(0,.2)}]a3.north) -- (i3m.north east) ;
      \path[b2f2,-] (i3m.north west) -- ([shift={(.05,0)}]i3.north) ;
      \path[b2f2,->] ([shift={(.05,0)}]i3.north) -- ([shift={(.25,0)}]e) ; 
      \path (i3m.north) node[above] {\CoGrowStart} ;
      \path (i2m.north) node[above] {\CoGrowStart} ;
    \end{tikzpicture}}
  \caption{\AlphaMax with infinite number of ultimately periodic paths.}
\end{figure}
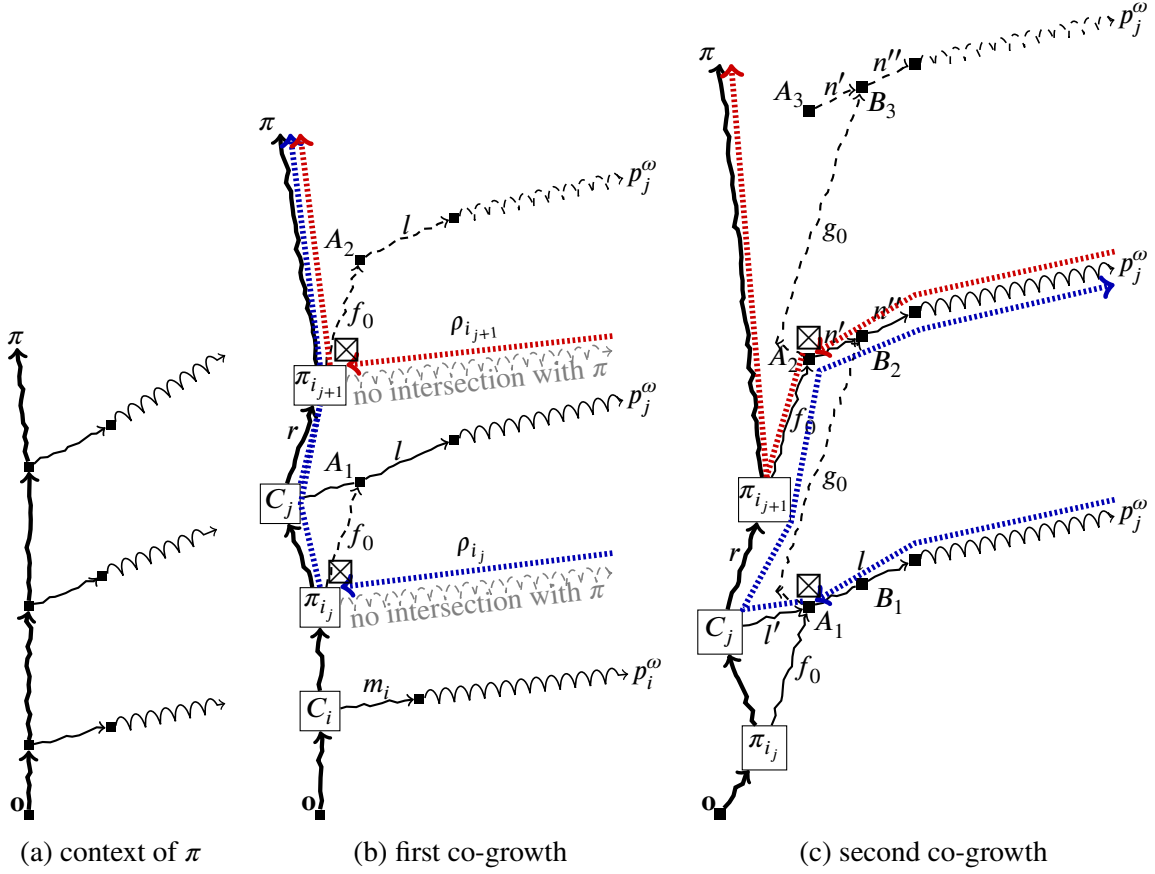

\begin{proof}
  Let $j>i$, and let $\Path_{i_{j}}$, $\Path_{i_{j+1}}$ be two points of interest such that $C_j$ is between them in \Path. 
  Because the cycles in the quipu corresponding to $p_k^\omega$ represent infinite non intersecting paths in \AlphaMax, by construction of quipu extension steps in the filtration $\Quipu_{\TAS}$, the backbones $C_j.m_jp_j^{\omega}$ do not intersect $\rho_{i_j}$ for all $j$.
  Without loss of generality (up to rotation and symmetry) we assume that the direction $d$ in which \Path is infinitely hit is \North.
  The situation is as depicted in \RefFig{fig:quipu:PoI}.

  Let $f=\CoGrow{\RevBack{\rho_{i_{j}}}}{\Path_{\IntegerInterval{i_{j}}{+\infty}}}
  {\RevBack{\rho_{i_{j+1}}}}{\Path_{\IntegerInterval{i_{j+1}}{+\infty}}}$.
  The two paths that are being superimposed with the co-growth are in red and blue in \RefFig{fig:quipu:PoI} with \CoGrowStart indicating the starting point of the co-growth.
  Observe that the infinite path $\rho_{i_{j+1}}$ belongs to both right regions 
  of $\RevBack{\rho_{i_{j}}}.\ZZOrigin.{\hat \Path_{\IntegerInterval{i_{j}}{+\infty}}}$ 
  and 
  $\RevBack{\rho_{{j+1}}}.\ZZOrigin.\hat\Path_{\IntegerInterval{i_{j+1}}{+\infty}}$, where ${\hat \Path_{\IntegerInterval{i_{j}}{+\infty}}}$ and 
  $\hat\Path_{\IntegerInterval{i_{j+1}}{+\infty}}$ are free paths corresponding to the indicated infinite segments of $\Path$. By \RefLem{inf-co-grow}, $f$ is infinite.

  The result of CoGrow, $f$, takes the `right most' 
  turns of the two paths that are being superimposed and necessarily extends `northward' crossing 
  $C_j.m_jp_j^\omega$.
  Therefore the path $\Path_{i_{j}}.f$ has to intersect $C_j.m_jp_j^{\omega}$.
  Let $f_0$ be the prefix of $f$ that leads $\Path_{i_{j}}$ to the first intersection (point $A_1$) with $C_j.m_jp_j^{\omega}$ and $l$ the path that leads from $A_1$ to the cycle $p_j^\omega$.

  The non-causal condition of \RefLem{lem:co-grow-tile} is satisfied, and due to confluence, $f$ is a path in \AlphaMax from both $\Path_{i_{j}}$ and $\Path_{i_{j+1}}$.
  Furthermore, both paths $\Path_{i_{j}}.f_0l p_j^{\omega}$ and $\Path_{i_{j+1}}.f_0l p_j^{\omega}$ exist in \AlphaMax.
  One being a shift from the other in direction non colinear to $\Vect{p_i}$, these two paths are not intersecting.
  Hence, in \AlphaMax there are two ultimately periodic paths with the same period, $p_j$, connected by a path.

  Let $A_2$ be the end vertex of the path $\Path_{i_{j+1}}.f_0$.
  Except meeting at points $A_1$ and $A_2$, the following paths are disjoint in \AlphaMax:
  \begin{enumerate}[(a)]
  \item $\Path_{\IntegerInterval{0}{i_{j-1}}}f_0$ (ends in $A_1$),
  \item $A_1.\RevBack{l'}rf_0$ (ends in $A_2$, $l'$ is the segment of $C_j.m_jp^\omega$ from $C_j$ to $A_1$, and $r$ is the segment of $\Path$ from $C_j$ to $\Path_{i_{j+1}}$),
  \item\label{path:c} $A_1.l.p_j^{\omega}$, and
  \item\label{path:d}$A_2.l.p_j^{\omega}$.
  \end{enumerate}

  Now consider
  $g=\CoGrow{^\omega\RevBack{p_j}\RevBack{l}}
  {\RevBack{l'}r f_0l p_j^{\omega}}{^\omega\RevBack{p_j}\RevBack{l}}
  {\RevBack{f_0}\Path_{\IntegerInterval{i_{j+1}}{+\infty}}}$ which co-grows paths with the same backward infinite paths, the reverse complement of paths \ref{path:c} and \ref{path:d}.
  The bi-infinite paths in co-grow are shown in red and blue respectively in
  \RefFig{fig:quipu:cg2}.
  The right regions of both paths contain $\ZZOrigin.l''p_j^{\omega}$ so by \RefLem{inf-co-grow}, $g$ is infinite.

  The path $l'.A_1$ is in \NonCausal{A_1} because $A_1$ can be reached though $f_0$.
  The path $A_1.\RevBack{f_0}$ is not part of the right region of 
  $^\omega\RevBack{p_j}\RevBack{l}
  .A_1
  .\RevBack{l'}r f_0l p_j^{\omega}$.
  By the definition of co growth, it follows that no part of $\RevBack{f_0}$ can be part of $g$.
  Although the non-causal condition of \RefLem{lem:co-grow-tile} does not completely hold (because $\RevBack{f_0}$ could be in the causal part of $A_2$ and is in the second forward infinite path in CoGrow) the two paths can still co-grow in \AlphaMax because $\RevBack{f_0}$ is in the left region of the first bi-infinite part (depicted in blue) and the co-growth of the two paths cannot take that route. The interior of the two right regions are entirely in the non-causal parts of $A_1$ and $A_2$. The start of $g$ must follow $l'$.

  Similarly to $f$, the co-growth $g$ necessarily
  goes `northward' and because $p_j^\omega$ is right most (compared to $\Path$) the co-growth $g$ ends with $p_j^{\omega}$. 
  The path $A_1.g$ has to intersect with $A_2.l p_j^{\omega}$ after possibly a finite prefix of $l$.
  Let $B_2$ be the first intersection of $g$ with 
  $A_2.l p_j^\omega$, and let $g_0$ the prefix of $g$ such that $A_1+\Vect{g_0}=B_2$.
  Let $n'$ be the prefix of $l$ such that $A_2+\Vect{n'}=B_2$ and let $n''$ be such that $l=n'n''$.
  Let $B_1=A_1+\Vect{n'}$.

  Consider the free path $q= \RevBack{n'}g_0$.
  Then $q^2$ is a free path (not a walk) because the only intersection between $B_1.q$ and $B_2.q$ is $B_2$ and $B_2=B_1+\Vect{q}$.
  By \RefLem{lem:double-pumpable}, $q^{\omega}$ is also a free path and due to confluence $B_1.q^{\omega}$ belongs to \AlphaMax.

  Therefore the infinite ultimately periodic path $\Path'=\Path_{\IntegerInterval{0}{i_{j-1}}}f_0g_0q^{\omega}$ belongs to \AlphaMax. Moreover, the path has $n''p_j^\omega$ as a tooth at each period.
  So $\Path'$ has a backbone $q^\omega$ of a comb (with teeth $p_j^{\omega}$).
\end{proof}

\begin{theorem}\label{main}
  For every confluent \TAS such that \AlphaMax is infinite, the quipu filtration $\Quipu_{\TAS}$ is finite. It ends either with $\Quipu[\text{grid}]$ or with $\Quipu[\text{halt}]$.
\end{theorem}

\begin{proof}
  Consider a set of comb representatives \SetCombRepresentative for $\AlphaMax$.
  When \SetCombRepresentative is finite, by \RefLem{lem:comb:incorporation} there is a $j$ such that $\Domain{\Quipu_j}$ contains a set of comb representatives. 
  If for all $i$, $P_i\not =\emptyset$, then for all $k>j$ the quipu $\Quipu_k$ is extended with backbone cycles of toothless combs.
  Hence there is a pseudo-comb $\PseudoComb$ in \AlphaMax associated to this filtration such that all backbones starting from some point are toothless. 

  But then, by \RefCor{cor:infinite-hit} $\PseudoComb$ is infinitely hit, and by \RefLem{lem:extra-comb}, for sufficiently large $k$, there are points of interest on the path \Path indicating that there is $\Quipu_k$ ($k>j$) that has a backbone of a comb with teeth added in $\Quipu_k$ that is not present in $\Quipu_j$. 
  This is a contradiction to the assumption that \AlphaMax has a finite set of comb representatives. 

  So, if \SetCombRepresentative is finite, there is $k>j$ such that $P_k=\emptyset$. By \RefTh{th:finite-decorations}, the filtration $\Quipu_{\TAS}$ stops in finite number of steps with $\Quipu[\text{halt}]$.

  If \SetCombRepresentative is infinite, then the set of paths $C_j.m_jp_j^\omega$ within \RefLem{lem:extra-comb} contains backbones with teeth, hence the analysis in the proof of \RefLem{lem:extra-comb} is done on a comb with a backbone $p_j^\omega$ with teeth. The analysis shows a comb whose one tooth $p_j^\omega$ is a backbone of another comb with teeth.
  Since this would introduce an intersection arbitrarily away in the periodic parts of the teeth, by \RefLem{lem:intersecting-period:unbounded}, \AlphaMax is a grid and by \RefLem{lem:grid-detection} its filtration is finite.
\end{proof}

We point out that the proof of the above theorem implies that when \AlphaMax is infinite, either \AlphaMax is a grid or it has a finite set of comb representatives.

\begin{corollary}
  \label{cor:not-compute}
  Confluent temperature 1 TAS has no universal computational power. 
\end{corollary}

\begin{proof}
  By \RefTh{main} for every confluent temperature 1 TAS \TAS, either \AlphaMax is a grid or there is a quipu \Quipu such that $\Domain{\AlphaQuipu}=\Domain{\AlphaMax}$.
  Moreover, the brute force build-up of \Quipu is always finite.
  Given \TAS, the construction of \Quipu has the following steps (suppose \TAS has at least two tiles):
  \begin{enumerate}
  \item Start with $\Quipu=\{\Root\}$ and consider the next $mp^\omega$ in the ordered list, starting with a single symbol $m$ and a single symbol $p$.
  \item\label{step:loop}
    Check whether there exist a \Quipu-vertex $x$ and $d\in D$ such that $\CoverBy{\Quipu}+d\not \subseteq\CoverBy{\Quipu}$ and there is $t$, a tile type in \TileSet, whose glue on $\Backward{d}$ side equals the glue $\VertexLabel(x)_d$. If no, halt.
  \item Check (test the tiles) whether $\ZZOrigin.mp^{|m|}$ is an assembly in \AlphaMax. The repetition of $p$ is done $|m|$ times so that the period exits possible blocking by $m$. If this is the case, then by \RefLem{lem:double-pumpable}, $\ZZOrigin.mp^\omega$ is in \AlphaMax. If no, continue with step~\ref{step:decorations}.
  \item Check whether inserting $mp^\omega$ to \AlphaQuipu detects a grid (condition \ref{eq:grid-detection} and \RefLem{lem:grid-detection}).
  \item If not, check whether $\Domain{\ZZOrigin.mp^\omega}\cap \Domain{\AlphaQuipu}$ is finite. Because both sets $\Domain{\ZZOrigin.mp^\omega}$ and $\Domain{\AlphaQuipu}$ are semi-linear, the intersection is computable. If the intersection is finite, extend \Quipu to $\Quipu[mp^{\omega}]$.
  \item\label{step:decorations}
    Add all possible decorations of length $<|m|+|p|$ (extend \Quipu to $\Quipu[r]$ with every free path $r$ such that $|r|<|m|+|p|$ and $\ZZOrigin.r\in\AlphaMax$). 
  \item Consider the next possible $mp^\omega$ in the ordered list and start from step~\ref{step:loop}.
  \end{enumerate}
  The construction intermittently adds finite decorations between cycle additions in \Quipu.
  This ensures that \Quipu ends with only \ZoneZero when \AlphaMax is finite. The decorations and cycle additions are added in parallel to ensure the end of quipu construction. By \RefTh{th:finite-decorations} and \RefTh{main} the process must stop because when \AlphaMax is infinite it is either a grid, or it has a finite set of comb representatives. 
  Therefore, if any computation can be encoded within \AlphaMax, because the \Domain{\AlphaMax} can be effectively described with semi-linear sets as \Domain{\AlphaQuipu}, presence (or absence) of a tile, or of some sub-structure within \AlphaMax, can be decided.
  Hence a computation with undecidable question cannot be encoded within \AlphaMax.
\end{proof}

\begin{example}
  Consider the TAS \TAS in Example~\ref{example:confl} depicted in \RefFig{fig:tas-example}. We can order the directions $D$ such that $
  \South<\East<\West<\North$. The quipu build-up, depicted in \RefFig{fig:quipu-example} starts with $\Quipu_0=\{\Root\}$. The symbol $\North$ is not in any ultimately periodic path (the tiles are reached by $\South$ first and $C$, and $D$, do not have glues east and west. 
  First we consider the path $\South\South^{\omega}$ and $\Quipu_1=\Quipu[\South\South^{\omega}]$, \RefFig{fig:quipu-example}(b).
  The transient part of $\ZZOrigin.\South\South^\omega$ in \AlphaMax consists of edge labeled $\South$ with endpoints labeled $A$ and $C$.
  The cycle $C_{\South}$ is the loop $(C,C)$.
  
  \begin{figure}[hbt]
    \includegraphics[width=\textwidth]{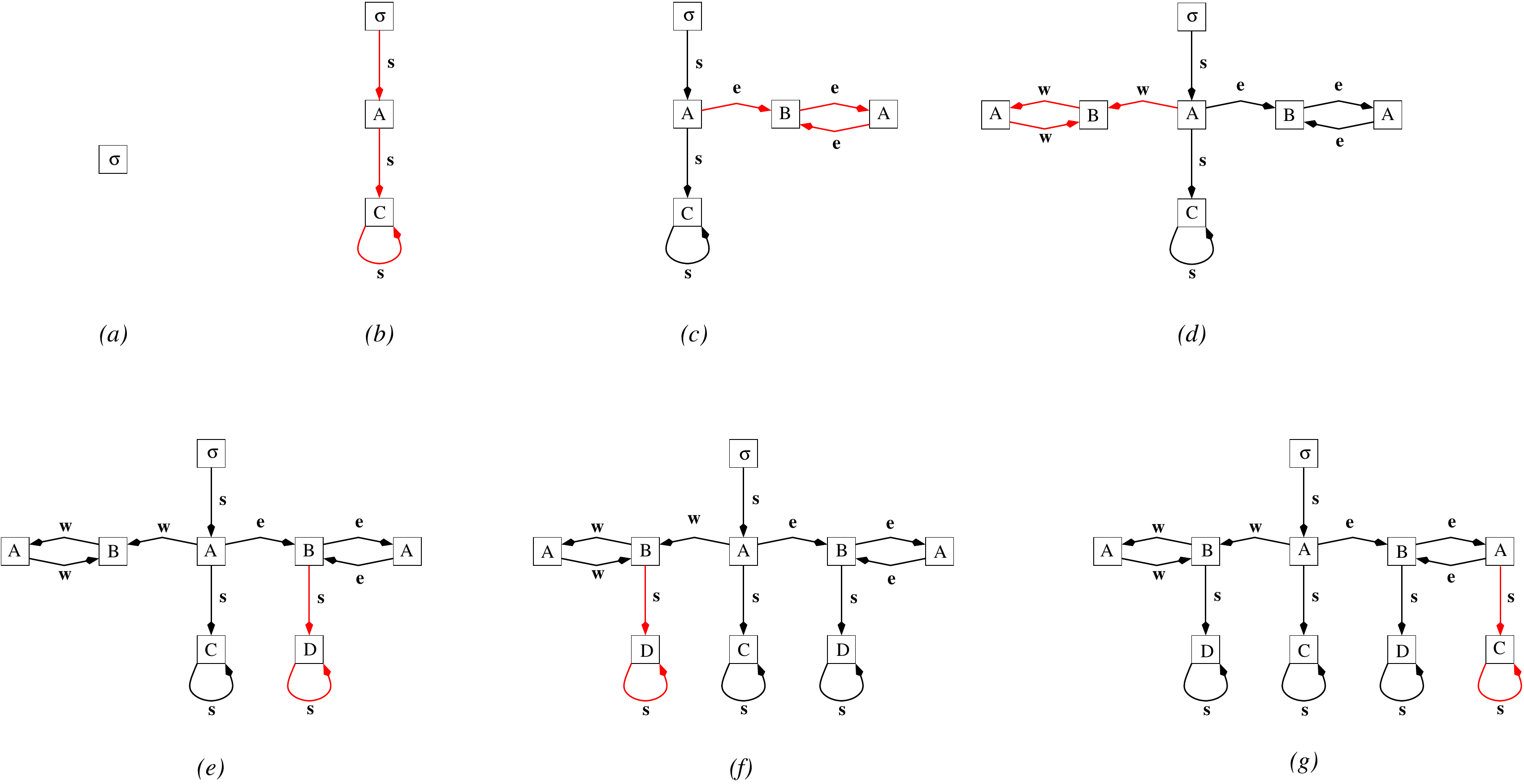}
    \caption{A build-up of a quipu for TAS \TAS from Example~\ref{example:confl}. (a) 
      \QuipuZero, (b) $\Quipu_1=\Quipu_0[\South\South^{\omega}]$,(c) $\Quipu_2=\Quipu_1[\South\East^{\omega}]$, (d) $\Quipu_3=\Quipu_1[\South\West^{\omega}]$, (e) $\Quipu_4=\Quipu_3[\South\East\South^{\omega}]$, (f) $\Quipu_5=\Quipu_4[\South\West\South^{\omega}]$, (g) $\Quipu_6=\Quipu_5[\South\East\East\South^{\omega}]$}
    \label{fig:quipu-example}
  \end{figure}
  
  In order of appearance of the next cycles are 
  $\South\East^\omega$, and $\South\West^\omega$ giving extensions $\Quipu_2=\Quipu_1[\South\East^{\omega}]$, and $\Quipu_3=\Quipu_1[\South\West^{\omega}]$.
  Note that in order to get $\Quipu_3$ the cycle $A\East B\East A$ must be unrolled. Because $\South(\South\East)^\omega$, $\South\South\East^\omega$ and $\South(\East\South)^\omega$ are not in \AlphaMax (similarly with using $\West$ instead of $\East$)
  the next path to consider is 
  $\South\East\South^\omega$. Only one vertex in $\Quipu_3$ corresponds to an end of a path in \AlphaMax with label $\South\East$. This is vertex labeled with $B$ (reached from the root by traveling $\South\East$). Hence, 
  $\Quipu_4=\Quipu_3[\South\East\South^{\omega}]$, and similarly $\Quipu_5=\Quipu_4[\South\West\South^{\omega}]$. 
  The next cycles in the order are $\South\East\East\South^\omega$ and $\South\West\West\South^\omega$, hence $\Quipu_6=\Quipu_5[\South\East\East\South^{\omega}]$ and $\Quipu_7$ is the quipu in \RefFig{fig:path-automaton}. The process stops because no other tile can be added. 
\end{example}

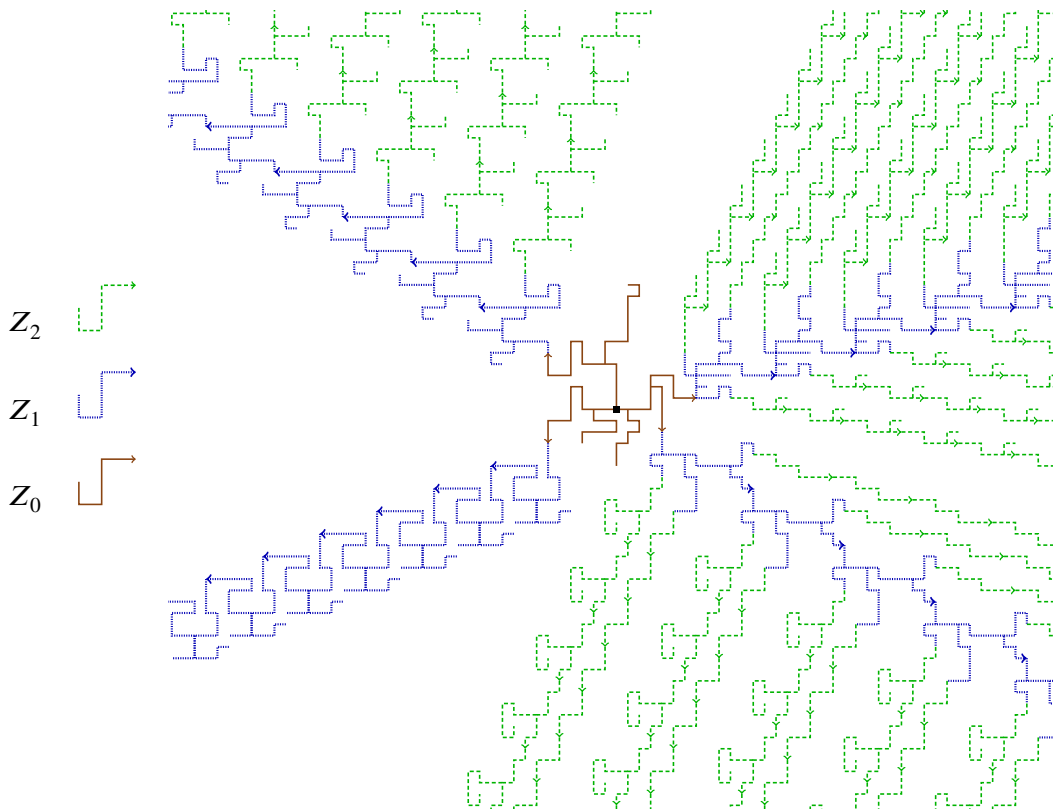
\begin{figure}[hbt]
  \newcommand{\ScaleFactor}{.5}
  \newcommand{\LibLine}[2]{
    #1 &
    \scalebox{\ScaleFactor}{
      \begin{tikzpicture}
        \draw[#2] (0,0) \PathLib ; 
      \end{tikzpicture}} \\[1em]
  }
  \PreparePath{Lib}{SSEENNNNEEE}\SetUnitlength{3mm}\small\centering\begin{tabular}[c]{cc}\LibLine{\ZoneTwo}{struct-q}\LibLine{\ZoneOne}{struct-p}\LibLine{\ZoneZero}{struct-m}\end{tabular}\begin{tabular}[c]{cc}\scalebox{\ScaleFactor}{\begin{tikzpicture}\path[clip] (-39.25,-35.25) rectangle (38.25,35.25) ;\draw (0,0) \CoorNode{D};\begin{scope}[struct-m,-]\Deco{-2,0}{SEESWWWS}\Deco{1,0}{SESWSWSS}\Deco{-1,4}{NNEENNNNENW}\end{scope}\CombBB[6]{0,0}{EEENNNEESSEE}{NNNEEEESEEE}{\Deco{0,2}{EEE}\Deco{0,1}{E}\ProtoTooth[9]{0,2}{WNN}{NNNNEE}{\Deco{0,4}{NENN}}\ProtoTooth[4]{2,3}{NENWNNEN}{NNENNENNENNE}{}\ProtoTooth[6]{0,0}{EENES}{EESEEE}{\Deco{4,-1}{NE}}}\CombBB[6]{3,2}{ESSSS}{SSEEESEEESESE}{\ProtoTooth[4]{6,-3}{ENNES}{EESEESEEE}{}\ProtoTooth[6]{0,-2}{WSES}{SWSSWSSWS}{\Deco{-2,-3}{WWNWSSSEN}} \ProtoTooth[7]{2,-2}{SSESSSWW}{SSWWSS}{} }\CombBB{0,0}{WWWNNWSSSWWSS}{SSSESSWWWWNNEENWWWW}{\Deco{-1,-5}{SSEENE}\Deco{-1,-5}{SSWW}}\CombBB[6]{0,0}{NNNNWWWNNWSSSWWNN}{NWWWWNNEENWWWW}{\Deco{-4,2}{WWWN}\Deco{-3,1}{SWWSE}\ProtoTooth{-2,4}{EEENNWSWWNN}{NNWNEEENNN}{\Deco{2,5}{EEEN}\Deco{2,3}{EES}}}\draw (0,0) node [rectangle,fill,inner sep=.3\unitlength] {} ;\end{tikzpicture}}\end{tabular}
  \caption{Example of a union of combs.}
  \label{fig:comb-union}
\end{figure}

\section{Conclusion}
\label{sec:conclusion}

The notion of para-periodicity can be considered in all aTAM systems.
We showed that every confluent (deterministic) temperature 1 TAS \TAS is para-periodic.
In this proof we used notions of `left' and `right' regions of a bi-infinite path.
Being in two dimensions, the Jordan curve theorem ensures that a bi-infinite path divides the plane in two regions.
In our case, the regions are connected subgraphs of $\IntegerSet^2$ and both regions contain the path itself.
The notions of left and right of a path are also used in \cite{meunier+regnault16arxiv,meunier+woods17stoc}, except in our case the paths are bi-infinite and are not necessarily part of \AlphaMax.
The other tool developed for the proof is the co-grow of two paths, which is a superposition of the two paths that takes the right most turns of the two paths.
This tool provided a way to identify, or construct, an ultimately periodic path in \AlphaMax.
In order to co-grow two paths we relied on the confluence of the system.

It may be of interest to characterize para-periodic systems that are not necessarily confluent, nor at temperature 1, or even in higher dimensions.
The co-grow can be applied in any confluent two-dimensional system.
We believe that the para-periodicity is not decidable in non-confluent systems, following the result of undecidability of growing infinite ribbons \cite{adleman+cheng+goel+huang+kempe+moisset+rothemund02stoc}.
Characterizing para-periodic temperature 2 systems might be of interest. 
Confluent temperature 2 systems may not be para-periodic (e.g. \cite{Patitz-fractal}), however, it would be quite interesting if para-periodic temperature 2 systems (confluent or not) exhibit terminal structures different than grids or combs. 

We identified two basic structures that an infinite \AlphaMax can have, a grid, or a finite union of combs.
These two structures were identified also in \cite{doty+patitz+summers11tcs} where the authors discussed the pumpability of paths. 
The notion of para-periodic (that is, the existence of an ultimate periodic path, rather than pumpability of the path itself) allowed us to distance ourselves from discussing one specific path within \AlphaMax, and to concentrate on the structure of the whole assembly. 
We observe that if \AlphaMax is not a grid, then every infinite path is pumpable (as it is part of a comb).
We believe this to be the case for the grid as well, although we do not focus on that here.
Hence, in the case of a finite union of combs, we have that every infinite path is pumpable, i.e., the domain of \AlphaMax is semi-linear.
Moreover, grids, by definition are semi-linear, hence all \AlphaMax of confluent temperature 1 systems are semi-linear sets. 

The proof that confluent temperature 1 systems do not have universal computational power follows from the construction of a finite quipu for every terminal maximal assembly in such systems.
Quipu is a specific automaton associated with a given \AlphaMax that contains a cycle for every ultimately periodic path.
It has a tree like structure such that each leaf is either a cycle or an end of a path such that \AlphaMax is the union of all paths generated by the quipu. 
The finiteness of the quipu construction implies that temperature 1 TAS \TAS cannot have universal computing power, because otherwise, something about its maximal quipu should not be decidable (\RefCor{cor:not-compute}). 
The quipu associated with \AlphaMax is obtained in a constructive way. 
Since \TAS is para-periodic, we start by adding a cycle to the quipu that corresponds to the shortest length transient and periodic part of a path.
The quipu extension step is algorithmic. 
At each stage, the quipu is finite, and the set of paths in $\IntegerSet^2$ that are generated by the quipu is a union of semi-linear sets.
Hence, the intersection of any new path with paths generated by the existent quipu can be computed.
The next ultimately-periodic path from \AlphaMax to incorporate within the quipu is obtained by breadth first search on the length of the transient and periodic parts of the path. 
In order to use the breadth first search, at every step we also include all of the decorations in \AlphaMax and keep track of their locations in $\IntegerSet^2$.
We show that this process necessarily stops, hence the quipu is finite.
If at any time during the process it is identified that the conditions for a grid are satisfied, the construction of the quipu stops and we conclude that \AlphaMax is a grid.
If a characterization of para-periodic temperature 2 confluent systems can be obtained, the concept of quipu may be generalized to capture the terminal maximal assemblies in these systems as well.

It will be of interest to see whether the construction of the quipu and the process of breadth first search can lead to an algorithm to determine whether a TAS \TAS is confluent at temperature 1.
It is clear that this is not decidable if TAS is in three dimensions, because such systems have universal computing power \cite{cook+fu+schweller11soda}.
However, the nice and relatively simple structures that \AlphaMax can achieve, may lead to decidability in two dimensions. 

\paragraph{Acknowledgement.}
NJ has been supported in part by the NSF grant CCF-1526485
and the NIH grant R01 GM109459.
We kindly acknowledge institutional support of the University of South Florida in 2017 to JDL and HJH whose visit initiated this work, and University of Orl\'eans visiting professorship grant to NJ in 2018.

\small

\providecommand{\href}[2]{#2}\makeatletter
  \@ifundefined{mathbb}{\long\def\mathbb{\mathsf}} \makeatother

\clearpage
\appendix

\tableofcontents

\JDLvocabularyTableofsymbols


\begin{thebibliography}{24}
\providecommand{\natexlab}[1]{#1}
\providecommand{\url}[1]{\texttt{#1}}
\expandafter\ifx\csname urlstyle\endcsname\relax
  \providecommand{\doi}[1]{doi: #1}\else
  \providecommand{\doi}{doi: \begingroup \urlstyle{rm}\Url}\fi

\bibitem[Adleman et~al.(2002{\natexlab{a}})Adleman, Cheng, Goel, Huang, Kempe,
  Moisset~de Espan{\'{e}}s, and
  Rothemund]{adleman+cheng+goel+huang+kempe+moisset+rothemund02stoc}
Leonard~M. Adleman, Qi~Cheng, Ashish Goel, Ming{-}Deh~A. Huang, David Kempe,
  Pablo Moisset~de Espan{\'{e}}s, and Paul W.~K. Rothemund.
\newblock Combinatorial optimization problems in self-assembly.
\newblock In John~H. Reif, editor, \emph{Proceedings on 34th Annual {ACM}
  Symposium on Theory of Computing ({STOC} 2002), May, 2002, Montr{\'{e}}al,
  Qu{\'{e}}bec, Canada}, pages 23--32. {ACM}, 2002{\natexlab{a}}.
\newblock \doi{10.1145/509907.509913}.

\bibitem[Adleman et~al.(2002{\natexlab{b}})Adleman, Kari, Kari, and
  Reishus]{adleman+kari+kari+reishus02focs}
Leonard~M. Adleman, Jarkko Kari, Lila Kari, and Dustin Reishus.
\newblock On the decidability of self-assembly of infinite ribbons.
\newblock In \emph{43rd Symposium on Foundations of Computer Science ({FOCS}
  2002), November 2002, Vancouver, BC, Canada, Proceedings}, pages 530--537.
  {IEEE} Computer Society, 2002{\natexlab{b}}.
\newblock \doi{10.1109/SFCS.2002.1181977}.

\bibitem[Adleman et~al.(2009)Adleman, Kari, Kari, Reishus, and
  Sos{\'{\i}}k]{adleman+kari+kari+reishus+sosik09siam}
Leonard~M. Adleman, Jarkko Kari, Lila Kari, Dustin Reishus, and Petr
  Sos{\'{\i}}k.
\newblock The undecidability of the infinite ribbon problem: Implications for
  computing by self-assembly.
\newblock \emph{{SIAM} J. Comput.}, 38\penalty0 (6):\penalty0 2356--2381, 2009.
\newblock \doi{10.1137/080723971}.

\bibitem[Behsaz et~al.(2012)Behsaz, Ma{\v n}uch, and
  Stacho]{stage-temp1-assembly}
Bahar Behsaz, J{\'a}n Ma{\v n}uch, and Ladislav Stacho.
\newblock Turing universality of step-wise and stage assembly at temperature 1.
\newblock In Darko Stefanovic and Andrew Turberfield, editors, \emph{DNA
  Computing and Molecular Programming}, pages 1--11, Berlin, Heidelberg, 2012.
  Springer Berlin Heidelberg.
\newblock ISBN 978-3-642-32208-2.

\bibitem[Brijder and Hoogeboom(2009)]{brijder+hoogeboom09tcs}
Robert Brijder and Hendrik~Jan Hoogeboom.
\newblock Perfectly quilted rectangular snake tilings.
\newblock \emph{Theoret Comp Sci}, 410\penalty0 (16):\penalty0 1486--1494,
  2009.
\newblock \doi{http://dx.doi.org/10.1016/j.tcs.2008.12.010}.

\bibitem[Chakraborty et~al.(2012)Chakraborty, Jonoska, and Seeman]{Banani}
Banani Chakraborty, Natasha Jonoska, and Nadrian~C. Seeman.
\newblock A programmable transducer self-assembled from {DNA}.
\newblock \emph{Chem. Sci.}, 3:\penalty0 168--176, 2012.

\bibitem[Cook et~al.(2011)Cook, Fu, and Schweller]{cook+fu+schweller11soda}
Matthew Cook, Yunhui Fu, and Robert~T. Schweller.
\newblock Temperature 1 self-assembly: Deterministic assembly in 3d and
  probabilistic assembly in 2d.
\newblock In Dana Randall, editor, \emph{Proceedings of the Twenty-Second
  Annual {ACM-SIAM} Symposium on Discrete Algorithms, {SODA} 2011, San
  Francisco, California}, pages 570--589. {SIAM}, 2011.
\newblock \doi{10.1137/1.9781611973082.45}.

\bibitem[Doty et~al.(2011)Doty, Patitz, and Summers]{doty+patitz+summers11tcs}
David Doty, Matthew~J. Patitz, and Scott~M. Summers.
\newblock Limitations of self-assembly at temperature 1.
\newblock \emph{Theoret Comp Sci}, 412\penalty0 (1--2):\penalty0 145--158,
  2011.
\newblock \doi{10.1016/j.tcs.2010.08.023}.

\bibitem[Doty et~al.(2012)Doty, Lutz, Patitz, Schweller, Summers, and
  Woods]{IntrUniver}
David Doty, Jack~H. Lutz, Matthew~J. Patitz, Robert~T. Schweller, Scott~M.
  Summers, and Damien Woods.
\newblock The tile assembly model is intrinsically universal.
\newblock In \emph{53rd Annual {IEEE} Symposium on Foundations of Computer
  Science, {FOCS} 2012, New Brunswick, NJ, USA, October, 2012}, pages 302--310.
  {IEEE} Computer Society, 2012.
\newblock ISBN 978-1-4673-4383-1.
\newblock \doi{10.1109/FOCS.2012.76}.

\bibitem[Evans(2015)]{Evans2015}
Constantine~Glen Evans.
\newblock \emph{Crystals that count! Physical principles and experimental
  investigations of DNA tile self-assembly}.
\newblock PhD thesis, California Institute of Technology, June 2015.

\bibitem[Fujibayashi et~al.(2008)Fujibayashi, Hariadi, Park, Winfree, and
  Murata]{Fujibayashi2008}
Kenichi Fujibayashi, Rizal Hariadi, Sung~Ha Park, Erik Winfree, and Satoshi
  Murata.
\newblock Toward reliable algorithmic self-assembly of {DNA} tiles: A
  fixed-width cellular automaton pattern.
\newblock \emph{NanoLetters}, 8:\penalty0 1791--1797, 2008.

\bibitem[Kari(2002)]{kari02dlt}
Jarkko Kari.
\newblock Infinite snake tiling problems.
\newblock In Masami Ito and Masafumi Toyama, editors, \emph{Developments in
  Language Theory, 6th Int. Conf., {DLT} 2002, Kyoto, Japan, September, 2002,
  Revised Papers}, volume 2450 of \emph{LNCS}, pages 67--77. Springer, 2002.
\newblock \doi{10.1007/3-540-45005-X_6}.

\bibitem[Karpenko(2015)]{Daria2015}
Daria Karpenko.
\newblock \emph{Active Tile Self-assembly and Simulations of Computational
  Systems}.
\newblock PhD thesis, University of South Florida, May 2015.

\bibitem[LaBean et~al.(2000)LaBean, Yan, Kopatsch, Liu, Winfree, Reif, and
  Seeman]{LaBean2000}
Thomas~H. LaBean, Hao Yan, Jens Kopatsch, Furong Liu, Erik Winfree, John~H.
  Reif, and Nadrian~C. Seeman.
\newblock Construction, analysis, ligation, and self-assembly of dna triple
  crossover complexes.
\newblock \emph{J. Am. Chem. Soc.}, 122:\penalty0 1848--1860, 2000.

\bibitem[Meunier and Regnault(2016)]{meunier+regnault16arxiv}
Pierre{-}{\'{E}}tienne Meunier and Damien Regnault.
\newblock A pumping lemma for non-cooperative self-assembly.
\newblock arXiv 1610.07908, 205 pages, 2016.
\newblock URL \url{http://arxiv.org/abs/1610.07908}.

\bibitem[Meunier and woods(2017)]{meunier+woods17stoc}
Pierre{-}{\'{E}}tienne Meunier and Damian woods.
\newblock The non-cooperative tile assembly model is not intrinsically
  universal or capable of bounded turing machine simulation.
\newblock In Hamed Hatami, Pierre McKenzie, and Valerie King, editors,
  \emph{Proceedings of the 49th Annual {ACM} {SIGACT} Symposium on Theory of
  Computing, {STOC} 2017, Montreal, QC, Canada, June, 2017}, pages 328--341.
  {ACM}, 2017.
\newblock \doi{10.1145/3055399.3055446}.
\newblock URL \url{http://doi.acm.org/10.1145/3055399.3055446}.

\bibitem[Patitz and Summers(2010)]{Patitz-fractal}
Matthew Patitz and Scott Summers.
\newblock Self-assembly of discrete self-similar fractals.
\newblock \emph{Natural Computing}, 9:\penalty0 135--172, 2010.

\bibitem[Patitz et~al.(2011)Patitz, Schweller, and Summers]{negative-glue}
Matthew~J. Patitz, Robert~T. Schweller, and Scott~M. Summers.
\newblock Exact shapes and turing universality at temperature 1 with a single
  negative glue.
\newblock In Luca Cardelli and William~M. Shih, editors, \emph{{DNA} Computing
  and Molecular Programming - 17th International Conference, {DNA} 17,
  Pasadena, CA, USA, September, 2011. Proceedings}, volume 6937 of \emph{LNCS},
  pages 175--189. Springer, 2011.
\newblock \doi{10.1007/978-3-642-23638-9\_15}.
\newblock URL \url{https://doi.org/10.1007/978-3-642-23638-9\_15}.

\bibitem[Patitz(2014)]{Patitz-survey}
MatthewJ. Patitz.
\newblock An introduction to tile-based self-assembly and a survey of recent
  results.
\newblock \emph{Natural Computing}, 13\penalty0 (2):\penalty0 195--224, 2014.
\newblock ISSN 1567-7818.

\bibitem[Rothemund and Winfree(2000)]{rothemund+winfree00stoc}
Paul W.~K. Rothemund and Erik Winfree.
\newblock The program-size complexity of self-assembled squares (extended
  abstract).
\newblock In Frances~F. Yao and Eugene~M. Luks, editors, \emph{Proceedings of
  the 32nd Annual {ACM} Symposium on Theory of Computing ({STOC})}, pages
  459--468. {ACM}, 2000.
\newblock \doi{10.1145/335305.335358}.

\bibitem[Rothemund et~al.(2004)Rothemund, Papadakis, and Winfree]{Sierpinski}
Paul~W.K. Rothemund, Nick Papadakis, and Erik Winfree.
\newblock Algorithmic self-assembly of {DNA} sierpinski triangles.
\newblock \emph{{PLoS} Biology}, 2\penalty0 (12):\penalty0 e424, 2004.

\bibitem[Wang(1975)]{Wang1975}
Hao Wang.
\newblock Notes on a class of tiling problems.
\newblock \emph{Fundamenta Mathematicae}, 82\penalty0 (4):\penalty0 295--305,
  1975.
\newblock URL \url{http://eudml.org/doc/214668}.

\bibitem[Winfree(1998)]{Winf98}
Erik Winfree.
\newblock \emph{Algorithmic Self-Assembly of DNA}.
\newblock PhD thesis, California Institute of Technology, June 1998.

\bibitem[Winfree et~al.(1998)Winfree, Liu, Wenzler, and Seeman]{Ferong}
Erik Winfree, Furong Liu, Lisa~A. Wenzler, and Nadrian~C. Seeman.
\newblock Design and self-assembly of two-dimensional {DNA} crystals.
\newblock \emph{Nature}, 394:\penalty0 539--544, 1998.

\end{thebibliography}
\end{document}